\documentclass[twocolumn,english,pra]{revtex4-1}
\usepackage{color}
\usepackage[utf8]{inputenc}
\usepackage{physics}
\usepackage{graphicx}
\usepackage[T1]{fontenc}
\usepackage{float}
\usepackage{natbib}
\usepackage{textcomp}
\usepackage{amsmath}
\usepackage{amssymb}
\usepackage{graphicx}
\usepackage{esint}
\usepackage{natbib}
\usepackage{xcolor}
\usepackage{multirow}
\usepackage{ulem}

\begin{document}
\title{Merging and oscillations of dipolar Bose-Einstein condensate droplets}
%structure and dynamics within a dipolar crystal for merged double well potential

\author{Wojciech Or\l{}owski and Bart\l{}omiej Szafran}

\affiliation{AGH University, Faculty of Physics and
Applied Computer Science,\\
 al. Mickiewicza 30, 30-059 Kraków, Poland}

\begin{abstract}
We investigate the dynamics of Bose-Einstein condensate droplets composed of $^{164}$Dy atoms
formed in a double-well potential following removal of the interwell barrier. By solving the dipolar
Gross-Pitaevskii equation, we determine phase diagrams of ground-state configurations as functions
of the atom number confined in the double-well potential.
{
For an interwell separation $2d=3\,\mu\mathrm{m}$, we consider the relative dipolar-to-contact interaction strengths
$\varepsilon_{dd}=1.4$, $1.45$, and $1.5$. Symmetry-broken lowest-energy configurations are found
for $\varepsilon_{dd}=1.5$ and, over a reduced range of atom numbers, for
$\varepsilon_{dd}=1.45$, whereas no symmetry-broken states are found for
$\varepsilon_{dd}=1.4$ up to $N=5\times10^4$.
}
We analyze the subsequent time evolution after removal of the central barrier, revealing both droplet
oscillations and merger events leading to the formation of larger droplets. The oscillations are driven
by the external potential and by the repulsive tails of the in-plane component of the dipolar interaction.
{
For the two-droplet states at $\varepsilon_{dd}=1.5$ and $2d=3\,\mu\mathrm{m}$, merging is found
for $N=8000$ and $8800$, whereas the droplets remain separated for the investigated cases with
$N\geq9000$. Increasing the initial separation to $2d=5\,\mu\mathrm{m}$ shifts the approximate
upper atom number for merging to $N\simeq1.65\times10^4$, demonstrating that the crossover is
controlled by the competition between the post-quench excess energy and the repulsive inter-droplet
barrier.
}

\end{abstract}
\maketitle

\section{Introduction}
In Bose–Einstein condensates of ultracold dipolar atomic gases, long-range dipole–dipole interactions give rise to inhomogeneous structures with modulated condensate density, up to the formation of supersolid phases composed of separated dipolar droplets \cite{naturedroplet, prxdroplet, silamu}.
Such droplets emerge in magnetic condensates in the regime of strong dipole–dipole interactions \cite{naturedroplet, prxdroplet, silamu}. They are stabilized by the attractive component of the dipolar interaction along the direction of spin polarization \cite{yahoo, santos}, which competes with the contact interaction.

The number of atoms self-bound within a droplet is limited by the repulsive character of the in-plane dipole–dipole interaction. For larger atom numbers, this leads to the formation of multiple-droplet states \cite{prldroplet, rosendroplet,striped2, striped, multidroplets, santos2}, including droplet crystals interacting via the long-range dipolar potential \cite{dropletcrystal, dr0, drm1, dropletcr4,drc1, dropletcr2, dropletcr3, drcr4}.
Extended systems with spatially ordered, density-modulated structures and long-range phase coherence form supersolid phases \cite{supersolid, supersolid2, supersolidlate, general, striped, review, review2, review3, merging_1d, gauss_sweep}.
In strongly dipolar condensates of the highly magnetic isotope $^{164}$Dy  \cite{164Dy}, with a magnetic dipole moment close to $10\mu_B$, coherence can be maintained for long times \cite{prldroplet}, reaching up to about one second in systems produced via direct evaporation \cite{death}.

The structure of droplet crystals has been extensively studied in recent years \cite{dropletcrystal, dropletcr2, dropletcr3, drcr4}, including configurations in multiple-well potentials \cite{multicond, merging} and interactions between separate droplet crystals \cite{multicond}.
The dynamics of dipolar systems, including droplet states, have also been investigated, for example in the context of semiclassical vibrations of linear chains \cite{oscila2}, forced oscillations induced by compression \cite{forced}, and periodic modulation of the optical trapping potential \cite{oscila}.
Transport of droplet “molecules” in a double-well potential driven by a time-dependent bias \cite{dropletsall}, as well as inelastic collisions between droplets in quasi-one-dimensional geometries, have recently been studied \cite{chiny, spania}.
Droplets can also be realized in attractive atomic mixtures \cite{dropmix1, dropmix2, dropmix3}, and their collisional properties have been explored \cite{dropzd, dropzd2}. In such systems, collisions occur without an inter-droplet barrier associated with the long-range anisotropic dipolar interaction.

In this work, we study droplet configurations formed in a double-well potential in the regime of strong dipolar interactions,
%$\varepsilon_{dd} \in [1.4,1.5]$, 
and analyze their dynamics following a sudden removal of the central barrier. The system is described by the extended Gross–Pitaevskii equation \cite{general, santos2, dropletsall}, including the Lee–Huang–Yang (LHY) correction \cite{lhy1, lhy0,lhy2} accounting for quantum fluctuations.

We characterize the lowest-energy configurations in the double-well potential, including droplet crystals and pillar-like droplet structures, as functions of the total atom number, thereby constructing phase diagrams of the system.
We further analyze the time evolution after barrier removal, including droplet merging processes that overcome the dipolar interaction barrier, as well as oscillatory dynamics driven by the repulsive tail of the dipolar interaction.
The oscillation dynamics and their periodicity depend on both the atom number and the initial droplet separation. We find that the lifetime of the oscillations is reduced by breathing modes of individual droplets, excited during near-contact droplet–droplet collisions.
The oscillation period is strongly influenced by the spatial symmetry of the initial droplet configuration and is generally more complex for systems with lower symmetry.

This paper is organized as follows. In Sec.~II, we introduce the extended Gross–Pitaevskii equation used in this work. Section~III presents the results and discussion. Finally, Sec.~IV contains the summary and conclusions.

\section{Theory}

We consider a three-dimensional confined condensate of ultracold $^{164}$Dy atoms, described by the extended Gross-Pitaevskii Hamiltonian \cite{general, santos2, dropletsall, prl25, supersolidlate}
\begin{eqnarray}
H_{\mathrm{eGP}} &=& -\frac{\hbar^2}{2m}\nabla^2 + V_{\mathrm{ext}}({\bf r},t) + g |\Psi({\bf r},t)|^2 \nonumber \\
&+& \gamma(\varepsilon_{dd}) |\Psi({\bf r},t)|^3 + V_{dd}({\bf r},t), \label{hgp}
\end{eqnarray}
where $\Psi({\bf r},t)$ is the condensate wave function.

The third term in Eq.~(\ref{hgp}) describes the contact interaction, with $g =  {\frac{4\pi \hbar^2 a}{m}}$, where $a$ is the $s$-wave scattering length tuned via a Feshbach resonance \cite{fes1, fes2, fes3}. We take the atomic mass of the $^{164}$Dy isotope to be $m = 163.929$ Da.

The term proportional to $\gamma$ accounts for the Lee-Huang-Yang (LHY) correction due to quantum fluctuations \cite{lhy1, lhy2}, with
\begin{equation}
\gamma(\varepsilon_{dd}) = \frac{128\sqrt{\pi a^5}}{3m}\left(1 + \frac{3}{2}\varepsilon_{dd}^2\right),
\end{equation}
where $\varepsilon_{dd} = a_{dd}/a$ and the dipolar length is defined as $a_{dd} = \frac{\mu_0 \mu_B m}{12\pi \hbar^2}$. For $^{164}$Dy, $a_{dd} = 131a_B$.

The dipole-dipole interaction term $V_{dd}$ in Eq.~(\ref{hgp}) is given by
\begin{equation}
V_{dd}({\bf r},t) = \int d^3{\bf s}\, n({\bf s},t)\, U_{dd}({\bf r}-{\bf s}),
\end{equation}
where the density $n({\bf s},t) = |\Psi({\bf s},t)|^2$ is normalized to the total number of atoms $N$. The dipolar interaction kernel reads
\begin{equation}
U_{dd}({\bf r}) = \frac{c_{dd}}{4\pi} \frac{1 - 3\cos^2\theta}{|{\bf r}|^3}, \label{uj}
\end{equation}
where $c_{dd} = { \frac{12\pi \hbar^2 a_{dd}}{m}}$ and $\theta$ is the angle between the polarization axis (taken along $z$) and the relative position vector.

The external confinement potential $V_{\mathrm{ext}}$ in Eq.~(\ref{hgp}) is assumed to be separable in the $x$, $y$, and $z$ directions,
\begin{equation}
V_{\mathrm{ext}} = V_x(x) + V_y(y) + V_z(z),
\end{equation}
with harmonic confinement in the $y$ and $z$ directions,
\begin{equation}
V_y = \frac{m}{2}\omega^2 y^2, \quad V_z = \frac{m}{2}\omega_z^2 z^2,
\end{equation}
where ${ \omega = 2\pi \cdot  60 \;\frac{1}{\mathrm{s}}}$ and ${\omega_z = 2\pi \cdot 120 \;\frac{1}{\mathrm{s}}}${, with the corresponding harmonic oscillator energies $\hbar\omega=2.88$ nK and $\hbar\omega_z=5.76$ nK, respectively.}

Along the $x$ direction, we consider a double-well potential of the form of a Mexican-hat potential,
\begin{equation}
V_x(x) = \alpha x^4 - \beta x^2,
\end{equation}
with $\beta = \frac{m\omega^2}{2}$ and $\alpha = \frac{m\omega^2}{4d^2}$, where $2d$ denotes the distance between the minima of the double-well potential.

\begin{figure}[htbp]
            \includegraphics[width=0.25\textwidth]{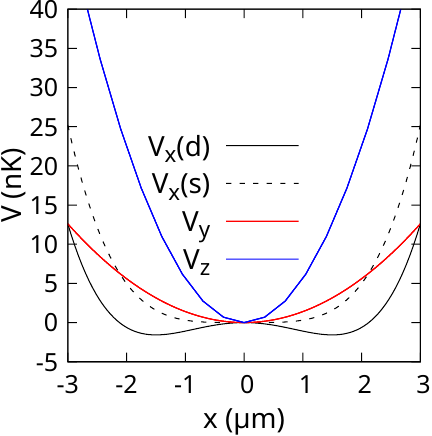}
            \caption{The applied model potentials for the $y$ and $z$ directions with ${\omega_y=2\pi \cdot 60 \frac{1}{\mathrm{s}}}$ (red line) and ${\omega_z=2\pi \cdot 120 \frac{1}{\mathrm{s}}}$ (blue line). The potential for $x$ is plotted for a single (dashed black line) and double well (solid black line) for $2d=3\mu$m.}
   \label{pot0}
   \end{figure}

Figure~\ref{pot0} shows the components of the trapping potential along the three spatial directions, including the double-well potential $V_x(d)$ (solid black line) and the corresponding single-well potential $V_x(s)$ (dashed black line) for $2d = 3\,\mu\mathrm{m}$.

For the initial condition, we take the ground state of the system in the double-well potential. It is obtained by solving the Gross-Pitaevskii equation
\begin{equation}
i\hbar \frac{\partial \Psi}{\partial t} = H_{\mathrm{eGP}} \Psi({\bf r},t)
\end{equation}
on a three-dimensional finite-difference grid using the imaginary-time evolution method.  

To study the system dynamics, we use the ground-state density of the double-well potential as the initial state and subsequently remove the interwell barrier by setting $\beta = 0$ in the potential $V_x$. This procedure is equivalent to changing the potential in the $x$ direction from the solid curve to the dashed curve in Fig.~\ref{pot0}.  
The time evolution is then obtained by solving the extended Gross-Pitaevskii equation in real time. We used both the Crank-Nicolson scheme and the split-step method. The former as an implicit scheme requires convergence to be reached at each time step between the interaction potential and the atom density and hence  multiple evaluation of the interaction potentials for the iterated density. The latter works faster due to its explicit nature, and we found that for our study that the time-step required for a precise iteration is similar for both these methods, so a potential advantage of the better stability of the Crank-Nicolson cannot be used. The details of the split-step method, and the calculation parameters as well as the precision tests are given in the appendices.

Below, we analyze the contributions to the condensate energy, defined as
\begin{equation}
E = T(t) + E_{\mathrm{ext}}(t) + E_{\mathrm{int}}(t),
\end{equation}
where the kinetic energy is given by
\begin{equation}
T = -\int d^3{\bf r}\, \Psi^*({\bf r},t)\frac{\hbar^2 \nabla^2}{2m}\Psi({\bf r},t),
\end{equation}
the external potential energy is
\begin{equation}
E_{\mathrm{ext}} = \int d^3{\bf r}\, n({\bf r},t)\, V_{\mathrm{ext}}({\bf r},t),
\end{equation}
and the interaction energy reads
\begin{eqnarray}
E_{\mathrm{int}} &=& \int d^3{\bf r} \left[ \frac{g}{2} {n({\bf r},t)^2} + \frac{2\gamma}{5} n({\bf r},t)^{5/2} \right. \nonumber \\
&& \left. + \frac{1}{2} V_{dd}({\bf r},t)\, n({\bf r},t) \right],
\end{eqnarray}
which includes, respectively, the contact interaction, the LHY correction, and the dipole-dipole interaction.

%trim=left bottom right top
\begin{figure}[htbp]
    \begin{tabular}{l}
        \includegraphics[width=0.35\textwidth]{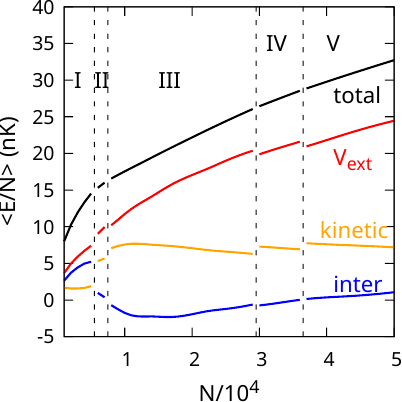}\put(-25,35){(a)} \\
        \includegraphics[width=0.35\textwidth]{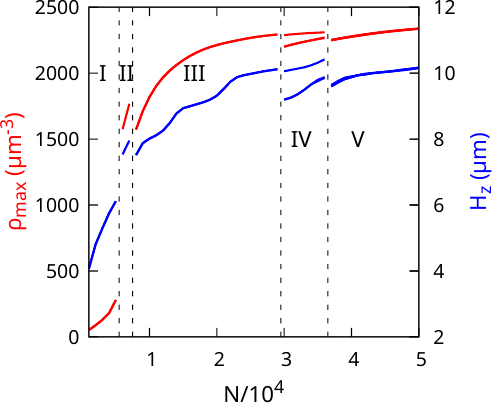} \put(-40,30){(b)}
    \end{tabular}  
    \begin{tabular}{lll}
        \includegraphics[width = 0.15\textwidth, trim={3.0cm 0cm 0cm 0cm}, clip]{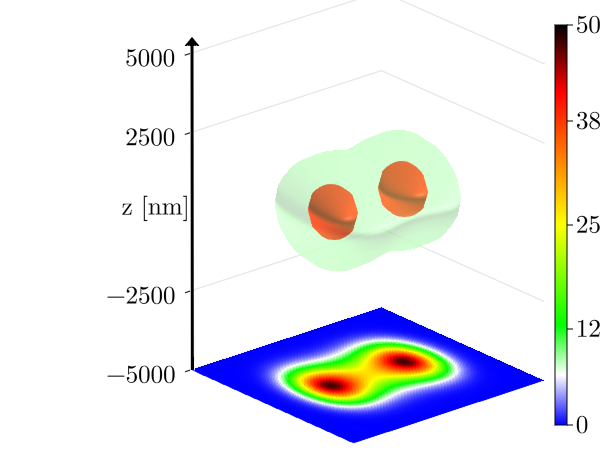} \put(-18.5, 1){(c)}&
        \includegraphics[width = 0.15\textwidth, trim={4cm 0cm 0cm 0cm}, clip]{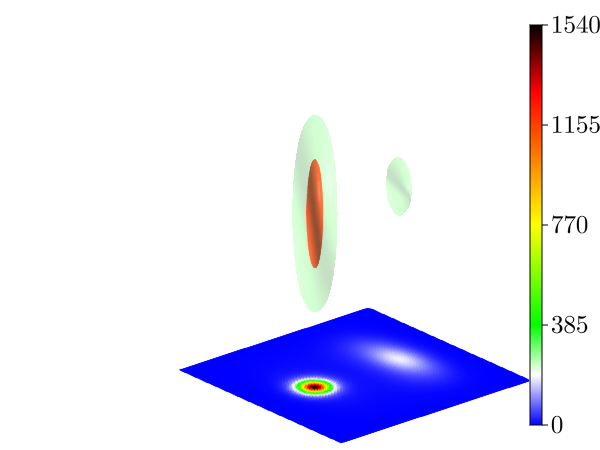} \put(-24.5,1){(d)}&
        \includegraphics[width = 0.15\textwidth, trim={4cm 0cm 0cm 0cm}, clip]{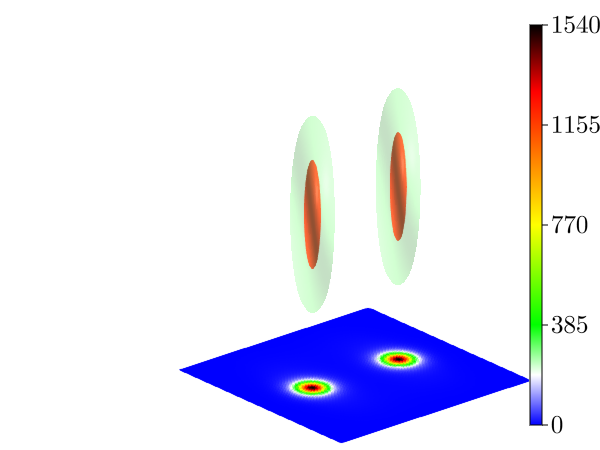} \put(-24.5,1){(e)}\\
        \includegraphics[width = 0.15\textwidth, trim={4cm 0cm 0cm 0cm}, clip]{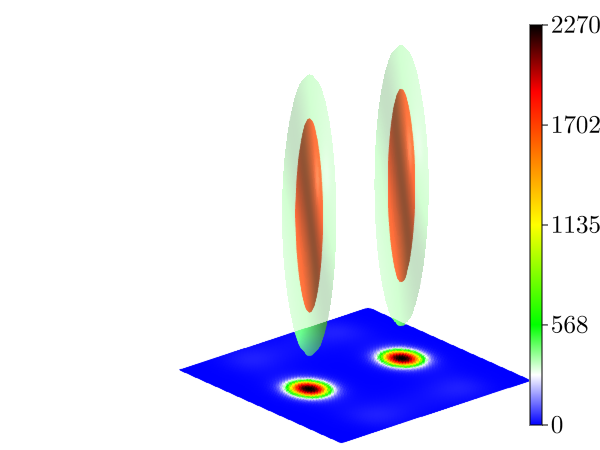} \put(-24.5,1){(f)}&
        \includegraphics[width = 0.15\textwidth, trim={4cm 0cm 0cm 0cm}, clip]{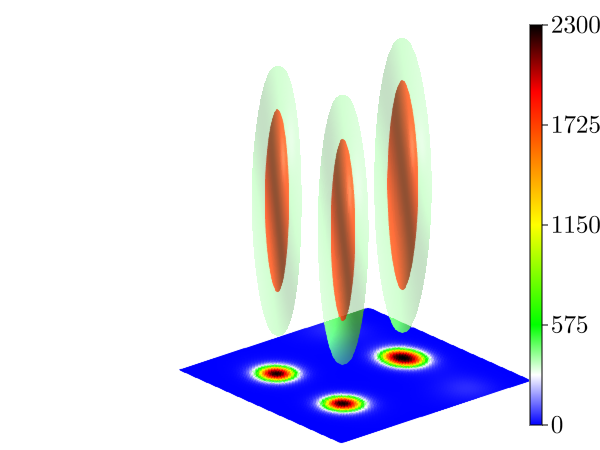} \put(-24.5,1){(g)}&
        \includegraphics[width = 0.15\textwidth, trim={4cm 0cm 0cm 0cm}, clip]{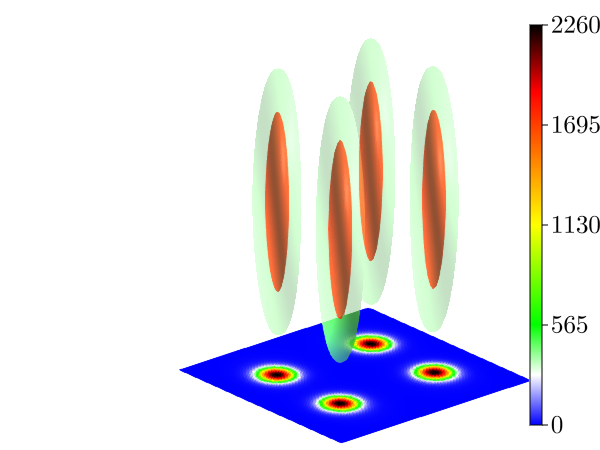} \put(-24.5,1){(h)}\\
        %\includegraphics[height=0.135\textwidth, trim={0.25cm .25cm 0 0},clip]{1500/1k5/kok1500_k5-crop.pdf}\put(-25,5){(a)} 
        %&\includegraphics[height=0.135\textwidth, trim={.25cm .25cm 0 0},clip]{1500/1k5/zkok1500_k6-crop.pdf}\put(-25,5){(b)} 
        %& \includegraphics[height=0.135\textwidth, trim={.25cm 0.25cm 0 0},clip]{1500/1k5/kok1500_k75-crop.pdf}\put(-25,5){(c)}  \\
        %\includegraphics[height=0.135\textwidth, trim={.25cm .25cm 0 0},clip]{1500/1k5/kok1500_2-crop.pdf}\put(-25,5){(d)}  &
        %\includegraphics[height=0.135\textwidth, trim={.25cm .25cm 0 0},clip]{1500/1k5/zkok1500_2k75-crop.pdf} \put(-25,5){(e)}  &
        %\includegraphics[height=0.135\textwidth,trim={.25cm .25cm 0 0},clip]{1500/1k5/kok_1500_4-crop.pdf}\put(-25,5){(f)} 
    \end{tabular}   
    \caption{
        (a) The  contributions to the total energy for the ground-state at $\varepsilon_{dd}=1.5$ as a function of $N$ for the double well  potential with minima spaced by {$2d=3 \,\mu$m};
        (b) maximal atom density at $z=0$ plane (red lines) and droplet height $H_z$ in $z$ direction (blue lines).
        The vertical lines mark the ranges of a varied localization type:
        I - a delocalized system (no droplet formation), II - an asymmetric system with a fully developed droplet in one of the wells, III - a symmetric system with droplets in both wells, IV -  two droplets in one well and one in the other, V - a symmetrical system with 4 droplets. 
        (c-h) Isosurfaces  of the BEC density for $N=10^3$ (c), $6\times 10^3$ (d), $8\times 10^3$ (e), $2.7\times 10^4$ (f), $3.8\times 10^4$ (g) and $4\times 10^4$ (h) in the lowest-energy configuration.
        The isosurfaces correspond to 20\% (transparent) and 80\% (opaque) of the maximum density value.
        The plots span the area of $x$ (horizontal direction) and $y$ from -3.6 $\mu$m to 3.6 $\mu$m.
        At the base of the plots, below the isosurfaces, a cross-section of the BEC density at $z=0$ is presented.
        The colorscale for the density is given in $\mu$m$^{-3}$ units.
        The plots correspond to regions marked by I (c), II (d), III (e,f),  IV (g) and V (h).
        }
    \label{difaz1k5_1500}
\end{figure}

\begin{figure}[tbp]
%trim=left botm right top
\begin{tabular}{l}
\includegraphics[width=0.35\textwidth]{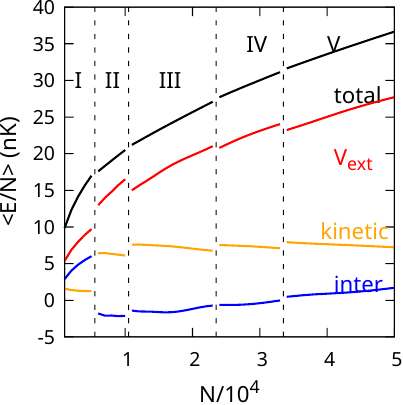} \put(-25,35){(a)} \\
\includegraphics[width=0.35\textwidth]{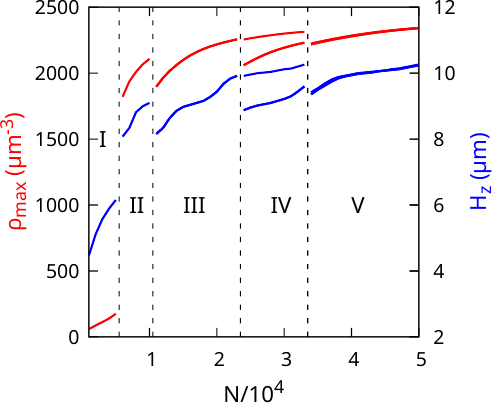} \put(-40,30){(b)} 
\end{tabular}
    \begin{tabular}{lll}
        \includegraphics[width = 0.15\textwidth, trim={3.0cm 0cm 0cm 0cm}, clip]{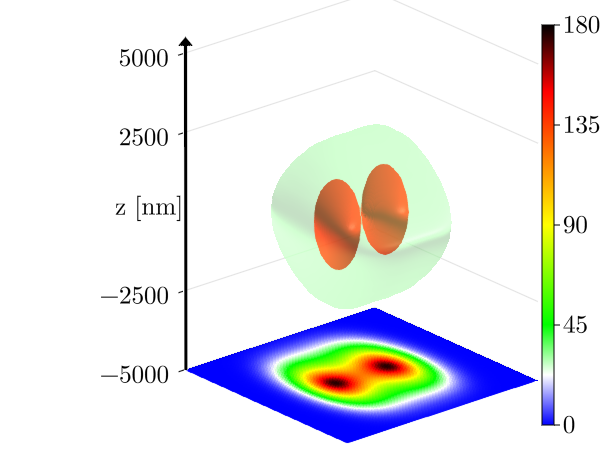} \put(-20.5, 1){(c)}&
        \includegraphics[width = 0.15\textwidth, trim={4cm 0cm 0cm 0cm}, clip]{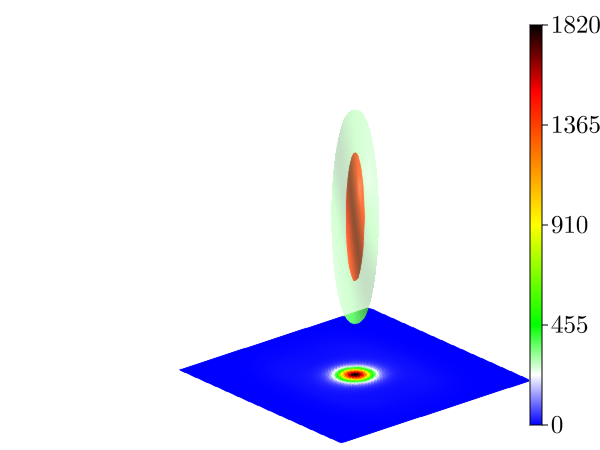} \put(-24.5,1){(d)}&
        \includegraphics[width = 0.15\textwidth, trim={4cm 0cm 0cm 0cm}, clip]{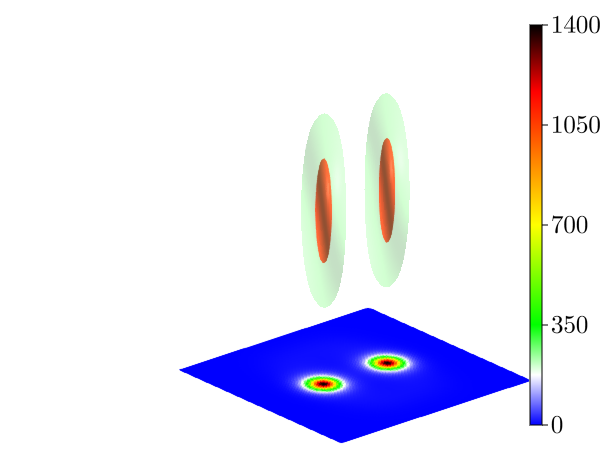} \put(-24.5,1){(e)}\\
        \includegraphics[width = 0.15\textwidth, trim={4cm 0cm 0cm 0cm}, clip]{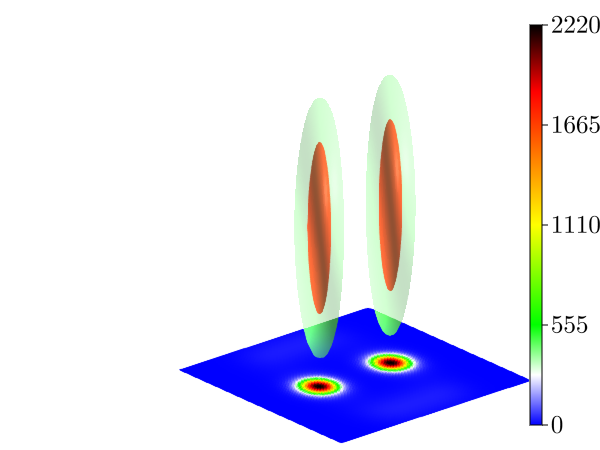} \put(-24.5,1){(f)}&
        \includegraphics[width = 0.15\textwidth, trim={4cm 0cm 0cm 0cm}, clip]{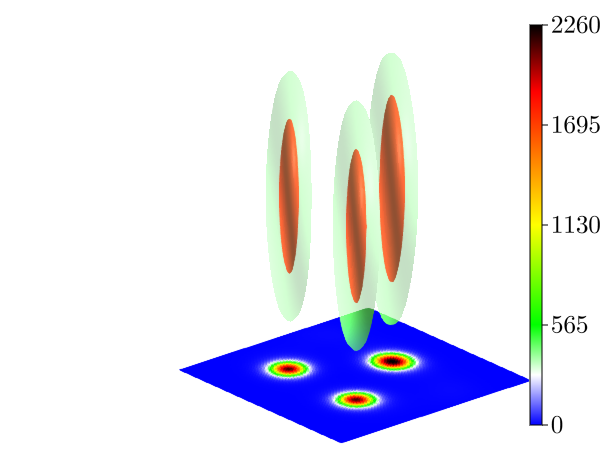} \put(-24.5,1){(g)}&
        \includegraphics[width = 0.15\textwidth, trim={4cm 0cm 0cm 0cm}, clip]{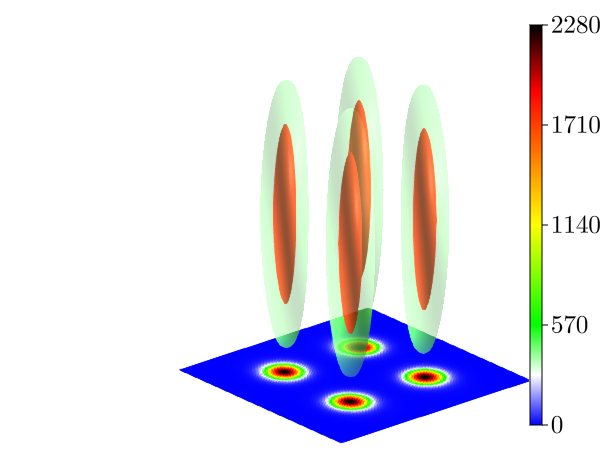} \put(-24.5,1){(h)}\\      

    \end{tabular}   

\caption{ (a) Total energy and contributions to the energy for the ground-state 
          at $\varepsilon_{dd}=1.5$ as a function of $N$ for an external potential with a single minimum. (b) the maximal gas density at $z=0$ plane (red lines) and droplet height $H_z$ in $z$ direction (blue lines).
          The vertical lines mark the ranges of a varied localisation type.
          I - a delocalized system, II - a single droplet in the center of the system, III - two droplets, IV - three droplets, V - four droplets.
          (c-h) Isosurfaces of the BEC density for $N=5\times 10^3$ (c), $6\times 10^3$  (d), $7.5\times 10^3$ (e),$2\times 10^4$ (f),  $2.75\times 10^4$(g) and $4\times 10^4$ (h) for $\varepsilon=1.5$ and the lowest-energy configuration for a potential with a single minimum.
          The isosurfaces correspond to 20\% (transparent) and 80\% (opaque) of the maximum density.
          The plots span the area of $x$ (horizontal direction) and $y$ from -3.6 $\mu$m to 3.6 $\mu$m.
          At the base of the plot, below the isosurfaces a cross-section of the BEC density at $z=0$ is presented.
          The colorscale for the density is given in $\mu$m$^{-3}$ units.
          The plots (c-h) correspond to regions marked by I, II, II (still), III, IV and V in (a) and (b).
        }

\label{difaz1k5_1500b}
\end{figure}

%Supersolid symmetry breaking from compressional
%oscillations in a dipolar quantum gas
%https://www.nature.com/articles/s41586-019-1568-6
% TO JEST PRACA GDZIE WIDZA NISKA I WYSOKA CZESTOSC
% OSCYLACJI NISKA JEST OD TLA W SUPERSOLID JEST W DROPET NIE MA

%PHYSICAL REVIEW A 96, 053630 (2017)
%Striped states in a many-body system of tilted dipoles
% - to tez niby ze formacja 1D dropletów
% coherebce: tracą wspólną każda ma własną

%O dynamice:

%Physics Letters A
%Volume 536, 15 March 2025, 130291 Zderzenia 1D

%Effect of beyond mean-field interaction on the structure and dynamics of the
%one-dimensional quantum droplet
%Sonali Gangwar,1 Rajamanickam Ravisankar,1, 2 Paulsamy Muruganandam,3 and Pankaj Kumar %Mishra https://arxiv.org/pdf/2303.01216
% dynamika bez zderzeń 

%tutaj sa zderzenia vortex droplets
% dwoch AIP Advances 13, 055130, 2023)
%Studies collision dynamics of moving 2D quantum droplets (with vorticity)

%Volume 535, 5 March 2025, 130279
%Physics Letters A
%Dynamics of quantum droplets falling under gravity on a reflector
%Phase-space analysis.
%- reflector

%https://arxiv.org/abs/0804.1836?utm_source=chatgpt.com
% kwestia spójności fazowej różnych dropletóþw

\section{Results and Discussion}
 { The presentation of the results is organized as follows. 
 In the main text of the paper we present the results for an interwell distance of $2d = 3\,\mu\mathrm{m}$ and the dipolar interaction parameter $\varepsilon_{dd}=a_{dd}/a=1.5$. 
 The results for $\varepsilon_{dd}=1.45$ and $\varepsilon_{dd}={1.4}$ are given and discussed in the appendices. We also present therein the results for larger interwell distances $2d=5\mu$m and $7\mu$m. }

For each value of $\varepsilon_{dd}$, we first present the phase diagram of the lowest-energy configurations as a function of the atom number $N$. We then analyze representative initial states from different regions of the phase diagram and discuss the underlying dynamical behavior.  
At the end of this section, we also provide selected results for $2d = 5\,\mu\mathrm{m}$ and $2d = 7\,\mu\mathrm{m}$, where the larger excess energy in the initial state leads to qualitatively different dynamics compared to the case of $2d = 3\,\mu\mathrm{m}$.

\subsection{The ground-state of the double well potential} 

Figure~\ref{difaz1k5_1500}(a) shows the total energy and its individual contributions for the lowest-energy configurations of the dipolar gas in the double-well potential with { $2d = 3\,\mu\mathrm{m}$} and $\varepsilon_{dd} = 1.5$, as functions of the atom number $N$. Discontinuities in the curves reflect changes in the lowest-energy configuration with increasing $N$.  

In Fig.~\ref{difaz1k5_1500}(b), we plot the maximum density in the $z = 0$ plane (red lines) and the height of the droplet structures along the polarization direction $z$ (blue lines), defined as the axial extent of the density profile above $1$ atom$/\mu\mathrm{m}^3$. Outside this region, only a negligible residual density is present.  

The condensate density distributions are shown in Fig.~\ref{difaz1k5_1500}(c)-(h). Each panel presents a cross section at $z = 0$ together with isosurfaces at $40\%$ and $80\%$ of the maximum density in that plane.  

For low atom numbers (region I), the collective dipolar interaction is too weak to support droplet formation.  
For $N \in (5060, 7410)$, there are enough atoms to form a droplet in a single potential minimum, but not in both wells simultaneously. As a result, the lowest-energy configuration becomes asymmetric. This corresponds to spontaneous symmetry breaking driven by the interaction energy, where a high-density droplet forms in one well, while the other well contains a dilute cloud below the droplet-formation threshold [see Fig.~\ref{difaz1k5_1500}(b)].  

For $N > 7410$, the system becomes symmetric again, forming two well-defined droplets that share the increasing number of atoms. At $N_{III/IV} = 24700$, a third droplet appears in the lowest-energy configuration, leading again to an asymmetric state { with one droplet in one of the wells and two in the other.
Symmetry is restored at $N_{IV/V} = 30750$, where a four-droplet configuration is formed.  }

As $N$ increases further, the droplets grow in size, enhancing the in-plane repulsive component of the dipolar interaction and increasing the external potential energy associated with elongation along the $z$ direction. These effects favour configurations with a larger number of droplets.  
The maximum density saturates for $N \simeq 2 \times 10^4$ [Fig.~\ref{difaz1k5_1500}(b)] at approximately $2250$ atoms$/\mu\mathrm{m}^3$.  
{ As $N$ grows, when the peak densities of the individual droplets become close to this value the ground-state switches to the next configuration with a larger number of droplets. 
In particular, for the asymmetric configuration (IV) 
one finds two values of the peak density [Fig. 2(b)] with larger value for the solitary droplet. 
As $N$ grows the lower value of the peak density corresponding to the droplet pair appears to converge to the peak density of the solitary droplet up to the point of the phase transition.}

In Fig.~\ref{difaz1k5_1500}(a), the total energy remains nearly continuous across phase boundaries, while more pronounced discontinuities appear in the individual contributions, particularly in the interaction and external potential energies. Typically, transitions to configurations with an additional droplet are accompanied by a decrease in interaction or potential energy, at the expense of increased kinetic energy due to stronger wave-function localization.  

For comparison, results for a single-well potential are shown in Fig.~\ref{difaz1k5_1500b}. { In the single-well case after the critical point of $N \simeq 5120$ atoms, one droplet forms in the center of the potential well. In the double-well case, beyond the critical point of $N\simeq 5060$ atoms, the droplet forms only in one of the two minima, which results in spontaneous symmetry breaking}.
%The first droplet forms abruptly at $N \simeq 5120$, without symmetry breaking, in contrast to the double-well case. In the double well, the density is initially distributed symmetrically between the two minima, but beyond a critical atom number, a droplet first forms in only one of them.  

For the single-well case, additional droplets appear at $N \simeq 10450$, $22800$, and $32800$, corresponding to the formation of the second, third, and fourth droplets, respectively. The density distributions in the single-well potential are qualitatively similar to those in the double-well case, but with reduced spacing between droplets along the $x$ direction due to the absence of the interwell barrier.

\subsection{The dynamics upon release of the interwell barrier}
We consider solutions of the time-dependent extended Gross-Pitaevskii equation for a system initially prepared in the lowest-energy state of the double-well potential (see Fig.~\ref{difaz1k5_1500}). The dynamics are induced by removing the interwell barrier, implemented by setting $\beta = 0$.

\subsubsection{Edge of localization}

\begin{figure}[tbp]

%trim=left botm right top
\begin{tabular}{lll}
\includegraphics[width=0.16\textwidth, trim={1.3cm 3cm 0 0}, clip]{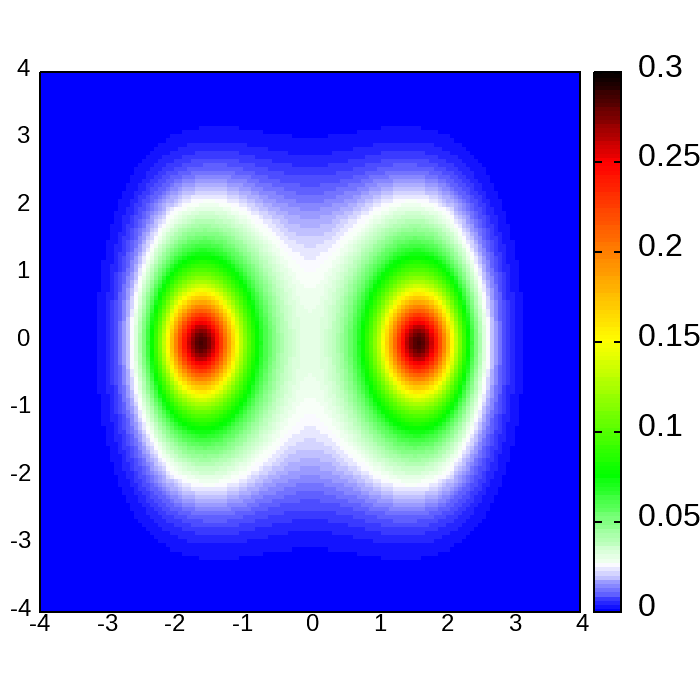}\put(-25,5){\color{yellow}(a)}&
  \includegraphics[width=0.16\textwidth, trim={1.3cm 3cm 0 0}, clip]{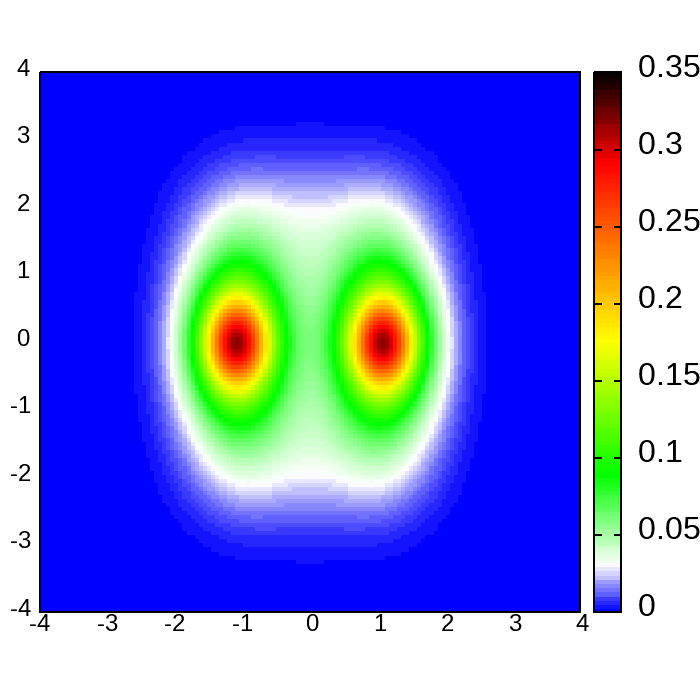} \put(-25,5){\color{yellow}(b)}&
  \includegraphics[width=0.16\textwidth, trim={1.3cm 3cm 0 0}, clip]{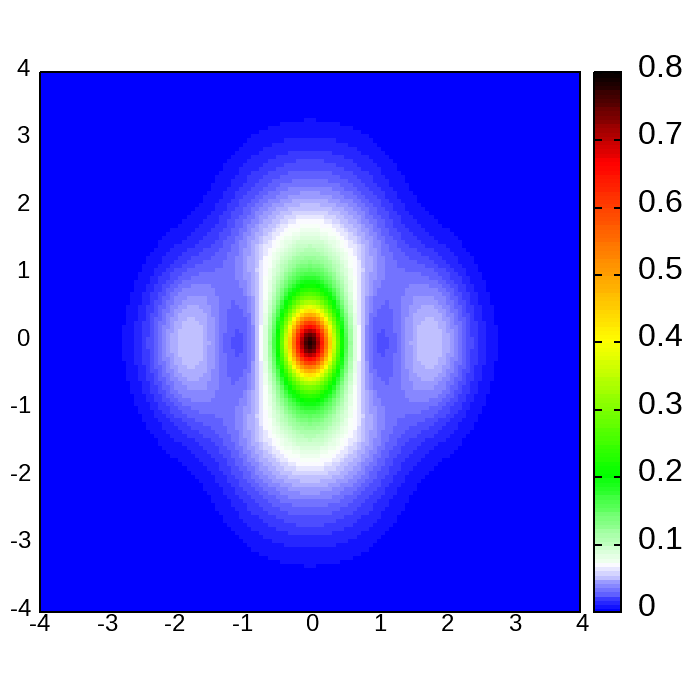}\put(-25,5){\color{yellow}(c)}\\
   \includegraphics[width=0.16\textwidth, trim={1.3cm 3cm 0 0}, clip]{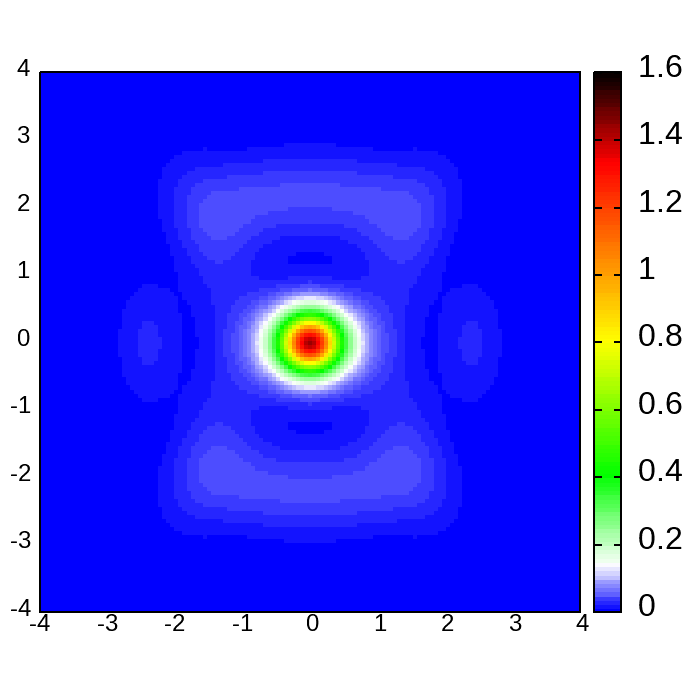}\put(-25,5){\color{yellow}(d)}    &
    \includegraphics[width=0.16\textwidth, trim={1.3cm 3cm 0 0}, clip]{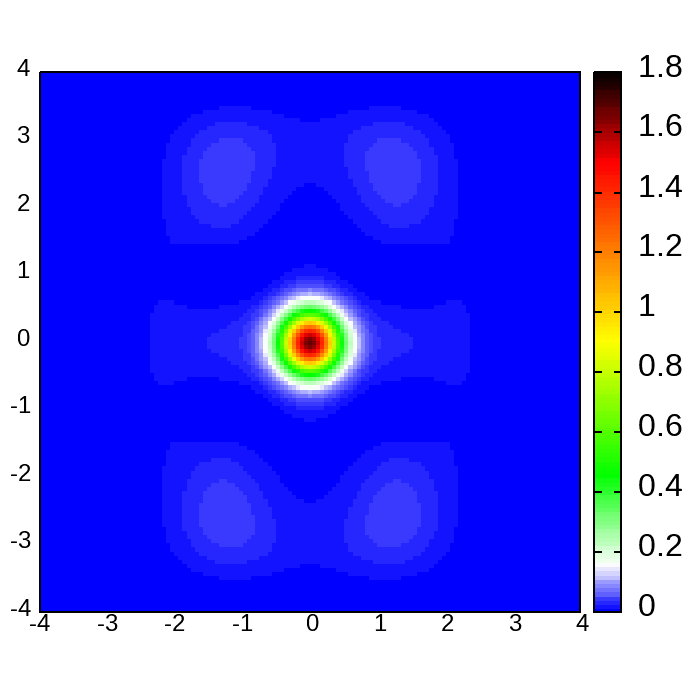} \put(-25,5){\color{yellow}(e)} &
   \includegraphics[width=0.16\textwidth, trim={1.3cm 3cm 0 0}, clip]{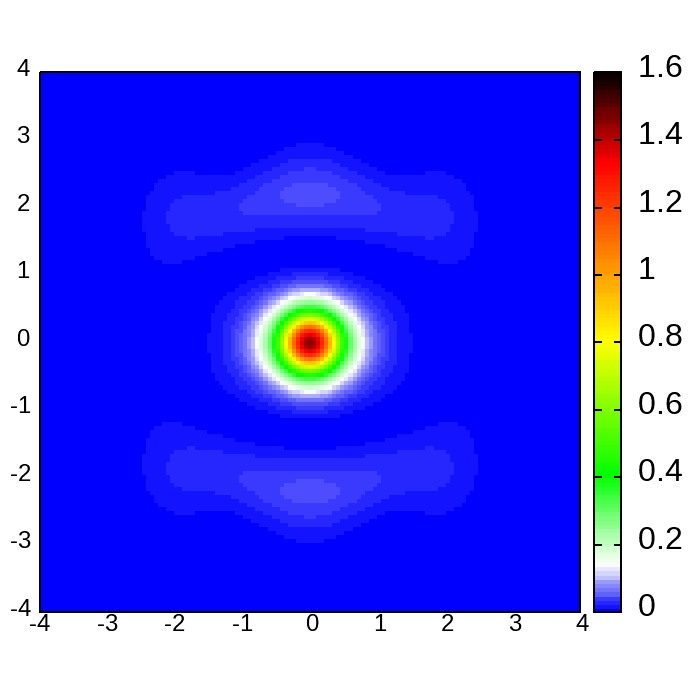}\put(-25,5){\color{yellow}(f)} \\
  \includegraphics[width=0.16\textwidth, trim={1.3cm 3cm 0 0}, clip]{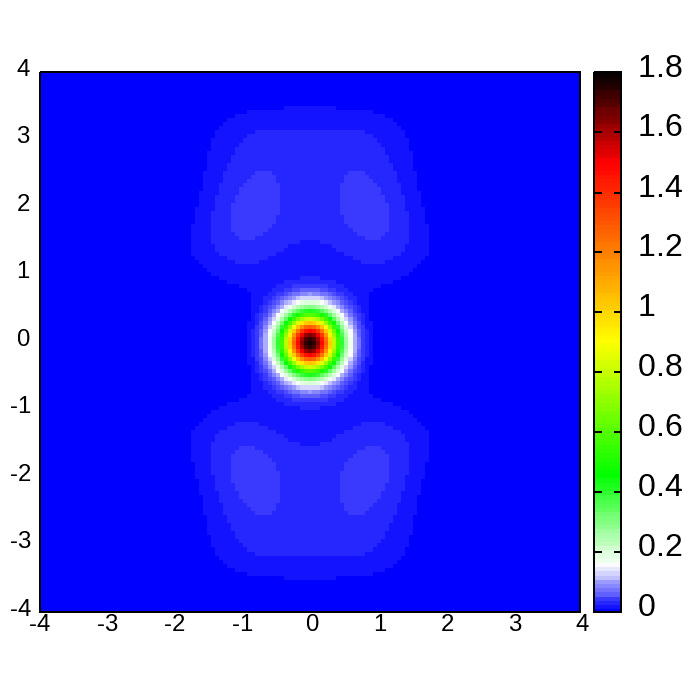} \put(-25,5){\color{yellow}(g)}& 
  \includegraphics[width=0.16\textwidth, trim={1.3cm 3cm 0 0}, clip]{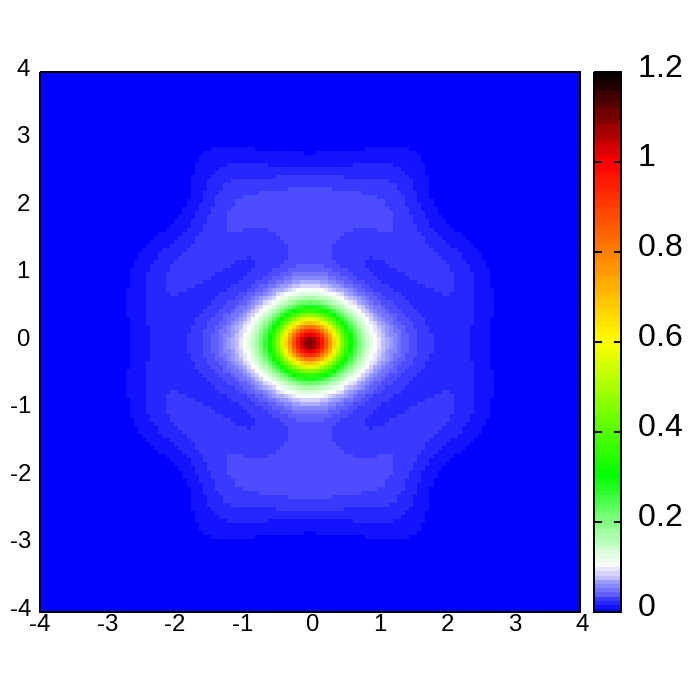} \put(-25,5){\color{yellow}(h)}&
   \includegraphics[width=0.16\textwidth, trim={1.3cm 3cm 0 0}, clip]{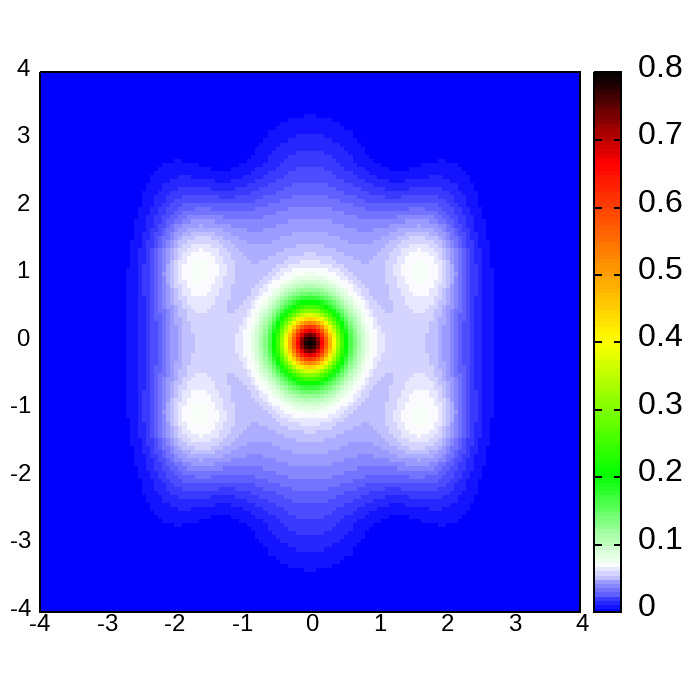} \put(-25,5){\color{yellow}(i)} \\
    \includegraphics[width=0.16\textwidth, trim={1.3cm 3cm 0 0}, clip]{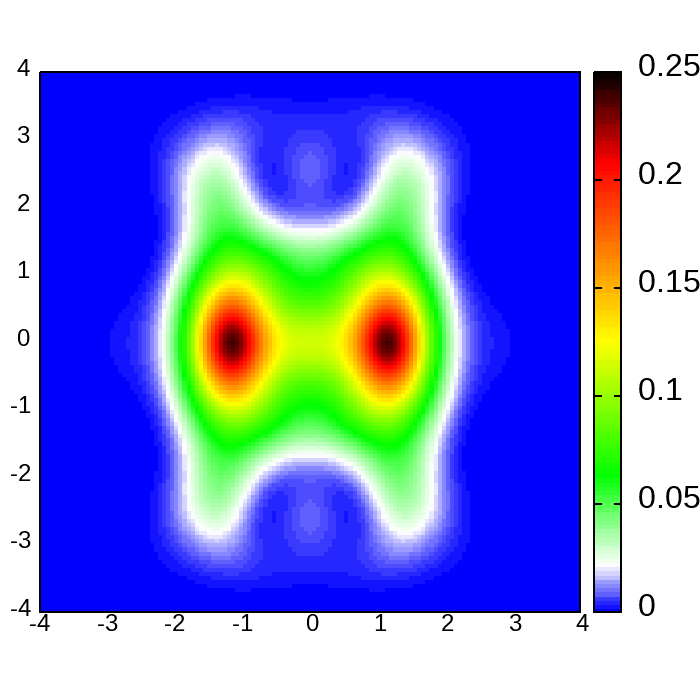} \put(-25,5){\color{yellow}(j)} &
    \includegraphics[width=0.16\textwidth, trim={1.3cm 3cm 0 0}, clip]{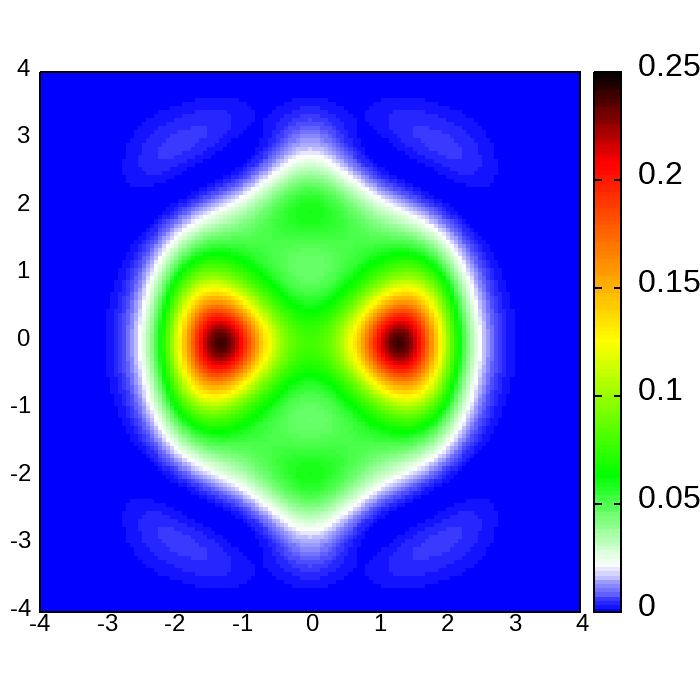} \put(-25,5){\color{yellow}(k)} &  
    \includegraphics[width=0.16\textwidth, trim={1.3cm 3cm 0 0}, clip]{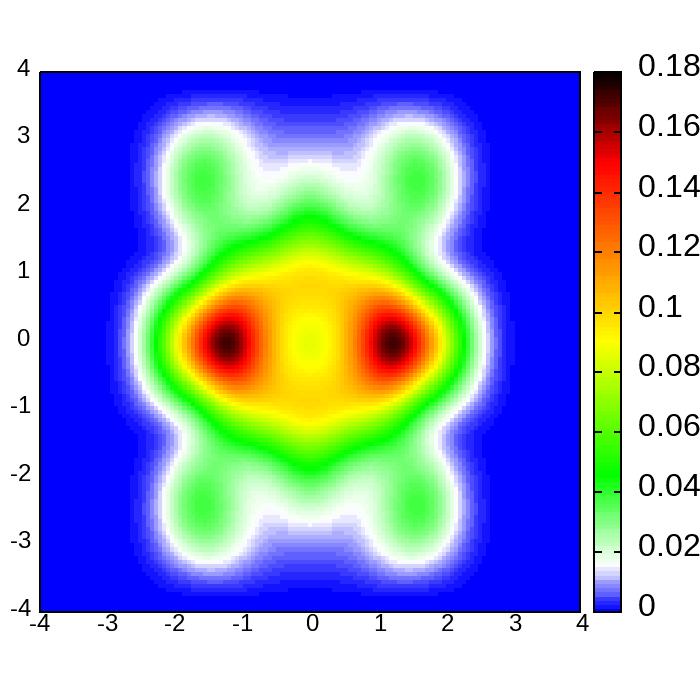} \put(-25,5){\color{yellow}(l)} \\
\end{tabular}
\includegraphics[width=0.35\textwidth]{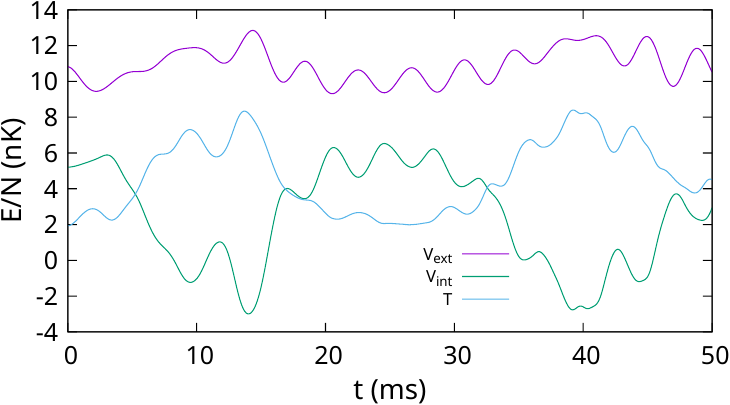}\put(-20,20){(m)}
\caption{Snapshots of the time evolution of the BEC density prepared for {$2d=3\,\mu\mathrm{m}$} $\varepsilon_{dd}=1.5$ for $N=5000$ atoms. The plots show the cross section taken at $z=0$ with $x\in[-4\mu$m,$4\mu$m] and $y\in[-4\mu$m,$4\mu$m]. The color scale for the density is given in units of $1000\mu$m$^{-3}$. Plots (a-l) show the moments $t$ after the confinement potential is set to a single well, with $t=0$ (a), 2.42 ms (b), 4.84 ms (c), 7.26 ms(d), 9.68 ms(e), 12.1 ms (f), 14.51 ms(g), 16.93 ms(h), 19.35 ms(i), 21.77 ms(j), 24.19 (k), 26.61 (l).
(m) Contributions to the energy per atom as functions of time.}
%densities, stepping: %25*400*1e10*2.4188e-14=2.419ms.}}
\label{gd2}
\end{figure}

Let us consider the time evolution for $N = 5000$ atoms (see Fig.~\ref{gd2}). This value is of particular interest, as it lies in region I for both the double-well system (Fig.~\ref{difaz1k5_1500}) and the single-well case [see Fig.~\ref{difaz1k5_1500}(b)]. On the other hand, the ground-state in a single well is just below the threshold for droplet formation in the single-well potential which is observed for 5060 atoms.

{ The double-well initial state is symmetric in space and corresponds to low extended density [Fig.~\ref{gd2}(a)] with two maxima near the minima of the potential wells. }
During the evolution, we observe the transient formation of a single central droplet [Fig.~\ref{gd2}(d)-(h)] localized in-plane with a high peak density, surrounded by a low-density halo extending away from the droplet after collapse. The droplet is not stable and persists for approximately $12$ ms before expanding back [Fig.~\ref{gd2}(d)-(l)] into a state resembling the initial configuration.  

The transient droplet formation is accompanied by a decrease in the interaction energy [Fig.~\ref{gd2}(m)] and a corresponding increase in kinetic energy due to enhanced wave-function localization. The formation and decay of the central droplet constitute a process that is reentrant in time.  

In Fig.~\ref{gd2}(m), we identify the energy signatures of the first cycle of droplet formation and decay for $t \in (0,25)\,\mathrm{ms}$ and the second cycle for $t \in (25,50)\,\mathrm{ms}$.

For a lower value, $N=2500$ - far below the threshold for droplet formation - the average values of the energy depend on time in a less dramatic manner while the number of the local density maxima changes from 1 to 3 (see Appendix \ref{sec:precondensation}).

\begin{figure}[htbp]
%trim=left botm right top
\begin{tabular}{lll}
\includegraphics[width=0.16\textwidth, trim={1.3cm 3cm 0 0}, clip]{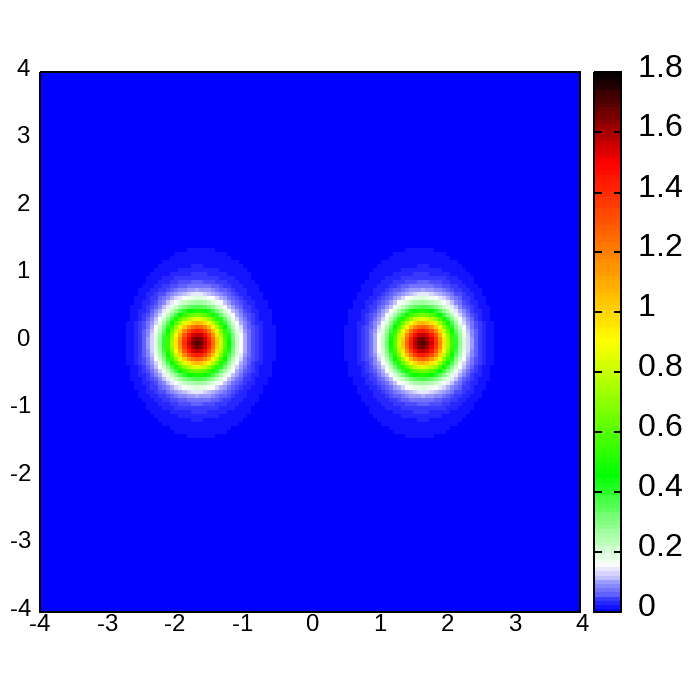}\put(-25,5){\color{yellow}(a)}&
\includegraphics[width=0.16\textwidth, trim={1.3cm 3cm 0 0}, clip]{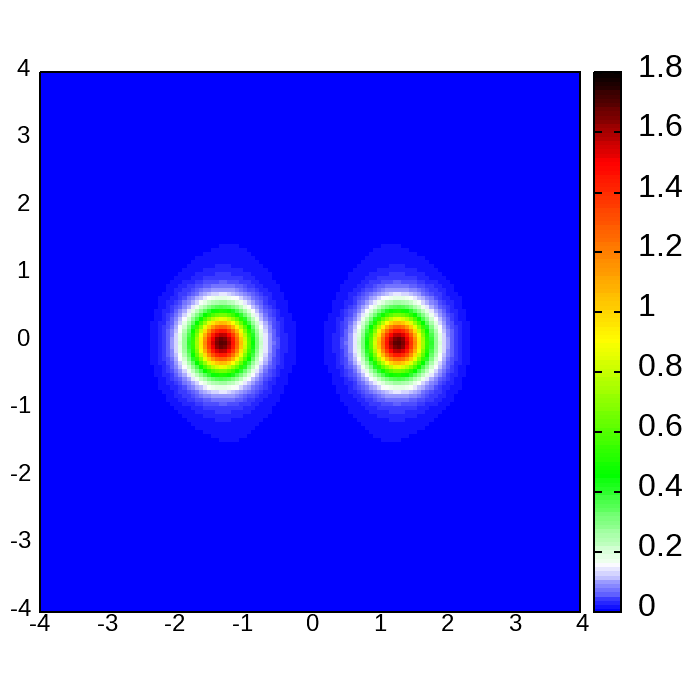}\put(-25,5){\color{yellow}(b)}&
\includegraphics[width=0.16\textwidth, trim={1.3cm 3cm 0 0}, clip]{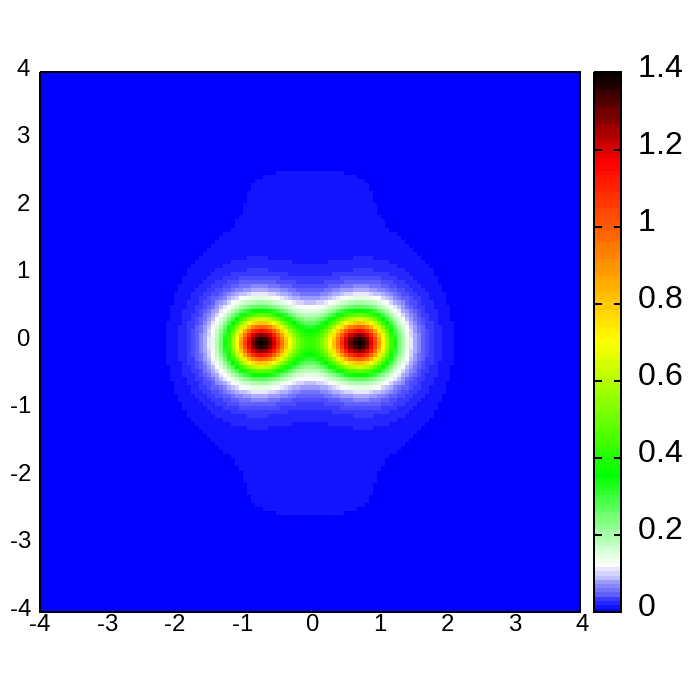}\put(-25,5){\color{yellow}(c)}\\
\includegraphics[width=0.16\textwidth, trim={1.3cm 3cm 0 0}, clip]{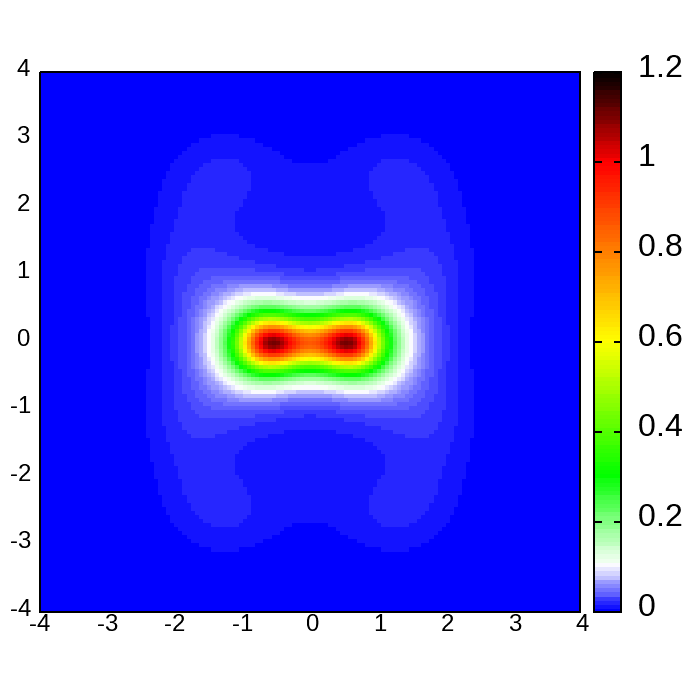}\put(-25,5){\color{yellow}(d)}&
\includegraphics[width=0.16\textwidth, trim={1.3cm 3cm 0 0}, clip]{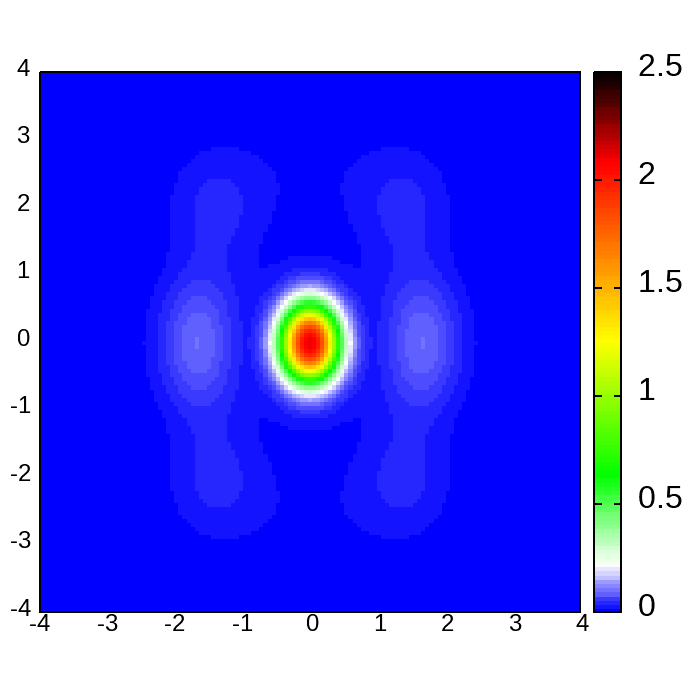}\put(-25,5){\color{yellow}(e)}&
\includegraphics[width=0.16\textwidth, trim={1.3cm 3cm 0 0}, clip]{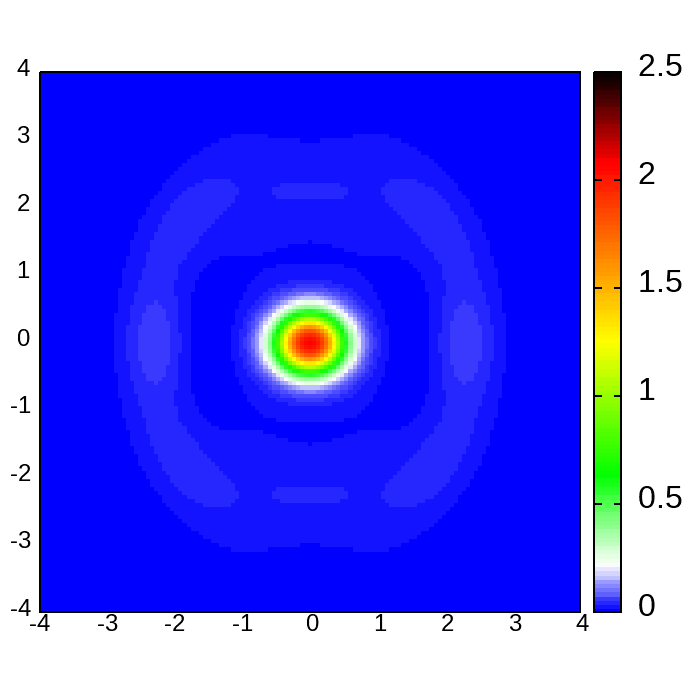}\put(-25,5){\color{yellow}(f)}\\
\includegraphics[width=0.16\textwidth, trim={1.3cm 3cm 0 0}, clip]{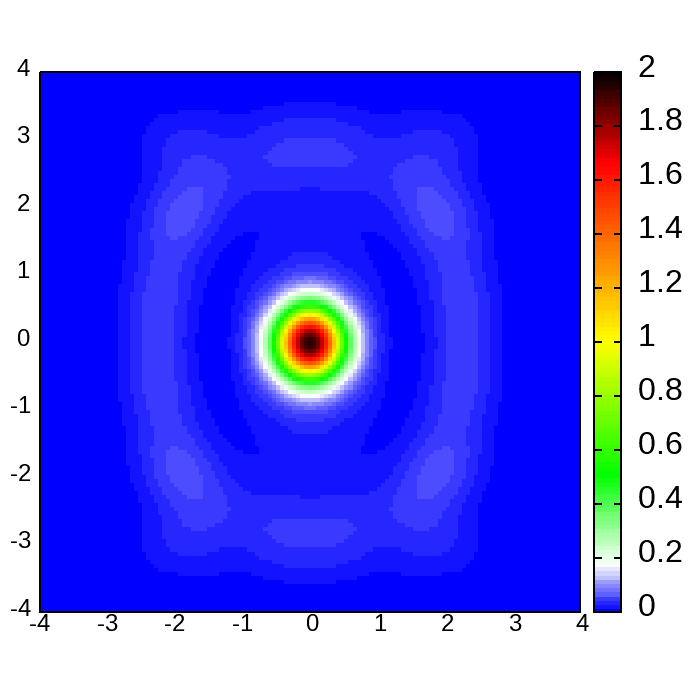}\put(-25,5){\color{yellow}(g)}&
\includegraphics[width=0.16\textwidth, trim={1.3cm 3cm 0 0}, clip]{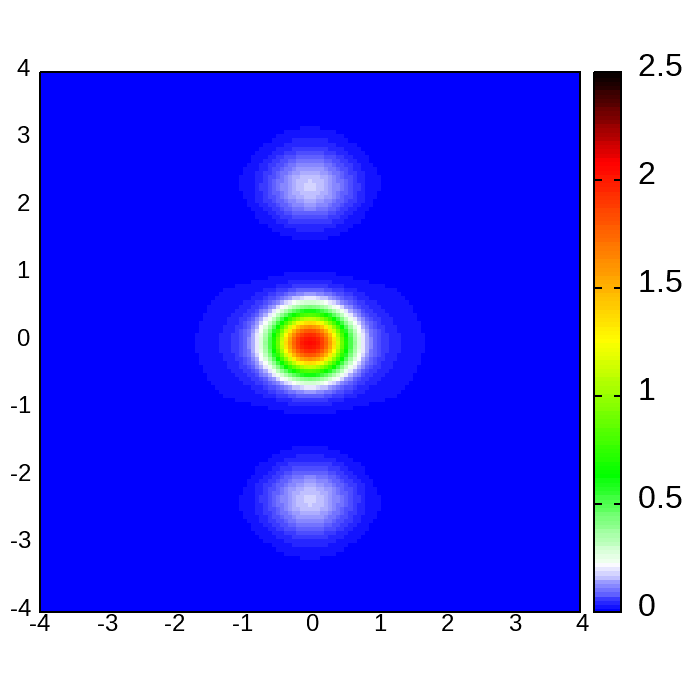}\put(-25,5){\color{yellow}(h)}&
\includegraphics[width=0.16\textwidth, trim={1.3cm 3cm 0 0}, clip]{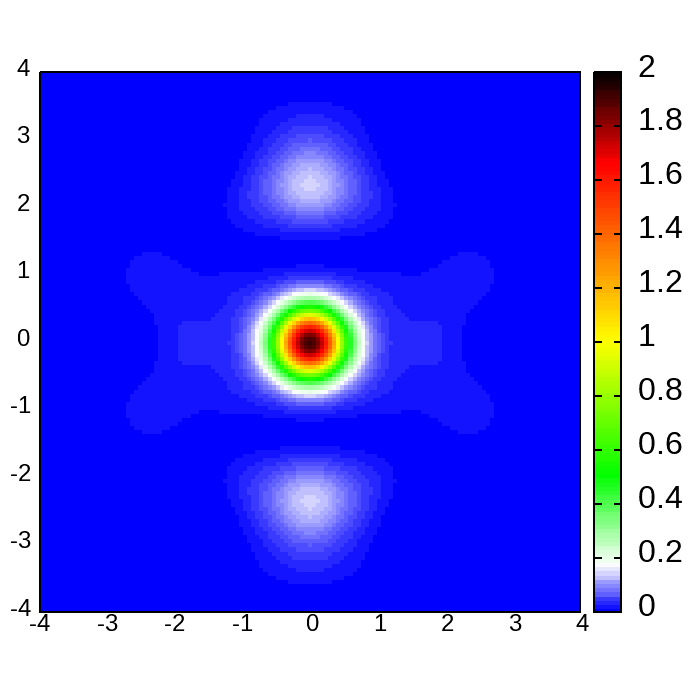}\put(-25,5){\color{yellow}(i)}\\
\includegraphics[width=0.16\textwidth, trim={1.3cm 3cm 0 0}, clip]{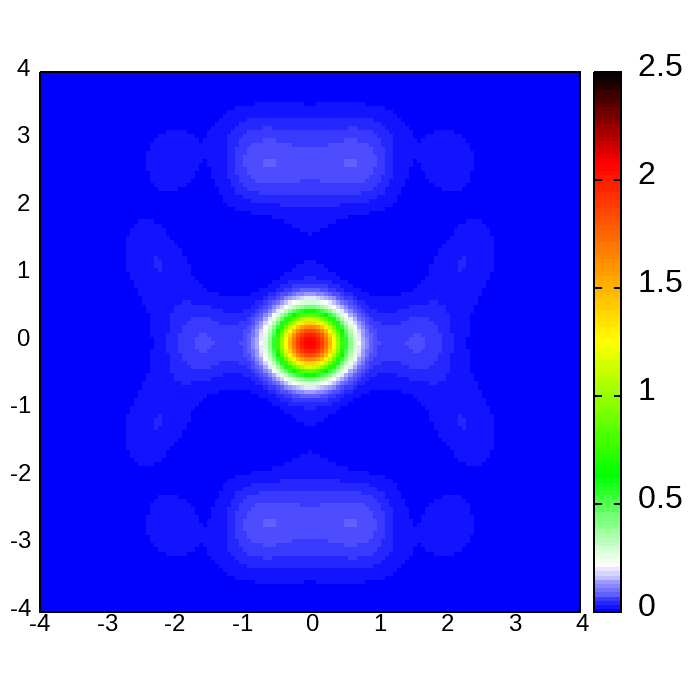}\put(-25,5){\color{yellow}(j)}&
\includegraphics[width=0.16\textwidth, trim={1.3cm 3cm 0 0}, clip]{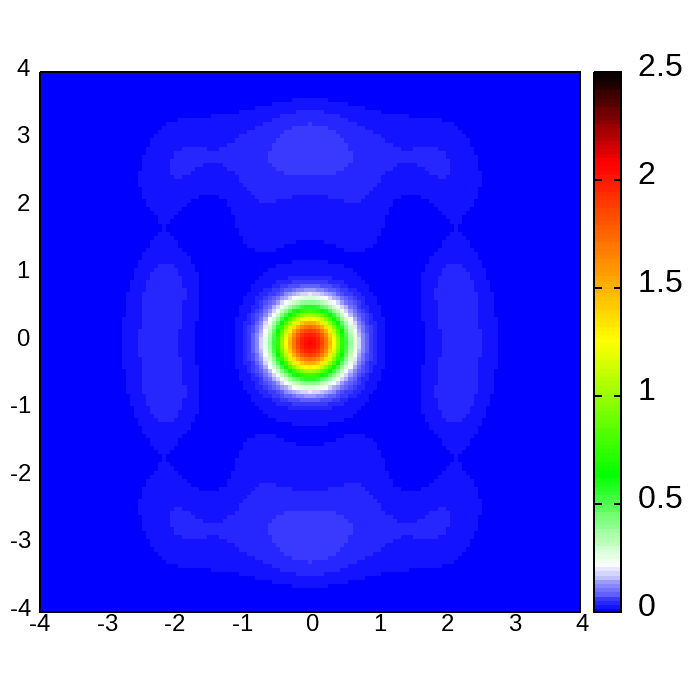}\put(-25,5){\color{yellow}(k)}&
\includegraphics[width=0.16\textwidth, trim={1.3cm 3cm 0 0}, clip]{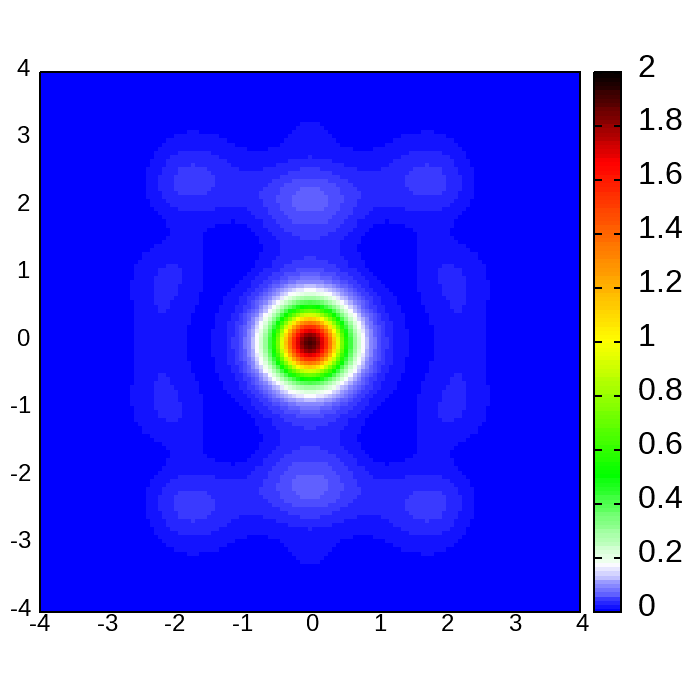}\put(-25,5){\color{yellow}(l)} \\
\end{tabular}

\begin{tabular}{ll}
%trim=left botm right top
%\includegraphics[width=0.3\textwidth, trim={3cm 3cm 0 0}, clip]{1500/1k5/zlepek/n88_131} 
\includegraphics[width=0.25\textwidth, trim={1.5cm 3cm 1.8cm 1cm}, clip]{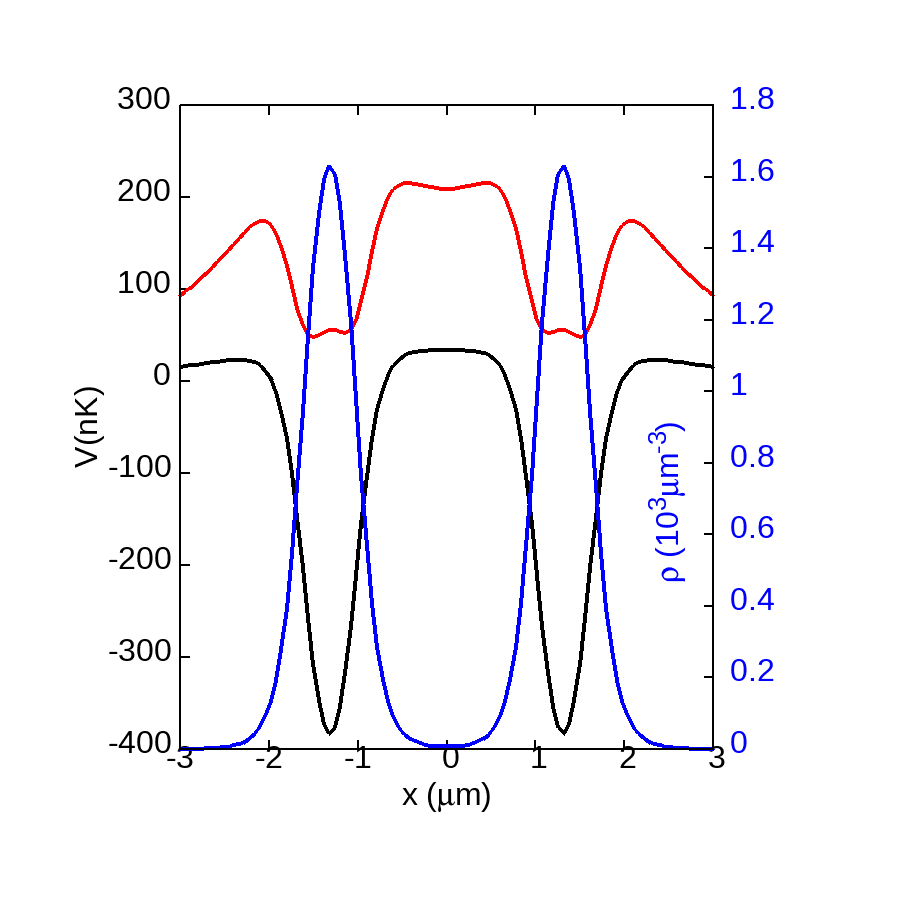}\put(-35,20){\color{black}(m)} &
\includegraphics[width=0.25\textwidth, trim={1.5cm 3cm  1.8cm 1cm}, clip]{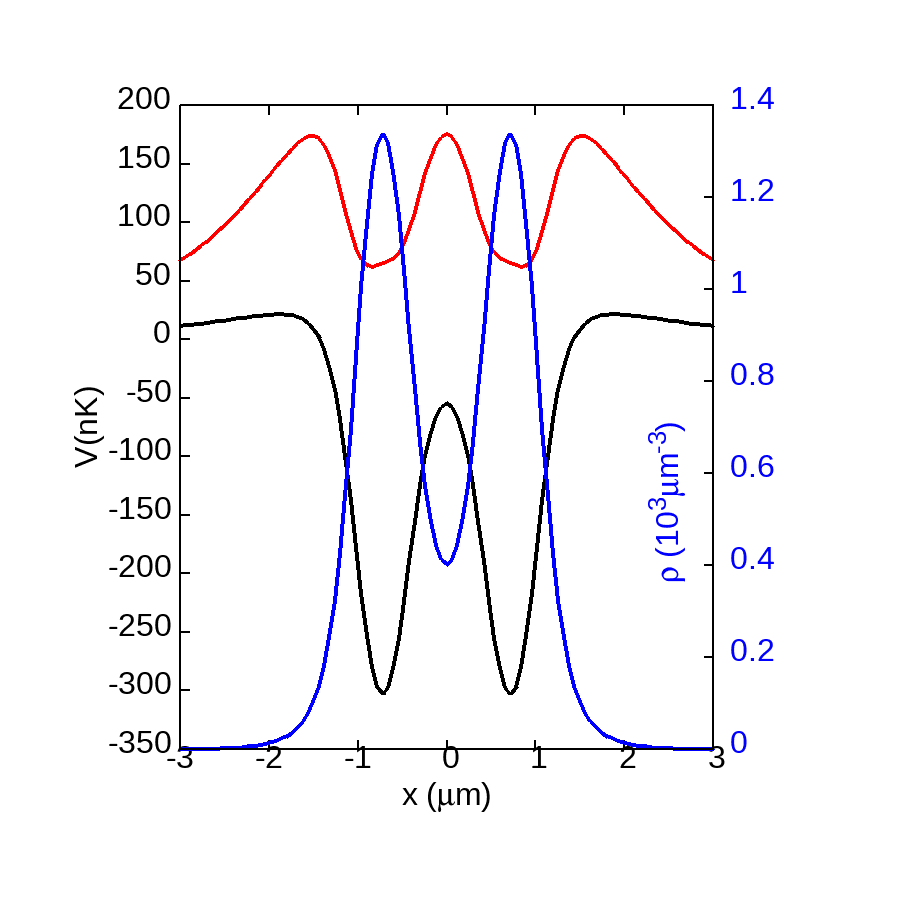}\put(-35,20){\color{black}(n)} \\
\includegraphics[width=0.25\textwidth, trim={1.5cm 3cm  1.8cm 1cm}, clip]{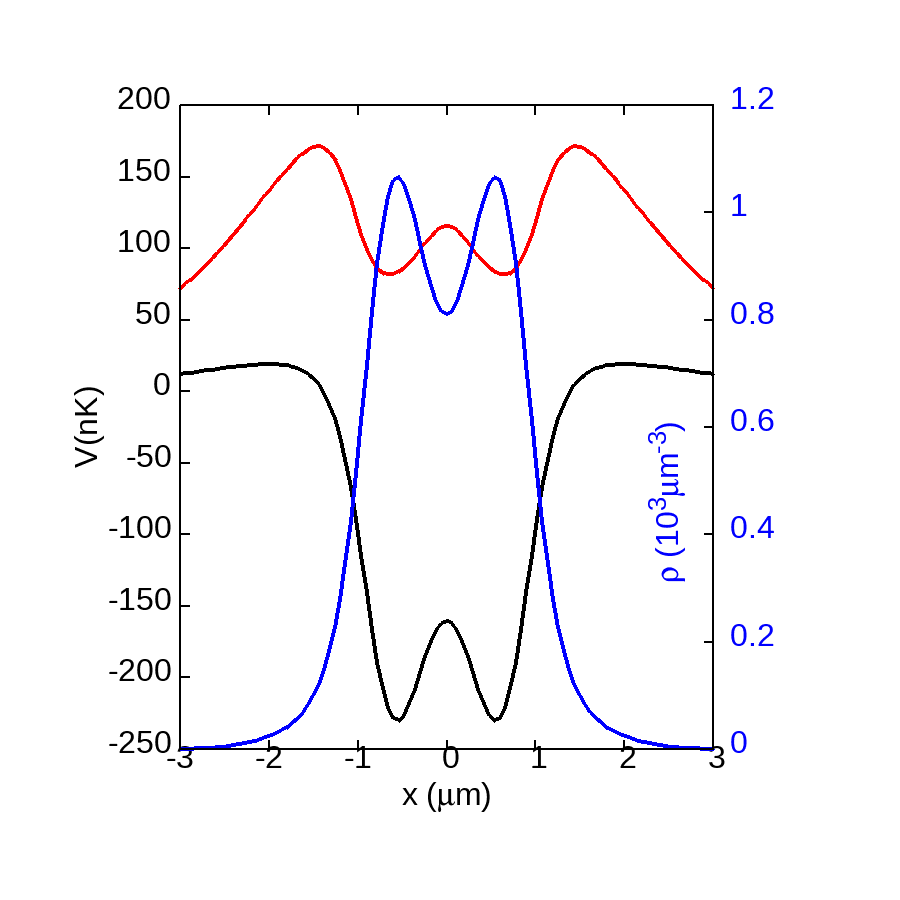}\put(-35,20){\color{black}(o)} &
\includegraphics[width=0.25\textwidth, trim={1.5cm 3cm  1.8cm 1cm}, clip]{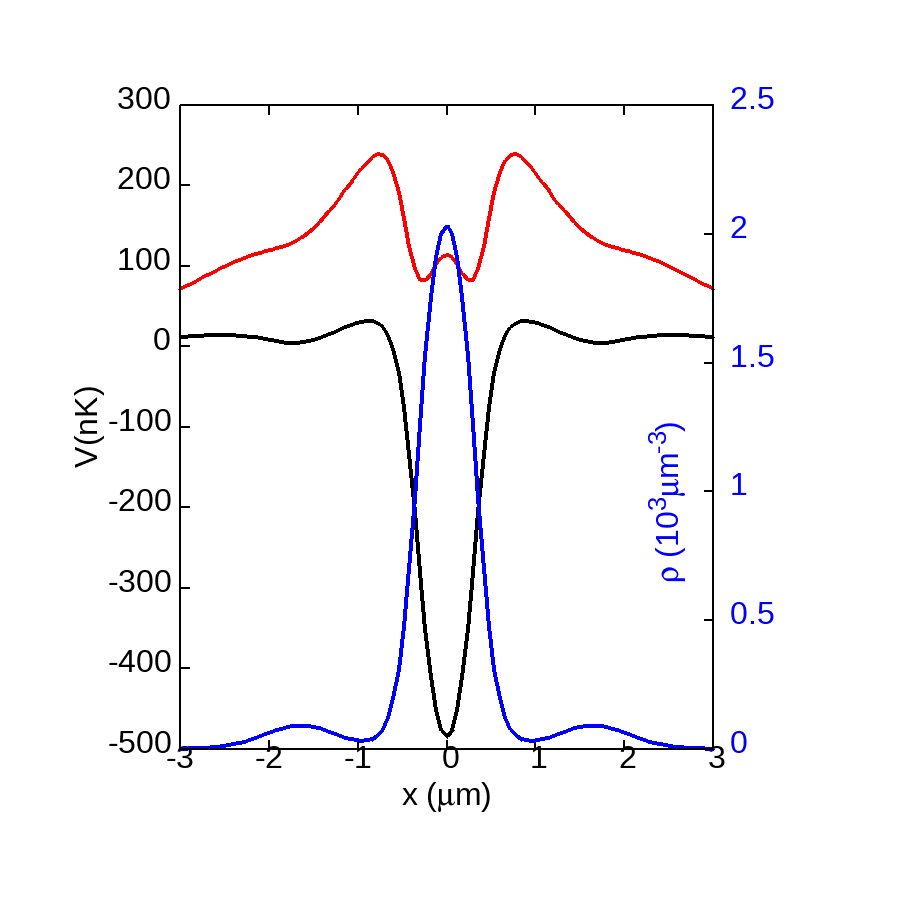}\put(-35,20){\color{black}(p)}
\end{tabular}
 \caption{
 (a-l) Snapshots of the BEC density for {$2d=3\,\mu$m}, $\varepsilon_{dd}=1.5$ and $N=8.8\times 10^3$. The plots show the cross-sections taken at $z=0$ with $x\in[-4\mu$m,$4\mu$m] and $y\in[-4\mu$m,$4\mu$m].
 The colorscale for the density is given in units of $1000\mu$m$^{-3}$ . Plots (a-l) show the moments $t$ after the removal of the inter-well barrier with $t=0$ (a), 1.93 ms(b), 5.56 ms(c), 8.71 ms(d), 10.4 ms(e), 11.6 ms(f), 13.49 ms(g), 16.44 ms(h), 17.68 ms(i), 19.54 ms(j), 20.8 ms(k) and 22.13 ms(l). (m-p)  The dipolar potential (black lines, left axis), the total interaction potential multiplied by 6 (red lines, left axis) and the density along the $z=0$ and $y=0$ line (blue lines, right axis). The plots (m-p)  correspond to $t=1.93$ ms, 5.56 ms, 8.71 ms, and 10.4 ms, respectively, i.e., to panels (b-e).}
\label{gd3}
\end{figure}

\subsubsection{The collisions of the two droplets: a giant droplet formation vs oscillations of a droplet pair}

Formation of a stable droplet is observed for in the system dynamics for $N = 8800$ atoms (Fig.~\ref{gd3})  with the initial state corresponding to region III of the double-well potential (see Fig.~\ref{difaz1k5_1500}), where two symmetric droplets are formed in both potential wells.  
In contrast, for the single-well potential (see Fig.~\ref{difaz1k5_1500b}), the value of $N=8800$ lies in region II, corresponding to a single droplet located at the center of the trap in the lowest-energy configuration.  

After switching to the single-well potential, the droplets are driven toward each other and move against the potential barrier formed by the repulsive tails of their dipolar interaction.  
Figure~\ref{gd3}(m)-(p) shows cross sections of the total interaction potential (red lines), its dipolar contribution (black lines), and the condensate density (blue lines) during the formation of a single droplet.  
The barrier due to the dipolar interaction is effectively reduced by the contact interaction and the LHY correction, which increase the local energy in regions of high density.  
For $N = 8800$, the droplets overcome the barrier and merge into a single large droplet [Fig.~\ref{gd3}(e)-(l)], surrounded by a low-density halo that evolves in time.  

\begin{figure}[htbp]
\begin{tabular}{ll}
%trim=left botm right top
\includegraphics[width=0.24\textwidth]{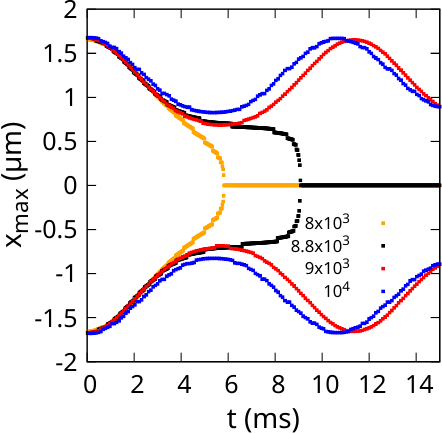} \put(-15,25){\color{black}(a)}&
\includegraphics[width=0.24\textwidth]{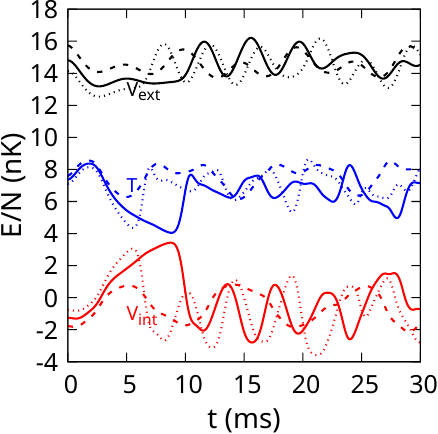} \put(-20, 25){\color{black}(b)}
\end{tabular}
\begin{tabular}{lll}
  \includegraphics[width=0.15\textwidth, trim={1.3cm 3cm 0 0}, clip]{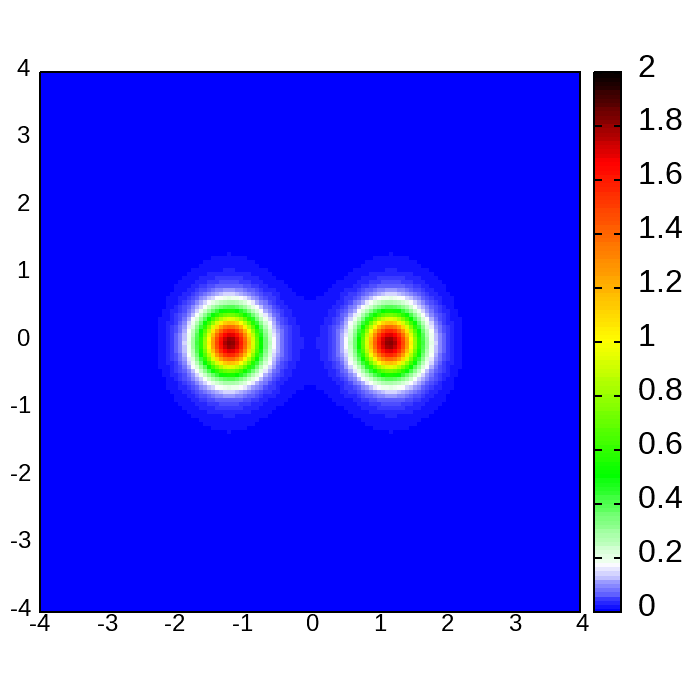}\put(-25,5){\color{yellow}(c)} &
  \includegraphics[width=0.15\textwidth, trim={1.3cm 3cm 0 0}, clip]{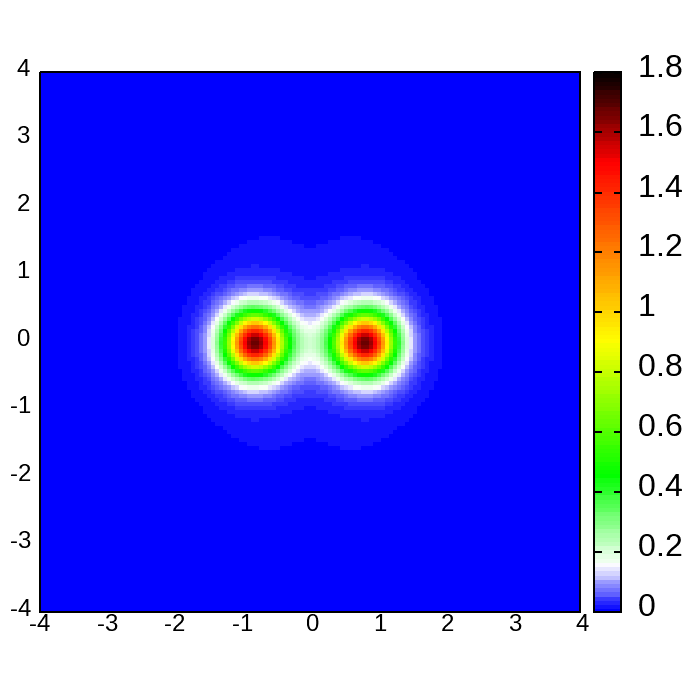}\put(-25,5){\color{yellow}(d)}&
   \includegraphics[width=0.15\textwidth, trim={1.3cm 3cm 0 0}, clip]{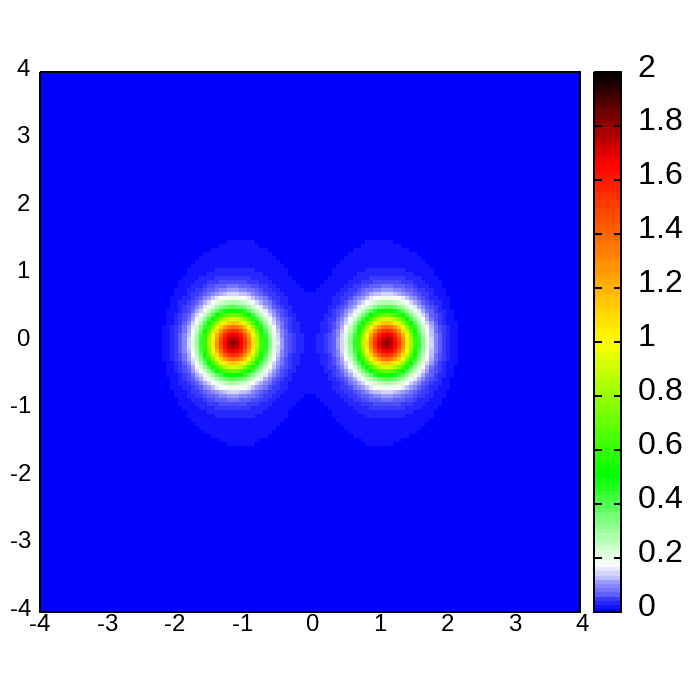} \put(-25,5){\color{yellow}(e)}  \\
    \includegraphics[width=0.15\textwidth, trim={1.3cm 3cm 0 0}, clip]{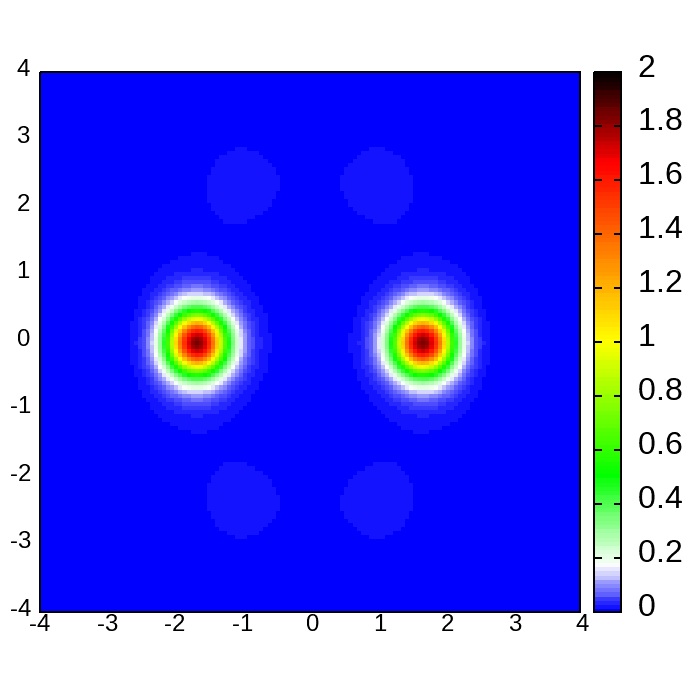} \put(-25,5){\color{yellow}(f)} &
   \includegraphics[width=0.15\textwidth, trim={1.3cm 3cm 0 0}, clip]{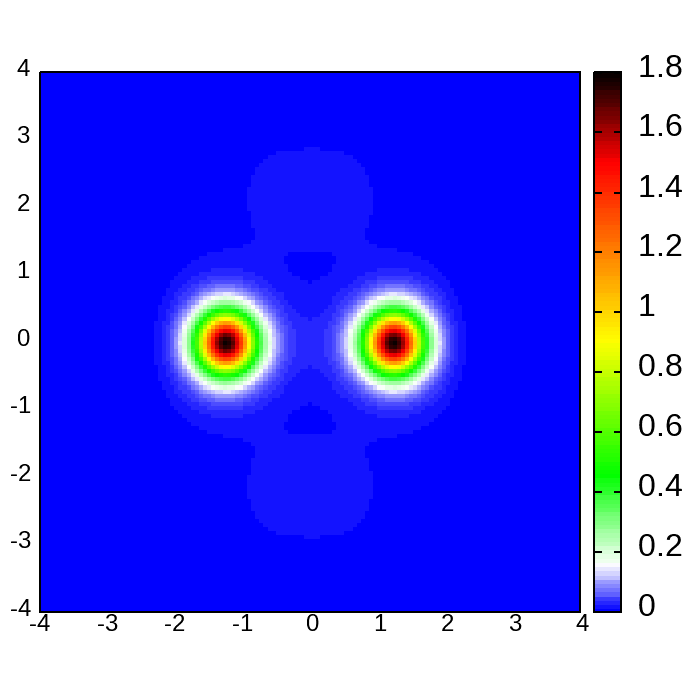}\put(-25,5){\color{yellow}(g)} &
  \includegraphics[width=0.15\textwidth, trim={1.3cm 3cm 0 0}, clip]{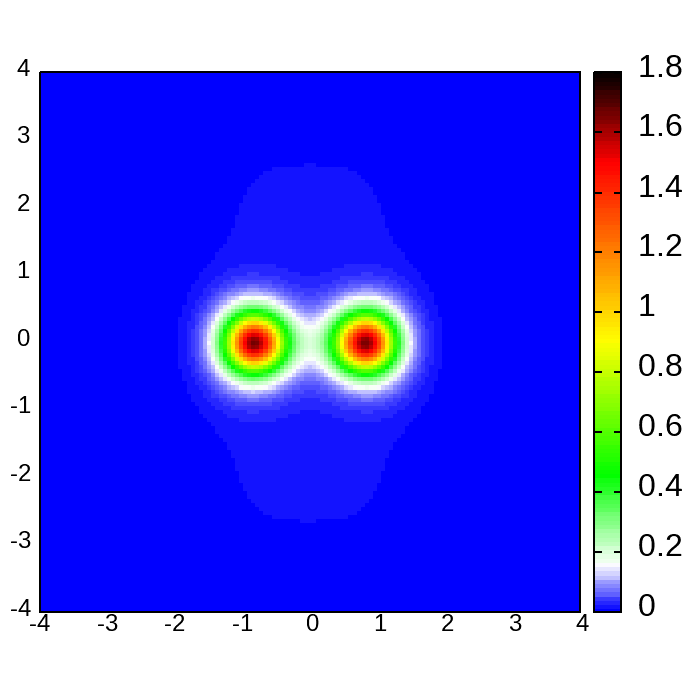}\put(-25,5){\color{yellow}(h)} 
\end{tabular}
\begin{tabular}{l}
\includegraphics[width=0.3\textwidth]{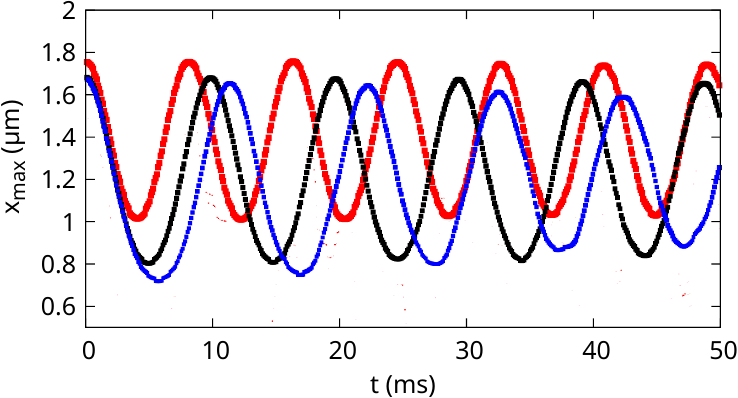}\put(-15,20){\color{black}(i)}\\
\includegraphics[width=0.3\textwidth]{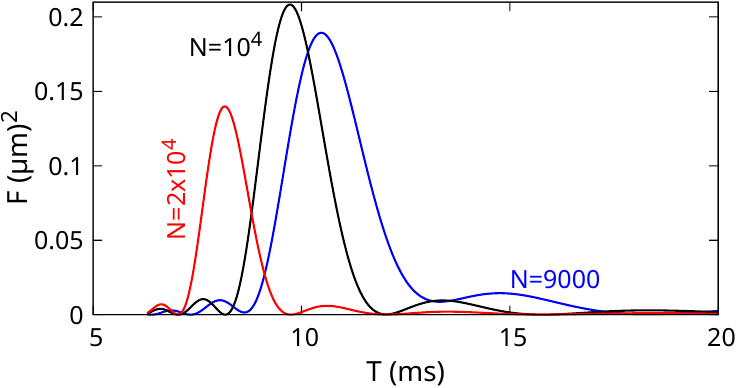}\put(-15,20){\color{black}(j)}\\
\end{tabular}
 \caption{
 (a-b) The black lines show the results for parameters of  Fig.\ref{gd3} with $N=8800$. Values for varied $N$ are plotted in different colors. 
 Panel (a) shows
 the positions of the centers of the two droplets for $N=8000$ (orange points), $N=8800$ (black points), $N=9\times 10^3$ (red points), and $N=10^4$ (blue points).
 Panel (b) shows contributions to the energy for $N=8000$ (dotted lines), $N=8800$ (solid lines) and $N=10^4$ (dashed lines).
 (c-h) Snapshots of the time evolution for $N=10^4$. The plots show the cross section taken at $z=0$ with $x\in[-4\mu$m,$4\mu$m] and $y\in[-4\mu$m,$4\mu$m]. The colorscale for the density is given in  units of $1000\mu$m$^{-3}$. Plots show the moments $t$ after the confinement potential is set to a single well, with 2.42 ms (c), 4.84 ms (d), 7.26 ms(e), 9.67 ms(f), 12.09 ms(g) and 14.5 ms(h). 
 (i) Position of the maximum of the droplet at the right-hand side of the well for $N=9000$ (blue line) $N=10^4$ (black line) and $N=2\times 10^4$ (red line). (j)
Fourier analysis of panel (i) with $F(T)=|\frac{2}{T_{max}}\int_0^{T_{max}} \left(x(t)-\langle x\rangle \right) \exp(i\frac{2\pi}{T})dt|^2 $. The maximum of $F(T)$ indicates the period of the oscillations, and its value is proportional to the square of their amplitude.
 }
\label{prog}
\end{figure}

Figure~\ref{prog}(a) shows the positions of the density maxima along the $x$ axis for $N = 8000$, $8800$, $9000$, and $10^4$ atoms.  
For the two lower values of $N$, the droplets merge into a single droplet. For the larger values, the potential barrier between the droplets is sufficiently strong to prevent a merger.
Notably, for the two larger values of $N$ considered, the lowest-energy configuration in the single-well potential corresponds to a single droplet [see Fig.~\ref{difaz1k5_1500b}(b)], but this state is not reached dynamically due to the barrier formed by the tails of the dipolar potential and the initial excess of the energy, which is not sufficient to overcome the barrier.  

Figure~\ref{prog}(b) shows the time evolution of the energy contributions for $N = 8000$ (dotted lines), $8800$ (solid lines), and $10^4$ (dashed lines).  
For $N = 8800$, the merging of the two droplets occurs between $t = 8.71$ ms [Fig.~\ref{gd3}(d)] and $t = 10.4$ ms [Fig.~\ref{gd3}(e)].  
Prior to merging, the interaction energy increases due to the overlap of the dipolar tails. It then drops once the single droplet is formed.  
The merging process is preceded by a decrease in kinetic energy, associated with the near-zero relative velocity of the droplets and reduced localization during coalescence. After merging, the kinetic energy increases again as the droplet becomes more localized.  
Subsequent oscillations of the interaction energy reflect breathing modes of the merged droplet.  

For $N = 8000$ (dotted lines), the merger occurs earlier, at $t \simeq 5$ ms.  
For both $N = 8000$ and $N = 8800$, where merger occurs, the interaction energy reaches values below its initial level due to formation of a droplet with attractive interaction along the droplet axis. This behavior is absent for $N = 10^4$, where no merging takes place.  

Figure~\ref{prog}(c)-(h) shows snapshots of the dynamics for $N = 10^4$. In this case, the droplets approach each other closely but do not merge.  
The energy contributions (dashed lines in Fig.~\ref{prog}(b)) exhibit oscillatory behavior without signatures of droplet formation.  
In general, for $\varepsilon_{dd} = 1.5$ and {$2d = 3\,\mu$m}, we find that for $N \geq 10^4$ the number of droplets remains constant during the evolution.  

Figure~\ref{prog}(i) shows the position of the density maximum at $x > 0$ as a function of time for $N = 9000$, $10^4$, and $2 \times 10^4$.  
As $N$ increases, the minimum distance between droplets increases due to the stronger inter-droplet barrier. Consequently, the oscillation frequency increases with $N$.  

The droplet motion is approximately periodic. Fitting the function
\begin{equation}
f(t) = c \exp(-t/\tau)\cos(2\pi t/T) + d
\end{equation}
to the trajectory of the density maximum yields oscillation lifetimes of $\tau = 92.5$ ms for $N = 9000$, $\tau = 328.8$ ms for $N = 10^4$, and $\tau = 775.23$ ms for $N = 2 \times 10^4$.   We can thus see that for $N = 9000$, which is close to the merging threshold [see Fig.~\ref{prog}(a)], the oscillation amplitude decays more rapidly.  
Close encounters between droplets deform their shape and transfer kinetic energy into internal (breathing) modes, leading to damping of the droplet oscillations.  

As the minimum distance between droplets increases during successive oscillations, the effective oscillation amplitude decreases and the apparent period shortens.  
For $N = 9000$, the intervals between successive maxima are $11.4$, $10.9$, $10.2$, and $9.8$ ms.  
In general, the effective period decreases as the oscillation amplitude decreases. This effect becomes weaker for larger $N$, where the droplets remain more separated.  
For $N = 10^4$, the corresponding intervals are $9.86$, $9.71$, $9.70$, and $9.63$ ms, while for $N = 2 \times 10^4$ the variation is barely noticeable.  The average oscillation periods are $T = 10.8$ ms, $9.79$ ms, and $8.17$ ms for $N = 9000$, $10^4$, and $2 \times 10^4$, respectively.  
In general, the agreement with the fitting function $f(t)$ improves for larger $N$.  

Figure~\ref{prog}(j) shows the Fourier transform of the oscillations of $x_{\mathrm{max}}(t)$
\begin{equation}
F(T) = \left| \frac{2}{T_{\mathrm{max}}} \int_0^{T_{\mathrm{max}}} \left(x_{\mathrm{max}}(t) - \langle x_{\mathrm{max}}\rangle \right)
\exp\left(\frac{2\pi i}{T} t\right) dt \right|^2,
\end{equation}
where the peak of $F(T)$ indicates the dominant oscillation period, and its magnitude is proportional to the squared amplitude.  
The largest amplitude is observed for $N = 10^4$ (see also Fig.~\ref{prog}(i)), while the oscillation period decreases with increasing $N$.

\begin{figure*}[tbp]

%trim=left botm right top
\begin{tabular}{llll}
\includegraphics[width=0.28\textwidth]{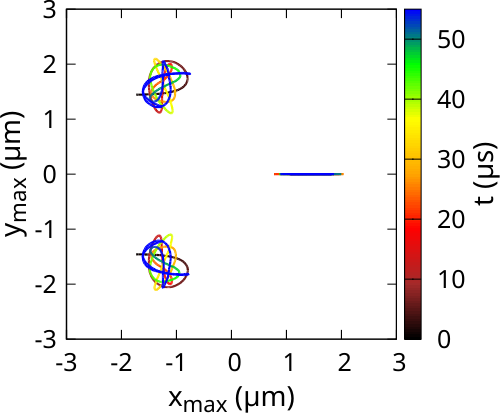}\put(-45,25){\color{black}(a)}&
\includegraphics[width=0.22\textwidth]{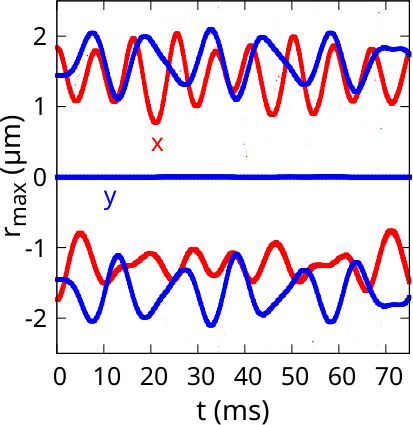}\put(-20,25){\color{black}(b)}&
\includegraphics[width=0.23\textwidth]{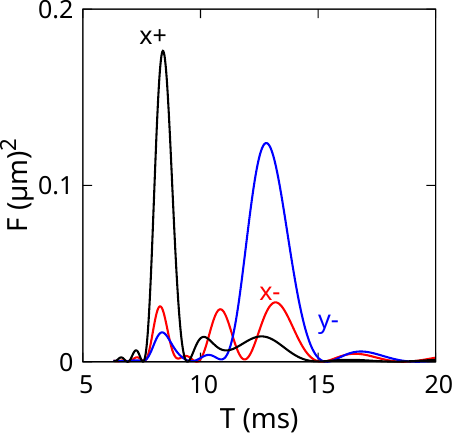}\put(-20,25){\color{black}(c)}&
\includegraphics[width=0.22\textwidth]{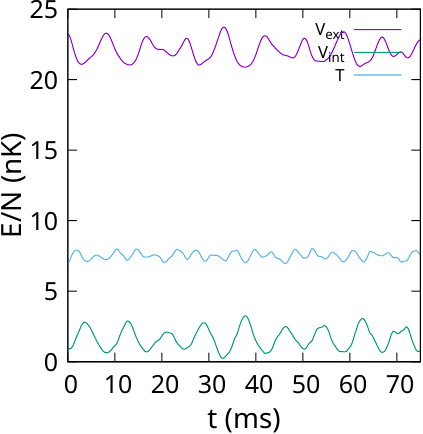}\put(-15,35){\color{black}(d)}\\
\end{tabular}
\caption{Results for $N=3\times 10^4$ and the initial state prepared for the double well with {$2d=3\,\mu$m} with $\varepsilon_{dd}=1.5$. (a) The trajectories of the droplets maxima. The time is given by the colorscale. (b) $x$ (red lines) and $y$ (blue lines) positions of the droplets maxima as functions of time. (c) Fourier transforms of the positions in time, with  $x+$ corresponding to the $x$ position of the droplet at $x>0$ side of the origin. $x-$ and $y-$ correspond to the positions of one of the droplets at the $x<0$ side of the origin. (d) Contributions to the energy}
\label{3e41k51k5}
\end{figure*}

\begin{figure*}[tbp]
\begin{tabular}{llll}
\includegraphics[width=0.3\textwidth]{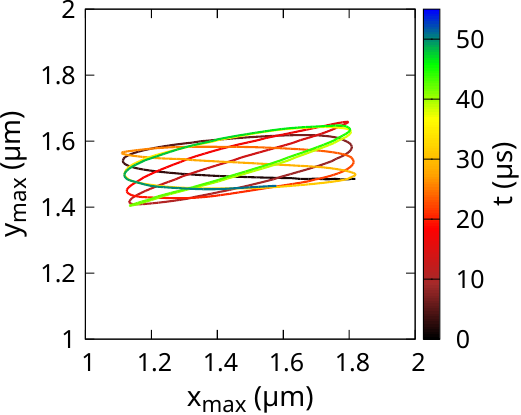}\put(-45,25){\color{black}(a)} &
\includegraphics[width=0.22\textwidth]{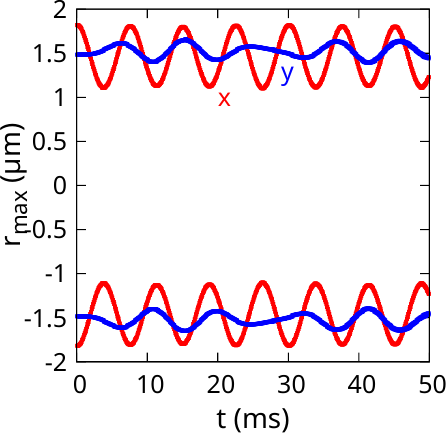} \put(-20,45){\color{black}(b)}&
\includegraphics[width=0.23\textwidth]{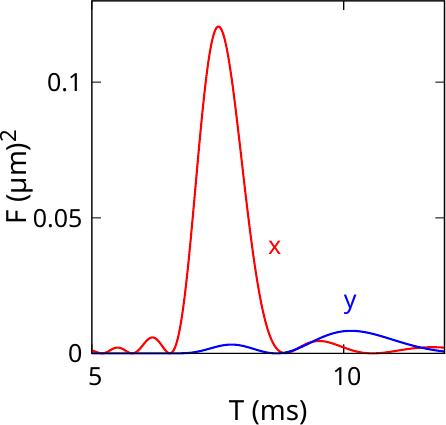}\put(-13,25){\color{black}(c)}&
\includegraphics[width=0.22\textwidth]{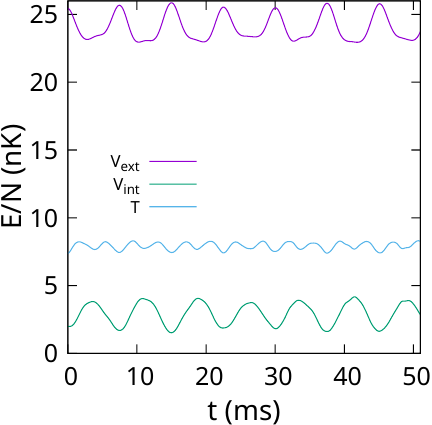}\put(-18,35){\color{black}(d)} \\
\end{tabular}
\caption{Same as Fig. \ref{3e41k51k5} but for $N=4\times 10^4$. The droplets are organized in a 2x2 array (see the initial state in Fig. \ref{difaz1k5_1500}(h)).
The trajectory of one of the droplets is shown in panel (a). Panel (b) shows the 
coordinates of the maxima of the droplets. (c) Fourier transform of the $x_{max}(t), y_{max}(t)$ positions of panel (b). (d) The contributions to the energy.}
\label{4e41k51k5}
\end{figure*}

 \subsubsection{asymmetric collisions: two droplets vs a single droplet}
 
Figure~\ref{3e41k51k5} shows the dynamics of the system for $N = 3 \times 10^4$, with an initial configuration corresponding to Fig.~\ref{difaz1k5_1500}(g), consisting of two droplets on one side of the origin and a single droplet on the other.  
The positions of the density maxima are shown in Fig.~\ref{3e41k51k5}(a). The droplet on the right-hand side moves along the symmetry axis, while the two droplets on the left execute confined motion within regions of area $\sim 1\,\mu\mathrm{m}^2$.  

The time dependence of the droplet positions along the $x$ and $y$ directions is shown in Fig.~\ref{3e41k51k5}(b). To analyze the periodicity, we compute the Fourier transform of the droplet trajectories. The function $F(T)$ reveals the dominant oscillation periods, with peak values proportional to the squared amplitudes.  
The Fourier spectra are shown in Fig.~\ref{3e41k51k5}(c). The components are labeled as $x-$ and $y-$ for one of the droplets on the left-hand side and $x+$ for the droplet on the right-hand side.  
The oscillations of the solitary droplet ($x+$) and the vertical motion of the left droplet ($y-$) exhibit well-defined dominant periods. In contrast, the horizontal motion of the left droplet ($x-$) shows multiple frequency components, including contributions similar to those observed in $x+$ and $y-$.  
Secondary peaks in the spectra reflect coupling between droplet motions mediated by the long-range dipolar interaction.  

 In spite of the complex motion of the centers of the droplets the oscillations of the energy contributions in time [Figure~\ref{3e41k51k5}(d)] are nearly periodic, with stable mean values and amplitudes.  

\subsubsection{Symmetric collisions: two droplets vs two droplets}

The dynamics for $N = 4 \times 10^4$ [Fig.~\ref{4e41k51k5}]  starts with an initial rectangular four-droplet configuration [Fig.~\ref{difaz1k5_1500}(h)]. Upon removal of the central barrier 
the droplets oscillate around their initial positions [Fig.~\ref{4e41k51k5}(a)], with a larger amplitude along the $x$ direction, where the trapping potential is modified at $t = 0$.  

The trajectory $x_{\mathrm{max}}(t)$ is well described by Eq. (13) with fitted parameters $d = 1.464\,\mu\mathrm{m}$, $T = 7.531$ ms, $c = 0.345\,\mu\mathrm{m}$, and a very long oscillation decay time of $\tau = 2.5$ s.  

The oscillation period in the $y$ direction is longer [Fig.~\ref{4e41k51k5}(b),(c)].  The external potential and interaction energies oscillate nearly periodically and in antiphase [Fig.~\ref{4e41k51k5}(d)], while the kinetic energy exhibits smaller-amplitude oscillations at approximately twice the frequency, reflecting contributions from motion of each  droplet and the degree of its internal localization.

\subsection{Results for $\varepsilon_{dd}<1.5$ and $d>3\mu$m}
{ The results for a smaller value of the dipolar interaction parameter $\varepsilon_{dd}=1.45$ and 1.4 are given in the Appendix \ref{sec:edd145} and \ref{sec:edd140}.
The phase diagrams corresponding to the results of Fig. 2 are simpler for lower values of $\varepsilon_{dd}$, e.g. contain a smaller number of phases, and for $\varepsilon_{dd}=1.4$ we do not find the states with the density of the condensate of lower symmetry than the confinement potential. Similar droplet systems appear for a larger number of atoms $N$ than for $\varepsilon_{dd}=1.5$.
For softer dipolar potential we find more pronounced halos of the droplets and a faster damping of the oscillations of the droplets.

The Appendix also contains dynamics for $\varepsilon_{dd}=1.5$ but for larger distance between the centers of the wells $2d$, e.g. 5 and 7$\mu$m. For larger $d$ the system acquires a larger initial excess of energy when the barrier is removed. In consequence, a pair of droplets upon removal of the optical potential barrier forms a giant droplet overcoming the long-range dipolar interaction barrier for larger values of $N$ than for $2d=3\mu$m. In general for values of $N$ that are large enough to prevent the merger, i.e. in the oscillatory regime, the oscillations of the droplet positions are damped faster than for $2d=3\mu$m due to closer encounters of the droplets during the oscillation. Also, for a 2+1 initial droplet configuration we observe coalescence of the droplet couple that are initially on the same side of the origin into a single droplet during the time evolution. A change of the oscillation axis for the droplet pair at $2d=7\mu$m is also found.}

\section{Summary and Conclusions}

We have
investigated the static configurations and
post-quench dynamics of dipolar $^{164}$Dy
condensates
in a double-well potential
within
the extended Gross-Pitaevskii
{model including}
the Lee-Huang-Yang correction.
The calculations cover $\varepsilon_{dd}=1.4$, $1.45$, and $1.5$ for an interwell separation
$2d=3\,\mu\mathrm{m}$, with additional dynamical results for $\varepsilon_{dd}=1.5$ at
$2d=5$ and $7\,\mu\mathrm{m}$.

{
For $2d=3\,\mu\mathrm{m}$, the lowest-energy configurations depend strongly on both the atom
number $N$ and the dipolar interaction strength $\varepsilon_{dd}$. Spontaneous symmetry breaking
is most pronounced for $\varepsilon_{dd}=1.5$: an asymmetric single-droplet configuration is obtained
for $5060<N<7410$, while an asymmetric $2+1$ configuration occurs approximately for
$2.47\times10^4<N<3.075\times10^4$. For $\varepsilon_{dd}=1.45$, the transition from the
extended cloud to the two-droplet state is continuous and symmetry preserving, although an asymmetric
$2+1$ configuration is energetically favorable for
$2.34\times10^4\lesssim N\lesssim3\times10^4$. For $\varepsilon_{dd}=1.4$, no symmetry-broken
ground-state configurations were found within the investigated range $N\leq5\times10^4$.
Thus, decreasing the dipolar interaction reduces both the occurrence and the atom-number range of
asymmetric lowest-energy configurations.
}

{
For well-developed droplets at $\varepsilon_{dd}=1.5$, the post-quench evolution is largely controlled
by the competition between the excess energy released by removal of the central barrier and the
repulsive inter-droplet barrier associated with the long-range dipolar interaction. For the initial
two-droplet states at $2d=3\,\mu\mathrm{m}$, merging is found for the calculated cases $N=8000$
and $8800$, whereas the droplets remain separated for $N=9000$, $10^4$, and $2\times10^4$.
The location of this crossover is not universal. Increasing the initial separation to
$2d=5\,\mu\mathrm{m}$ increases the approximate upper atom number for two-droplet merging to
$N\simeq1.65\times10^4$, while for $2d=7\,\mu\mathrm{m}$ and $N=2\times10^4$ a transient
giant droplet is formed and subsequently splits into two droplets aligned along the $y$ direction.
}

{
In the non-merging two-droplet regime at $\varepsilon_{dd}=1.5$ and $2d=3\,\mu\mathrm{m}$,
the droplet motion is approximately periodic. The average oscillation period decreases from $10.8$ ms
for $N=9000$ to $9.79$ ms for $N=10^4$ and $8.17$ ms for $N=2\times10^4$.
Close droplet encounters excite internal, predominantly breathing-like modes and provide a mechanism
for damping of the droplet motion. Consistently, the fitted decay time increases from $92.5$ ms
at $N=9000$ to $328.8$ ms at $N=10^4$ and $775.23$ ms at $N=2\times10^4$, where the
droplets remain farther apart. At $2d=5\,\mu\mathrm{m}$ and $N=2\times10^4$, the stronger
deformation induced by the larger quench energy reduces the fitted lifetime to $125$ ms.
}

{
The energetic signature of merging is also configuration dependent. For the two-droplet mergers
considered at $\varepsilon_{dd}=1.5$ and $2d=3\,\mu\mathrm{m}$, formation of the merged droplet
is accompanied by a decrease of the interaction energy together with a redistribution of kinetic
energy. In contrast, for the asymmetric $1+2$ configuration at $2d=5\,\mu\mathrm{m}$, the
interaction energy after merging exceeds its initial value. The detailed energy evolution therefore
depends on both the initial droplet configuration and the energy deposited by the quench.
}

{
The regular two-droplet dynamics does not extend uniformly to weaker dipolar interactions or to
configurations of lower symmetry. For $\varepsilon_{dd}=1.45$, irregular merging and splitting occur
at $N=10^4$, whereas two main droplets undergo damped oscillations at $N=2\times10^4$, an
asymmetric $2+1$ state at $N=2.75\times10^4$ exhibits several characteristic frequencies, and the
four-droplet configuration at $N=4\times10^4$ shows more regular oscillations with periods of
$8.24$ ms and $11$ ms along the $x$ and $y$ directions, respectively. For
$\varepsilon_{dd}=1.4$, strong density redistribution and a changing number of maxima are found for
$N=2\times10^4$, whereas the droplet number remains fixed and approximately periodic damped motion
is obtained for $N=4\times10^4$.
}

{
Beyond the specific parameter values considered here, the numerical results identify three more
general control mechanisms. First, the occurrence of symmetry-broken lowest-energy configurations
is governed by the relative strength of the dipolar interaction, and their stability range is strongly
reduced as $\varepsilon_{dd}$ is decreased from $1.5$ to $1.4$. Second, the merger-oscillation
crossover is controlled by the competition between the post-quench excess energy and the effective
inter-droplet repulsive barrier: increasing the initial separation raises the available quench energy
and shifts the crossover toward larger atom numbers. Third, in the non-merging regime the oscillation
time scale is set by the combined confinement and inter-droplet interaction, whereas damping is
enhanced by close encounters that transfer the droplet energy into internal droplet excitations.
These mechanisms, expressed in terms of energy and interaction scales rather than the specific
$^{164}$Dy parameters, are expected to provide the most transferable characterization of the
dynamics considered here.
} \color{black}

\section*{Acknowledgments}
This research was partially supported by a subsidy from the Polish Ministry of Science and Higher Education.
We gratefully acknowledge the Polish high-performance computing infrastructure PLGrid (HPC Center: ACK Cyfronet AGH) for providing computer facilities and support within computational grant no. PLG/2026/019337.

\appendix

\section{Methods}
\label{sec:methods}

In this section we describe numerical methods used in this work.
We considered a time-dependent 3D extended Gross-Pitaevskii equation
\begin{equation}
i\hbar \frac{\partial \Psi}{\partial t} = \hat{H}_{\mathrm{eGP}} \Psi({\bf r},t)
\label{tegpe},
\end{equation}
with the Hamiltonian
\begin{equation}  
\label{hgp_appendix}
\begin{aligned}  
\hat{H}_{\mathrm{eGP}} =
& -\frac{\hbar^2}{2m}\nabla^2 + V_{\mathrm{ext}}({\bf r},t) \\
& + g |\Psi({\bf r},t)|^2 + \gamma(\varepsilon_{dd}) |\Psi({\bf r},t)|^3 \\
& + V_{dd}({\bf r},t),
\end{aligned}
\end{equation}
where the third term describes the contact interaction, $\gamma(\varepsilon_{dd}) |\Psi({\bf r},t)|^3$ is the LHY correction accounting for quantum fluctuations beyond the mean-field approximation and $V_{dd}$ is the potential stemming from dipole-dipole interaction.

\subsection{Ground state}

For the determination of the ground state  the technique of the evolution in imaginary time has been used.
General solution of time dependent equation \eqref{tegpe} can be presented by using the evolution operator
\begin{equation}
    \Psi(\vb{r}, t) = \exp\left[-\frac{it}{\hbar}\hat{H}_{\mathrm{eGPE}}\right]\Psi(\vb{r}, 0).
\end{equation}
 After Wick's transformation $\tau = it$ into imaginary time, evolution takes the form
\begin{equation}
    \Psi(\vb{r}, \tau) = \exp\left[-\frac{\tau}{\hbar}\hat{H}_{\mathrm{eGPE}}\right]\Psi(\vb{r}, 0).
\end{equation}
As any state can be expanded in the basis of eigenstates of the operator $\hat{H}_{\mathrm{eGPE}}$
\begin{align}
    \Psi(\vb{r}, \tau) &= \exp\left[-\frac{\tau}{\hbar}\hat{H}_{\mathrm{eGPE}}\right]\sum_n c_n \psi_n(\vb{r}, 0) \\
    & \nonumber = \sum_n \exp\left[ - \frac{\tau}{\hbar} E_n\right] c_n \psi_n(\vb{r}, 0), \\
    & \nonumber = \exp\left[-\frac{\tau}{\hbar} E_0 \right]\sum_n  \exp\left[ - \frac{\tau}{\hbar} (E_n - E_0)\right] c_n \psi_n(\vb{r}, 0),
\end{align}
where $c_n \psi_n(\vb{r}, 0)$ is the contribution of $n$-th state to the initial state of the evolution in imaginary time and $E_n$ the  n-th state energy.
Considering that the state energies form an ordered increasing sequence, the coefficient for $E_n > E_0$ are negligible in the limit $\tau \to \infty$.
As $\tau \to \infty$, all excited-state components decay exponentially, while the ground state component remains.
Consequently, by evolving any wavefunction in imaginary time and normalizing at each step, one can project onto the ground state of the system.

Used iteration in imaginary time applies forward Euler method to equation \eqref{tegpe} transformed into imaginary time, thus getting an approximation
\begin{equation}
    \frac{\partial \Psi(\vb{r}, \tau)}{\partial \tau} \approx \frac{\Psi(\vb{r}, \tau + \Delta \tau) - \Psi(\vb{r}, \tau)}{ \Delta \tau}, 
\end{equation}
which results in
\begin{equation}
    \Psi(\vb{r}, \tau + \Delta \tau)  = \Psi(\vb{r}, \tau) - \Delta \tau \cdot \hat{H}_{\mathrm{eGPE}} \Psi(\vb{r}, \tau).
\end{equation}

\subsection{Real-time evolution}

For the real-time evolution split-step method has been used. We used both the Crank–Nicolson scheme and the split-step method. The latter proved preferable due to its faster computational time and the existing FFT implementation in the code. Additionally, it is more convenient for nonlinear problems, since the Crank–Nicolson scheme introduces an extra parameter — the number of iterations required by the nonlinear solver.
The results obtained by both methods are consistent with each other.
 The Hamiltonian \eqref{hgp_appendix} can be split into two parts
\begin{align}
\hat{H}_{\mathrm{eGP}} &= \hat{K} + \hat{P}, \\
    \hat{K} &\equiv -\frac{\hbar^2}{2m}\nabla^2,\\
    \hat{P} &\equiv g |\Psi({\bf r},t)|^2 + \gamma(\varepsilon_{dd}) |\Psi({\bf r},t)|^3 | \nonumber
    \\ & \;\;\;+ V_{\mathrm{ext}}({\bf r},t) + V_{dd}({\bf r},t),
\label{hsplit}
\end{align}
where $\hat{K}$ and $\hat{P}$ are kinetic and potential operators,  respectively.
The time-dependent equation \eqref{tegpe} has a general solution in the form
\begin{equation}
\begin{aligned}
    \Psi(\vb{r}, t)  
    &= \exp\left[-\frac{it}{\hbar} \hat{H}_{\mathrm{eGP}}\right]\Psi(\vb{r}, 0) \\
    &= \exp\left[-\frac{it}{\hbar} (\hat{K} + \hat{P})\right]\Psi(\vb{r}, 0).
\end{aligned}
\end{equation}
In order to approximate the solution for small time steps $\Delta t$ one can use the Trotter-Suzuki approximation
\begin{equation}
\begin{aligned}
    \exp\left[-\frac{i\Delta t}{\hbar} (\hat{K} + \hat{P})\right]
    \approx &
    \exp\left[ -\frac{i\Delta t}{2\hbar} \hat{P} \right] 
    \exp\left[ -\frac{i\Delta t}{\hbar} \hat{K} \right] \cross \\
    & \exp\left[ -\frac{i\Delta t}{2\hbar} \hat{P} \right]
    +
    \mathcal{O}(\Delta t ^3).
\end{aligned}
\end{equation}
In order to evaluate the evolution due to the kinetic part, we used the Fourier transform, as a nabla operator in reciprocal space is a simple multiplication operator
\begin{equation}
    \mathcal{F}\left\{ \exp\left[-\frac{i\Delta t}{\hbar}\hat{K}\right]\Psi \right\} = \exp\left[-\frac{i\Delta t}{\hbar}\frac{\hbar^2 k^2}{2m}\right]\mathcal{F}\{\Psi\},
\end{equation}
where $k$ is a wave vector in reciprocal space, and $\mathcal{F}$ denotes Fourier transform.
The whole kinetic step can be written as
\begin{equation}
\begin{aligned}
    \exp\left[-\frac{i\Delta t}{\hbar}\hat{K}\right]\Psi(\vb{r}, t) 
    &= \mathcal{F}^{-1}\left\{\mathcal{F}\left\{ \exp\left[-\frac{i\Delta t}{\hbar}\hat{K}\right]\Psi \right\}\right\} \\
    & = \mathcal{F}^{-1}\left\{ \exp\left[-\frac{i\Delta t}{\hbar}\frac{\hbar^2 k^2}{2m}\right]\mathcal{F}\{\Psi\}\right\}.
\end{aligned}
\end{equation}

Each time step can be described as a sequence
\begin{itemize}
    \item apply \textit{potential half-step} to the wavefunction,
    \item transform wavefunction to reciprocal space using FFT,
    \item apply kinetic kernel by multiplying every point in reciprocal space by $\frac{\hbar^2 k^2}{2m}$,
    \item transform wavefunction back to position space by applying inverse FFT,
    \item apply the second \textit{potential half-step} to the wavefunction.
\end{itemize}

\subsection{Dipolar potential}

Every time step (both in ground-state calculations and in real-time evolution) requires calculating the dipole-dipole interaction potential.
The dipole-dipole interaction between atoms in the system is given by a convolution 
\begin{equation}
    V_{dd}(\vb{r}, t) =  \int \dd^3\vb{s}\; n(\vb{s}, t) U_{dd}(\vb{r} - \vb{s}),
    \label{dd_pote}
\end{equation}
where $n(\vb{r}, t) = |\Psi(\vb{r}, t)|^2$ is normalised to the number of atoms in the system, and $U_{dd}(\vb{r})$ is the dipole-dipole interaction kernel defined as
\begin{equation}
    U_{dd}({\bf r}) = \frac{c_{dd}}{4\pi} \frac{1 - 3\cos^2\theta}{|{\bf r}|^3}, 
    \label{uj_appendix}
\end{equation}
where $c_{dd} = \frac{12\pi \hbar^2 a_{dd}}{m}$ and $\theta$ is the angle between the polarization axis (taken along $z$) and the relative position vector.
We used the Fourier transform in order to find $V_{dd}(\vb{r}, t)$.
The analytical form of the Fourier transform of the $U_{dd}$ kernel  is
\begin{equation}
    \mathcal{F}(U_{dd}) = c_{dd}\left(\frac{k_z^2}{k^2} - \frac{1}{3}\right).
\end{equation}
Now in order to find the convolution in real space, one uses the product in the reciprocal space using the identity $(\mathcal{F}(A * B) = \mathcal{F}(A) \cdot \mathcal{F}(B))$, where~$*$ is convolution.
Applying this to \eqref{dd_pote} results in

\begin{align}
    V_{dd}(\vb{r}, t)
    &= \mathcal{F}^{-1}\left\{\mathcal{F}(U_{dd}(\vb{r})) \cdot \mathcal{F}(n(\vb{r}, t))\right\} \nonumber \\
    &= \mathcal{F}^{-1}\left\{ c_{dd}\left(\frac{k_z^2}{k^2} - \frac{1}{3}\right)\cdot \mathcal{F}(n(\vb{r}, t))\right\} \nonumber \\
    &= \mathcal{F}^{-1}\left\{ c_{dd}\frac{k_z^2}{k^2} \mathcal{F}(n(\vb{r}, t)) - \frac{c_{dd}}{3} \mathcal{F}(n(\vb{r}, t))\right\} \nonumber \\
    &= \mathcal{F}^{-1}\left\{ c_{dd}\frac{k_z^2}{k^2}  \mathcal{F}(n(\vb{r}, t))\right\} - \mathcal{F}^{-1}\left\{\frac{c_{dd}}{3} \mathcal{F}(n(\vb{r}, t))\right\} \nonumber \\
    &= \mathcal{F}^{-1}\left\{ c_{dd}\frac{k_z^2}{k^2}  \mathcal{F}(n(\vb{r}, t))\right\} - \frac{c_{dd}}{3} n(\vb{r}, t).
\label{dpdp}
\end{align}

This method implicitly applies periodic boundary conditions.
For this particular step, the grid size is doubled using zero padding.
By adding additional padding, we mitigate the effects of periodic boundaries on the dipolar potential values.

\subsection{Grid and numerical validation}

A series of ground-state calculations was performed to determine the minimal required grid resolution.
In these calculations, the box size was kept constant at 15360 $\times$ 15360 $\times$ 19200 nm.

\begin{figure}[htbp]
\centering
\includegraphics[width=\linewidth]{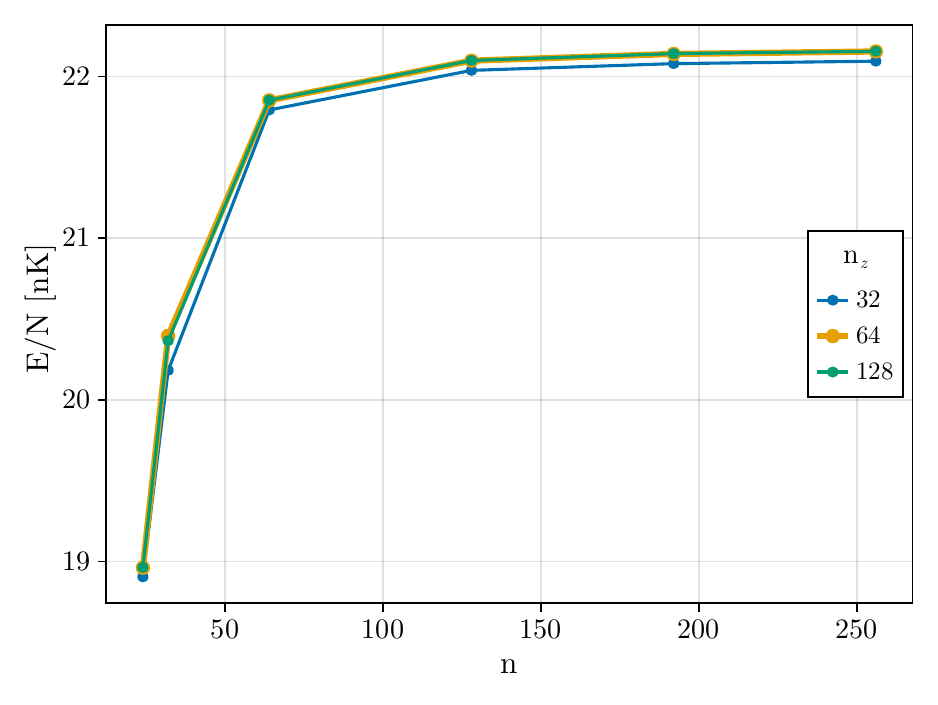}
\caption{Energy as a function of grid sizes used for grid size evaluation. $n = n_x = n_y$ is the number of nodes in $x$ and $y$ direction.}
\label{fig:gridxyz}
\end{figure}

The computational box was divided into $n_x \times n_y \times n_z$ nodes, forming a grid.
In the $xy$ directions, the box is square, with the same number of nodes, whereas in the $z$ direction, there are fewer nodes.
Higher accuracy was required in the $xy$ directions; therefore, divisions into 32, 64, 128, 192, and 256 nodes were investigated.
The nodes were chosen to be powers of two due to the discrete Fourier transform algorithm being used, which significantly accelerates the calculations.
In the $z$ direction, divisions into 32, 64, and 128 nodes were investigated.
Fig. \ref{fig:gridxyz} presents the ground state energies for $N = 20000$ atoms and $\varepsilon_{dd} = 1.5$.
As can be seen, the energy changes asymptotically with increasing resolution in the $xy$ directions.
Between 192 and 256 nodes, the energy change is approximately 0.06\% for each of the studied node counts in the $z$ direction.
In the calculations, $256 \times 256$ nodes were retained in the $xy$ plane, due to the computational speed advantage of a grid being a power of two.
In the $z$ direction, between 64 and 128 nodes, the change in the energy value was approximately 0.006\%.
In this case, the smaller grid was chosen. 

\subsection{Computational box size}

\begin{figure}[htbp]
\centering
\includegraphics[width=\linewidth]{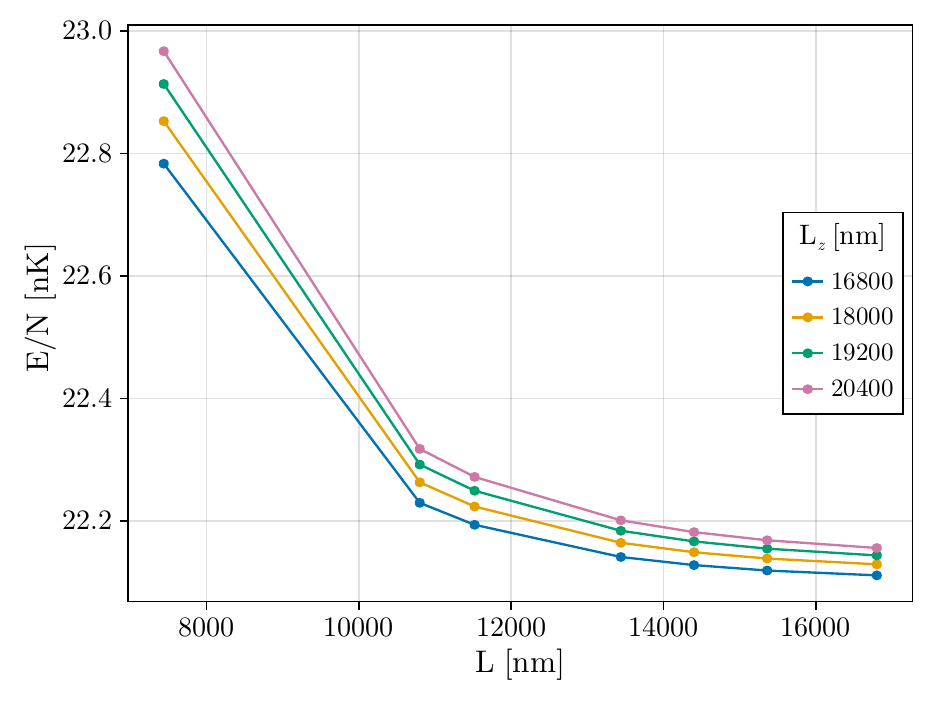}
\caption{Energy as a function of computational box size used for computational box size evaluation. $L$ is the length of the box in $x$ and $y$ direction, and $L_z$ is the length of the box in $z$ direction.}
\label{fig:sizes}
\end{figure}

Analogously to the convergence tests performed to determine the sufficient grid resolution, additional calculations were carried out to estimate the appropriate computational box size.
The grid used in these test calculations had dimensions $256 \times 256 \times 64$.
The results are presented in Fig. \ref{fig:sizes}.

As can be seen, the results converge to a single value; however, the differences with changing grid size are smaller than in the case of the resolution analysis.
In this case, a box size of $15360 \times 15360 \times 19200$ nm was chosen.
Enlarging the computational box in $z$ dimension to $20400$ nm changes energy by $\sim0.06$\% and enlarging computational box both in $x$ and $y$ dimensions to $16800$ nm changes energy by $\sim0.05$\%.

\begin{figure*}[tbp]
\begin{tabular}{ll}
    \centering
    \includegraphics[width=0.4\linewidth]{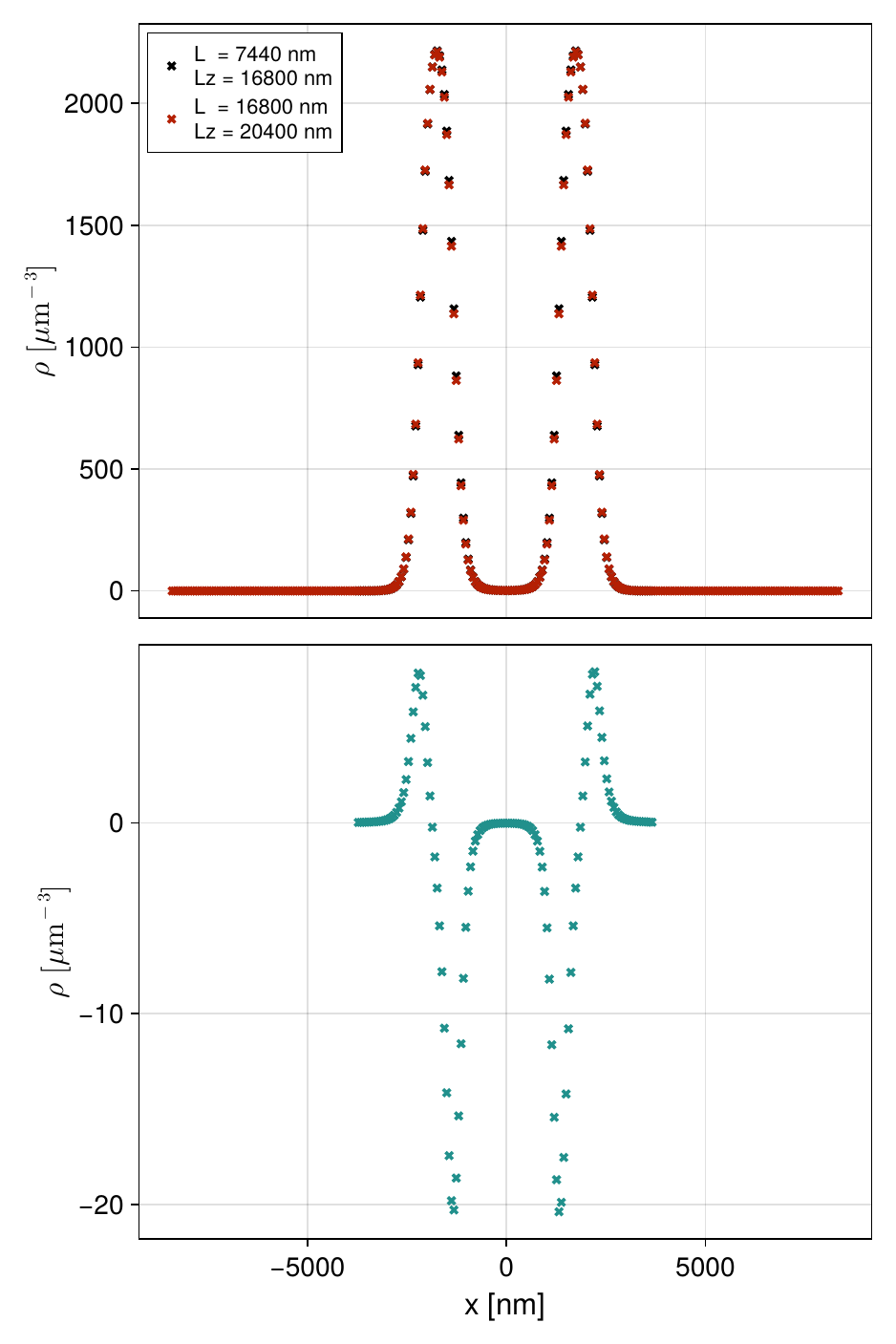}\put(-20, 200){a)}\put(-20, 38){c)} & 
    \includegraphics[width=0.4\linewidth]{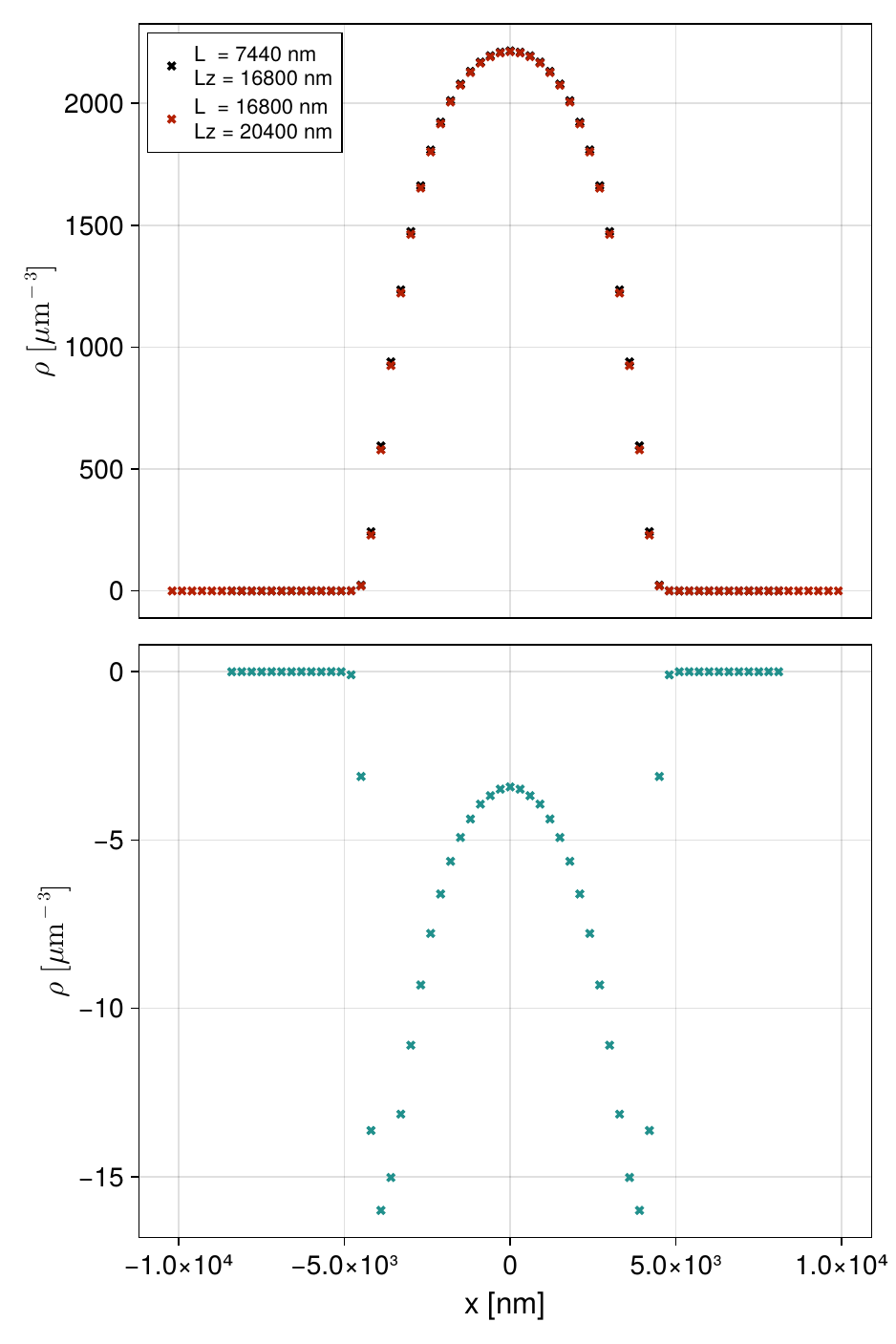}\put(-20, 200){b)}\put(-20, 38){d)}
\end{tabular}
    \caption{Density profile comparison between a small box ($L = 7440$ nm, $L_z = 16800$ nm, black markers) and a large box ($L = 16800$ nm, $L_z = 20400$ nm, brown markers). a) Profile at $z = 0$ and $y = 0$, b) $z$-axis profile through the left droplet of the two-droplet system, c) difference between larger box and smaller box across $x$ axis profile, d) difference between larger box and smaller box across $z$ axis profile.}
    \label{fig:sizes_densities}
\end{figure*}

Comparison between density profiles for a bigger and a smaller computational box size is shown in Fig. \ref{fig:sizes_densities}.
The comparisons, however, suggests that the density does not change within the range of the studied sizes of the computational box.

\subsection{Parameters}

Parameters used in the program are shown in table \ref{tab:params}.

\begin{table}[htb]
    \caption{Constant parameters used in calculations.
    \label{tab:params}}
    \begin{ruledtabular}
    \begin{tabular}{p{45mm}ll}
        Parameter  &    symbol    & Value   \\
        \hline
        grid size in x direction & $n_x$ & 256 \\
        grid size in y direction & $n_y$ & 256 \\
        grid size in z direction & $n_z$ & 64 \\
        distance per node in x direction & $d_x$ & 60 nm\\
        distance per node in y direction & $d_y$ & 60 nm\\
        distance per node in z direction & $d_z$ & 300 nm\\
        imaginary time iterations & $n_{\tau}$ & 20000 \\
        imaginary time time step & $\tau$ & 1.25e11 a.u. \\
        real time time step & $t$ & 1.00e10 a.u. \\
        trap frequency in x direction & $\omega_x$ & $2\pi \times 60 \; \frac{1}{\mathrm{s}}$ \\
        trap frequency in y direction & $\omega_y$ & $2\pi \times 60 \; \frac{1}{\mathrm{s}}$ \\
        trap frequency in z direction & $\omega_z$ & $2\pi \times 120 \; \frac{1}{\mathrm{s}}$ \\
        mass of single atom & $m$ & 163.929 Da
    \end{tabular}
    \end{ruledtabular}
\end{table}

All calculations are performed in atomic units, with all non-a.u. values converted before computation begins.
The code used for these calculations (split-step algorithm version) is available on GitHub \cite{github}.

\section{Dynamics: delocalized low-density gas for $\varepsilon_{dd}=1.5$ and $2d=3\mu\mathrm{m}$}
\label{sec:precondensation}

\begin{figure}[tbp]

%trim=left botm right top
\begin{tabular}{lll}
\includegraphics[width=0.165\textwidth, trim={1.3cm 3cm 0 0}, clip]{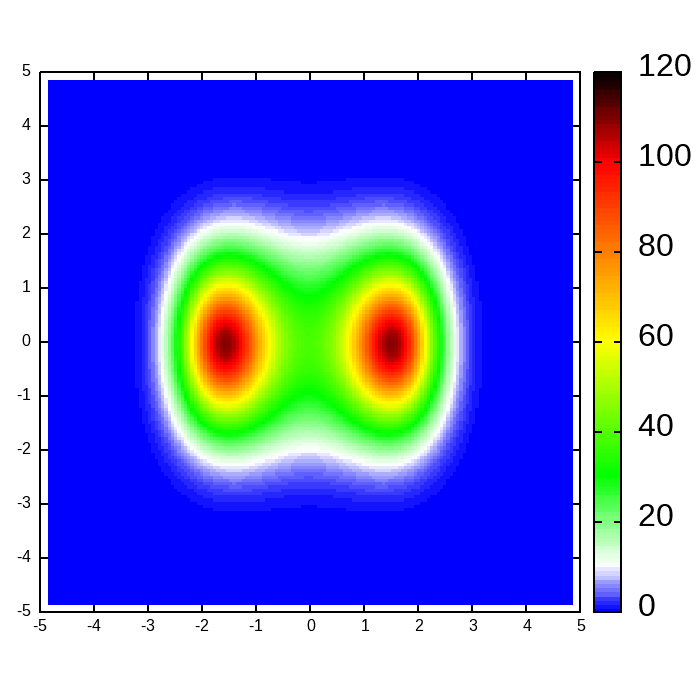}\put(-25,5){\color{yellow}(a)}&
  \includegraphics[width=0.165\textwidth, trim={1.3cm 3cm 0 0}, clip]{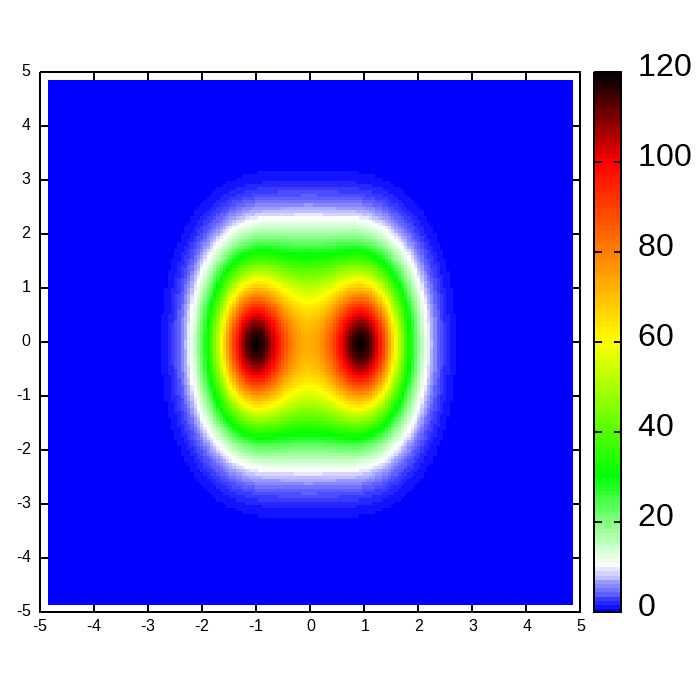} \put(-25,5){\color{yellow}(b)}&
  \includegraphics[width=0.165\textwidth, trim={1.3cm 3cm 0 0}, clip]{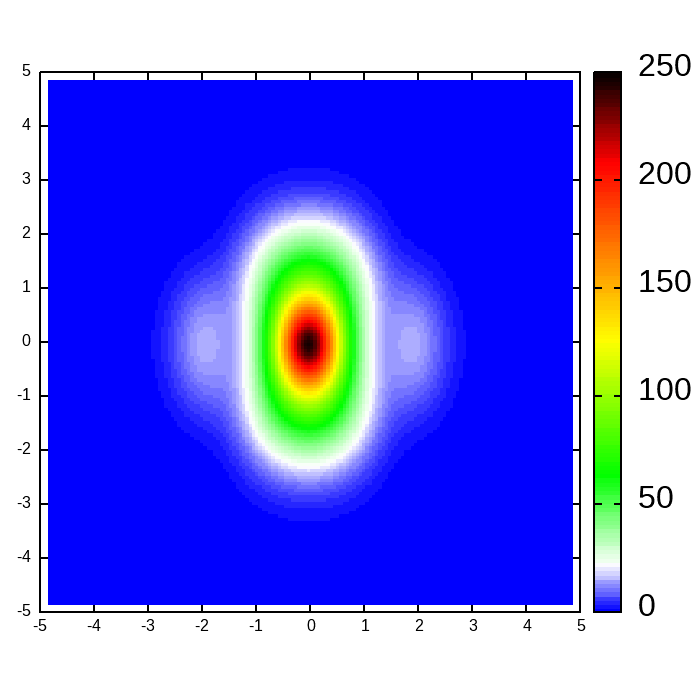}\put(-25,5){\color{yellow}(c)}\\
   \includegraphics[width=0.165\textwidth, trim={1.3cm 3cm 0 0}, clip]{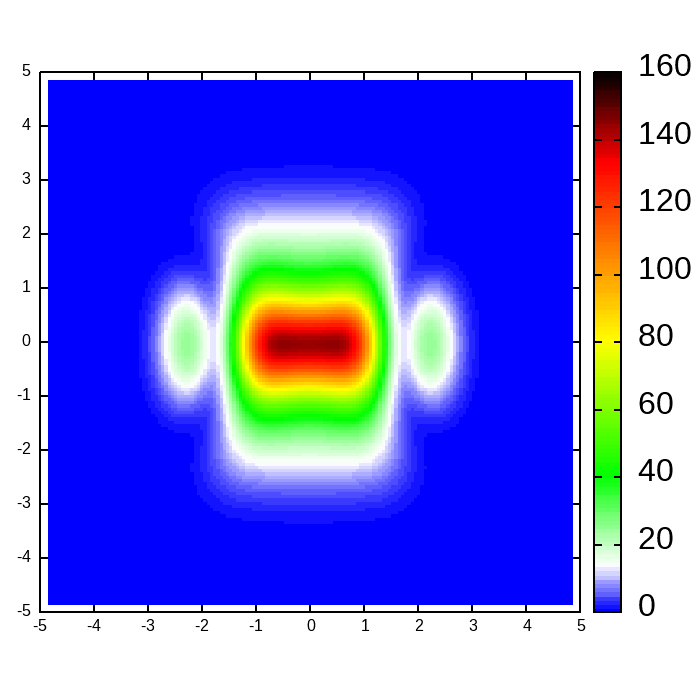}\put(-25,5){\color{yellow}(d)}    &
    \includegraphics[width=0.165\textwidth, trim={1.3cm 3cm 0 0}, clip]{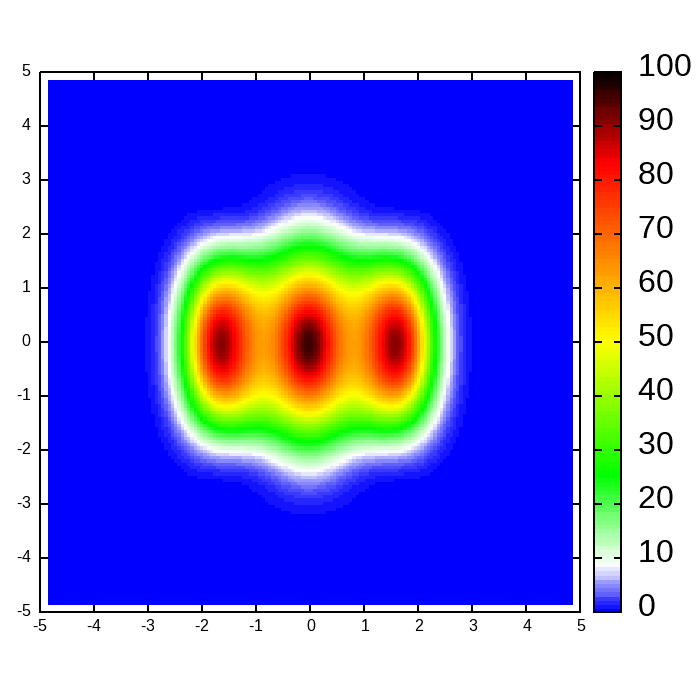}\put(-25,5){\color{yellow}(e)}  &
   \includegraphics[width=0.165\textwidth, trim={1.3cm 3cm 0 0}, clip]{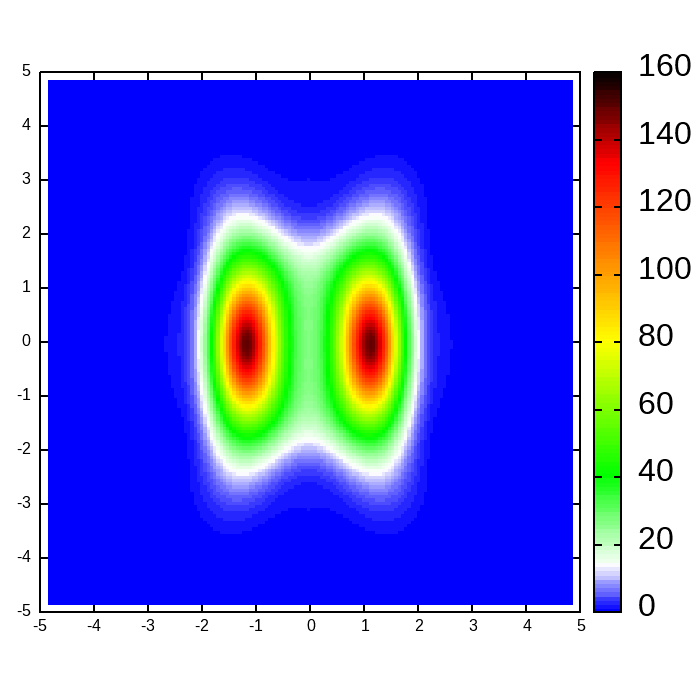} \put(-25,5){\color{yellow}(f)}\\
  \includegraphics[width=0.165\textwidth, trim={1.3cm 3cm 0 0}, clip]{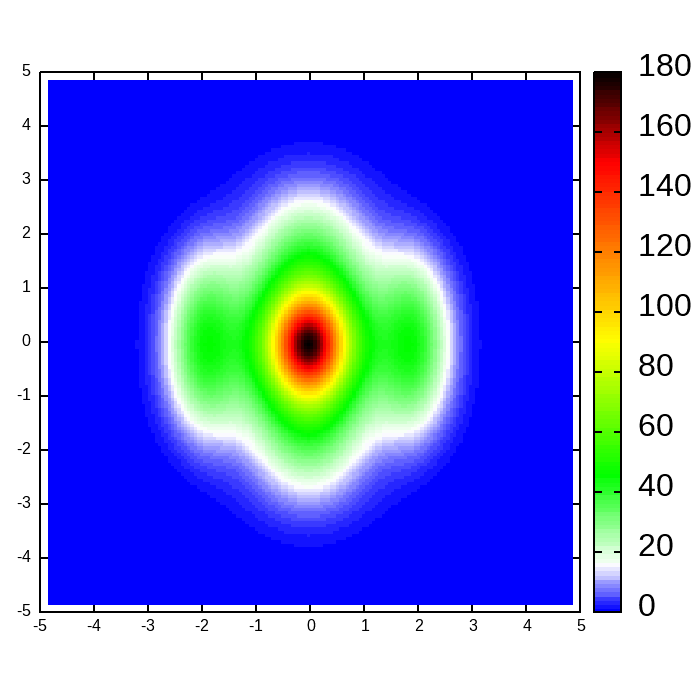} \put(-25,5){\color{yellow}(g)}& 
  \includegraphics[width=0.165\textwidth, trim={1.3cm 3cm 0 0}, clip]{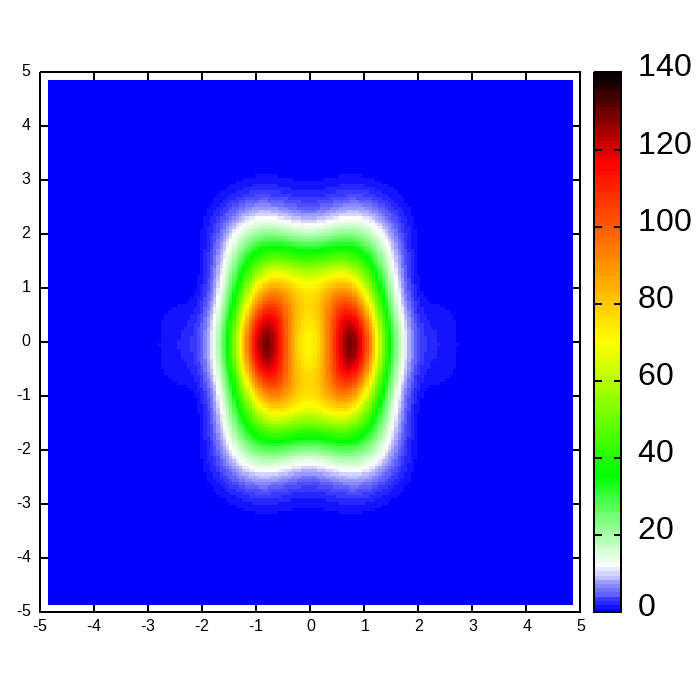} \put(-25,5){\color{yellow}(h)}&
   \includegraphics[width=0.165\textwidth, trim={1.3cm 3cm 0 0}, clip]{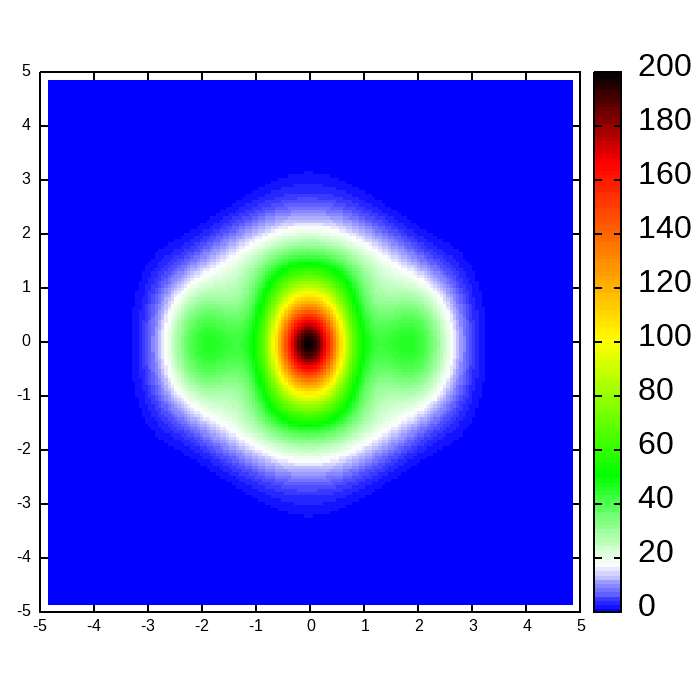} \put(-25,5){\color{yellow}(i)} 

\end{tabular}
\includegraphics[width=0.35\textwidth]
{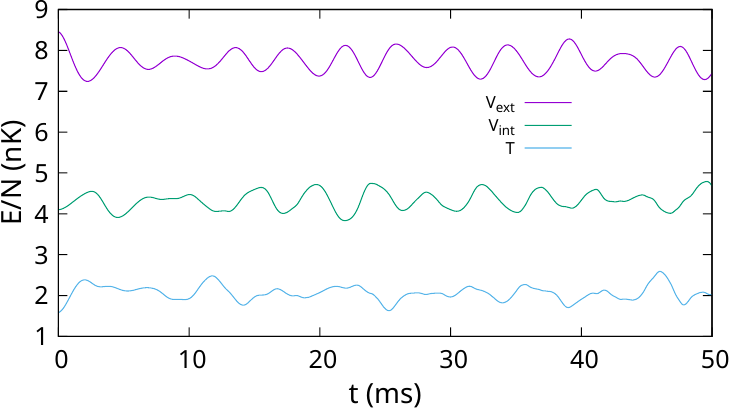}\put(-15,20){\color{black}(j)} 
\caption{Snapshots of the time evolution of the BEC density for $2d=3 \,\mu$m $\varepsilon_{dd}=1.5$ for $N=2500$ atoms. The plots show the cross section taken at $z=0$ with $x\in[-4\mu$m,$4\mu$m] and $y\in[-4\mu$m,$4\mu$m]. The color scale for the density is given in $\mu$m$^{-3}$ units. Plots (a-j) show the moments $t$ after the confinement potential is set to a single well, with $t=0$ (a), 2.42 ms (b), 4.84 ms (c), 7.26 ms (d), 9.68 ms (e), 12.1 ms (f), 14.51 ms (g), 16.93 ms (h), 19.35 ms (i). (j) Contributions to the energy per atom as functions of time.}
%densities, stepping: %25*400*1e10*2.4188e-14=2.419ms.}
\label{gd1}
\end{figure}

The paper focuses on a symmetric system with  $\varepsilon_{dd}=1.5$ and $2d=3\mu$m.
Here we provide the results for the dynamics with lower atom number than in the paper. Figure~\ref{gd1} shows the density evolution for a low atom number, $N = 2500$. For this value of $N$, no localized droplet is formed, and the system corresponds to region I in both the double-well [see Fig.2(b) of the paper] and single-well cases [see Fig.3(b)].

Snapshots of the density in the $z = 0$ plane (Fig.~\ref{gd1}) show that, after removal of the interwell barrier, the density undergoes oscillations within the trap. The number of local density maxima is not conserved during the evolution and varies between one and three.  

Figure~\ref{gd1}(j) presents the contributions to the total energy per atom. Immediately after barrier removal, the external potential energy decreases, accompanied by an increase in both the kinetic and interaction energies. During subsequent evolution, all energy components exhibit low-amplitude oscillations around their mean values, which are lower (for the external potential energy) and higher (for the kinetic and interaction energies) than their initial values.

\section{Interwell distance $2d = 3\,\mu\mathrm{m}$ and $\varepsilon_{dd} = 1.45$}
\label{sec:edd145}

\subsection{Lowest-energy configurations}

\begin{figure*}[tbp]

%trim=left botm right top
    \begin{tabular}{ll}
         \includegraphics[height=0.27\textheight]{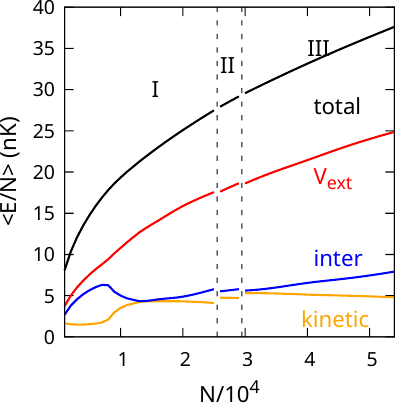} \put(-12,35){(a)}&
        \includegraphics[height=0.27\textheight]{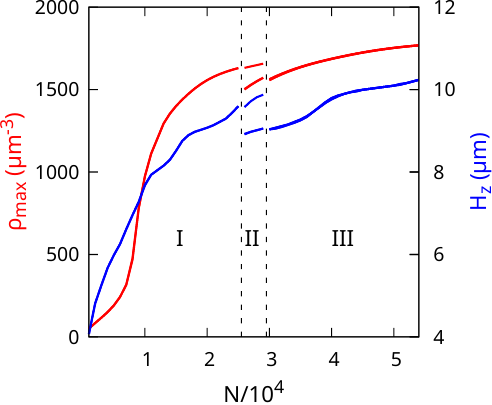} \put(-45,35){(b)} 
    \end{tabular}
    \begin{tabular}{llll}
        \includegraphics[width = 0.15\textwidth, trim={3.2cm 0cm 0cm 0cm}, clip]{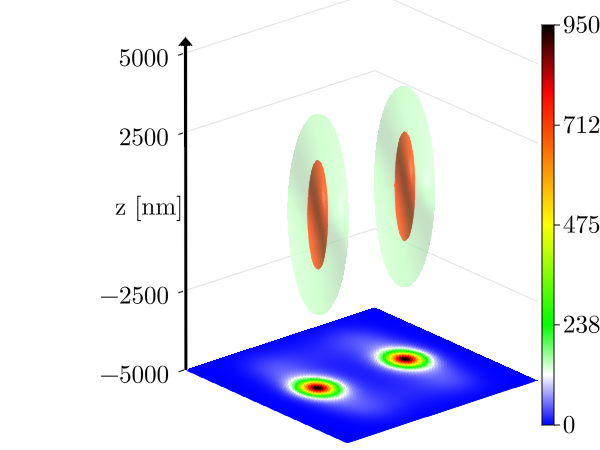} \put(-20.5,1){(c)}&
        \includegraphics[width = 0.15\textwidth, trim={3.5cm 0cm 0cm 0cm}, clip]{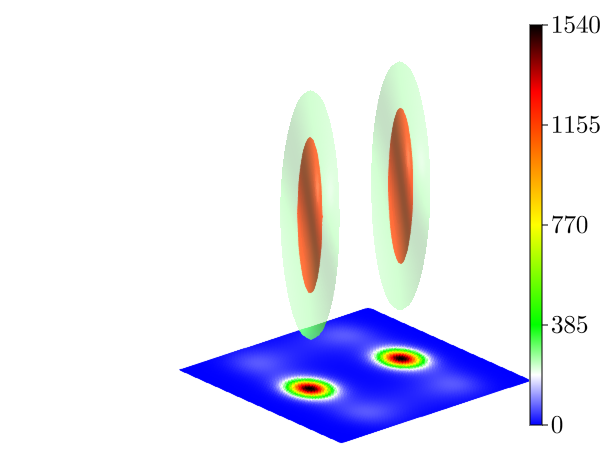} \put(-24.5,1){(d)}&
        \includegraphics[width = 0.15\textwidth, trim={3.5cm 0cm 0cm 0cm}, clip]{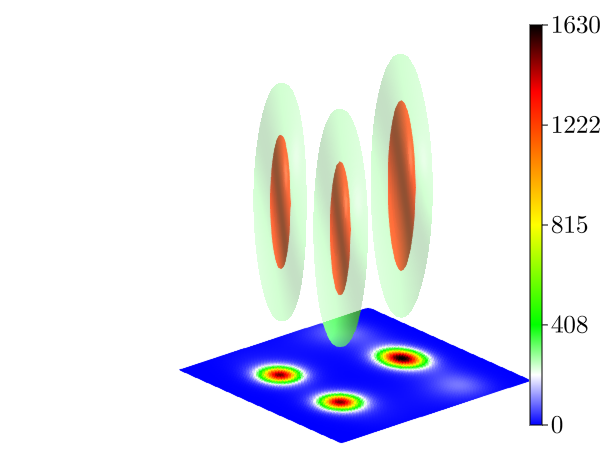} \put(-24.5,1){(e)}&
        \includegraphics[width = 0.15\textwidth, trim={3.5cm 0cm 0cm 0cm}, clip]{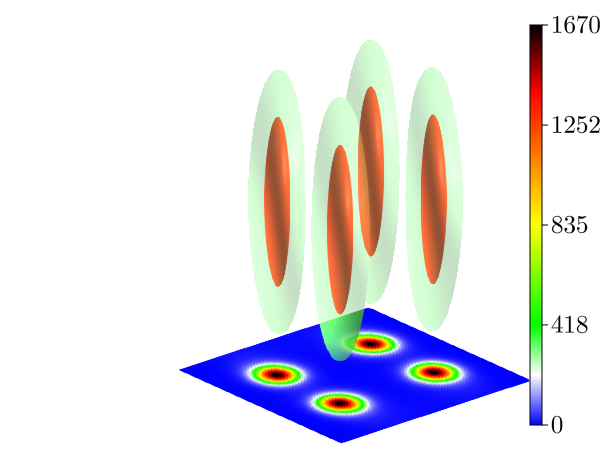} \put(-24.5,1){(f)}\\
    \end{tabular}
\caption{ (a) Total energy and contributions to the energy for the ground-state 
          at $\varepsilon_{dd}=1.45$ as a function of the number of atoms for the double-well potential with $2d=3\,\mu\mathrm{m}$. (b) maximal atom density at $z=0$ plane (red lines)  and droplet height $H_z$ in $z$ dimension (blue lines). 
          The vertical lines mark the ranges of a varied localization type. For each type, the number of well-developed droplets is larger by 1 than the phase number.
          (c-g) Isosurfaces  of the BEC density for $N/10^4=1$ (c), 2(d), 2.75(e) and 4(f)
          The isosurfaces correspond to 20\% (transparent) and 80\% (opaque) of the maximum value.
          The plots span the area of $x$ (horizontal direction) and $y$ from -3.6 $\mu$m to 3.6 $\mu$m.
          At the base of the plots, below the isosurfaces, a cross section of the BEC density at $z=0$ is presented. 
          The colorscale for the density is given in $\mu$m$^{-3}$ units.
          Plots correspond to regions marked by I (c,d), II (e), III (g).
         }
\label{1k45gs}
\end{figure*}

For a study of softer dipolar interaction we took the  value of the parameter $\varepsilon_{dd} = 1.45$. The energy contributions and peak densities with the height of the droplets are shown in Fig.~\ref{1k45gs}(a) and (b), respectively.  
The configurations labeled I, II, and III correspond to a two-droplet state, a $2+1$ configuration, and a symmetric four-droplet configuration.  

The transition from an extended atom cloud phase to the two-droplet phase [Fig.~\ref{1k45gs}(c)] is continuous and does not involve symmetry breaking, in contrast to the $\varepsilon_{dd} = 1.5$ case [Fig. \ref{difaz1k5_1500}].  
The energy contributions and density parameters as functions of $N$ evolve within the region of the droplet formation in a smooth way, without the discontinuities observed above for stronger interactions.  
Abrupt changes in the energy contributions and density parameters occur at $N = 2.34 \times 10^4$, where a $2+1$ configuration emerges, and at $N \simeq 3 \times 10^4$, where a four-droplet configuration becomes energetically favourable.  

Compared to the $\varepsilon_{dd} = 1.5$ case [Fig. \ref{difaz1k5_1500}], the droplets are more extended, with additional low-density halos and local maxima. The axial extent of the droplets is slightly larger, while the peak density decreases from $\sim 2300$ to $\sim 1700$ atoms$/\mu\mathrm{m}^3$.  

\subsection{System dynamics}

\begin{figure}[htbp]
%trim=left botm right top
\begin{tabular}{lll}
\includegraphics[width=0.15\textwidth, trim={1.3cm 3cm 0 0}, clip]{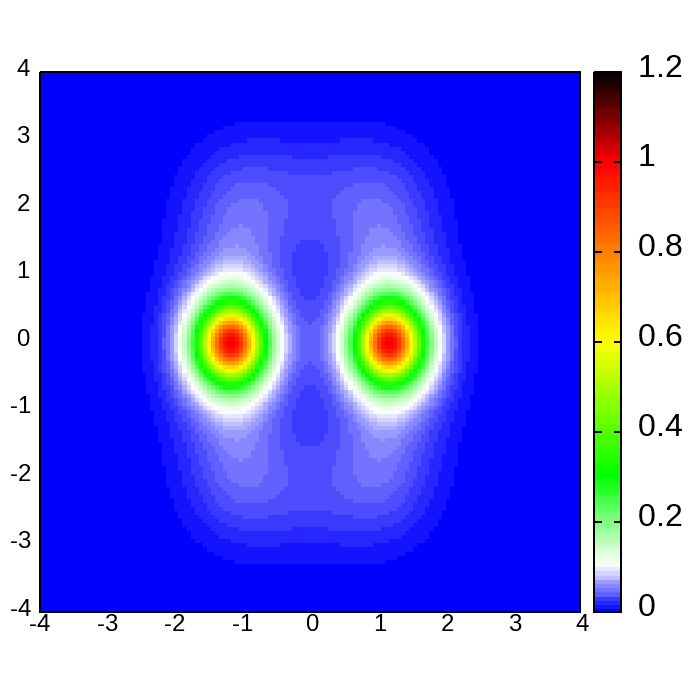} \put(-25,5){\color{yellow}(a)}&
\includegraphics[width=0.15\textwidth, trim={1.3cm 3cm 0 0}, clip]{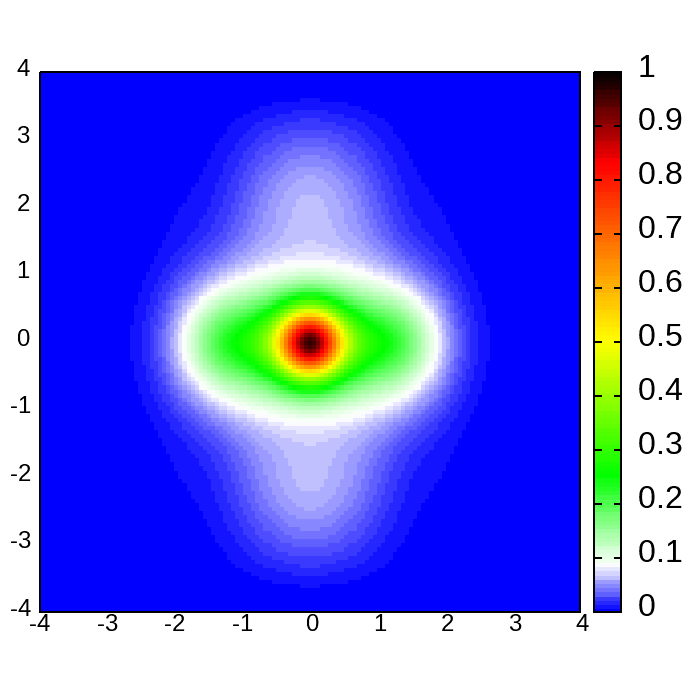} \put(-25,5){\color{yellow}(b)}&
\includegraphics[width=0.15\textwidth, trim={1.3cm 3cm 0 0}, clip]{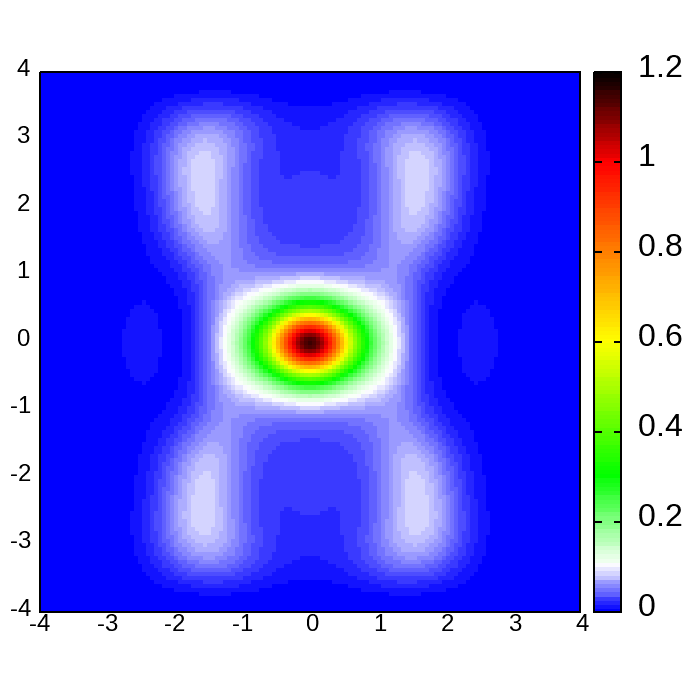} \put(-25,5){\color{yellow}(c)}\\
\includegraphics[width=0.15\textwidth, trim={1.3cm 3cm 0 0}, clip]{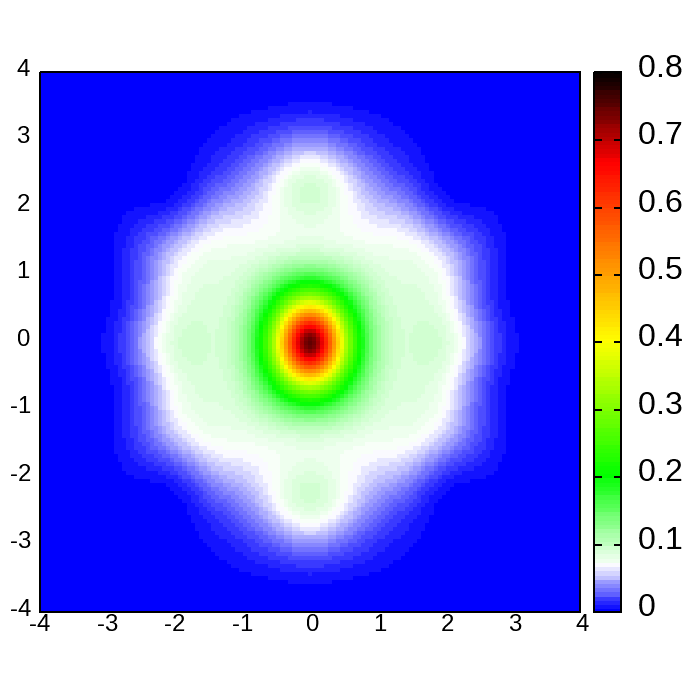} \put(-25,5){\color{yellow}(d)}&
\includegraphics[width=0.15\textwidth, trim={1.3cm 3cm 0 0}, clip]{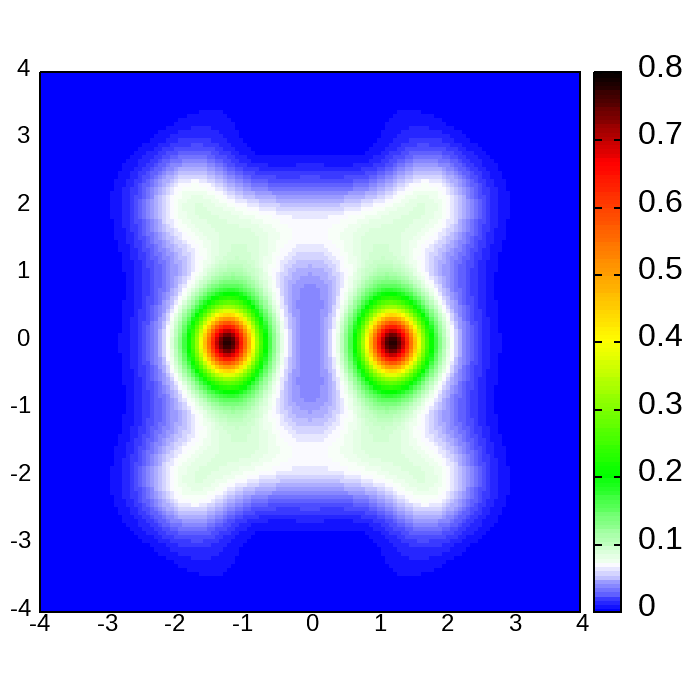} \put(-25,5){\color{yellow}(e)}&
\includegraphics[width=0.15\textwidth, trim={1.3cm 3cm 0 0}, clip]{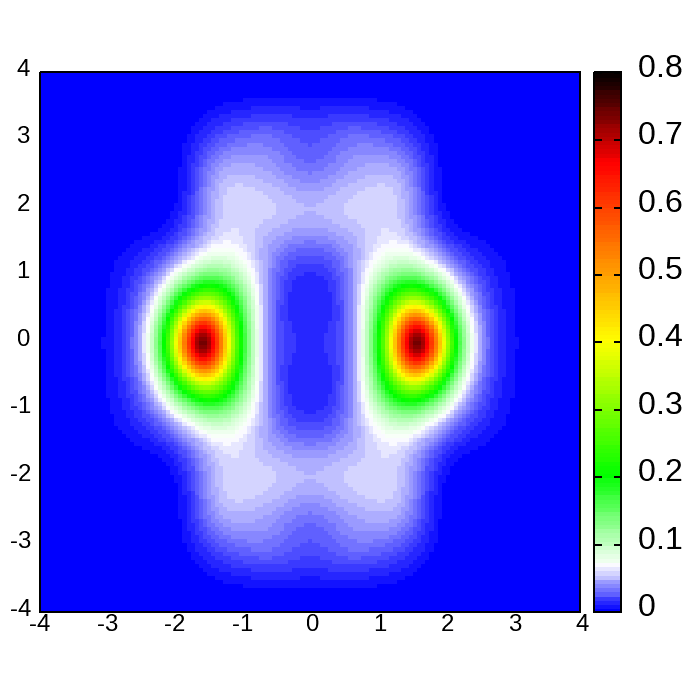} \put(-25,5){\color{yellow}(f)}
\end{tabular}
\includegraphics[width=0.43\textwidth]{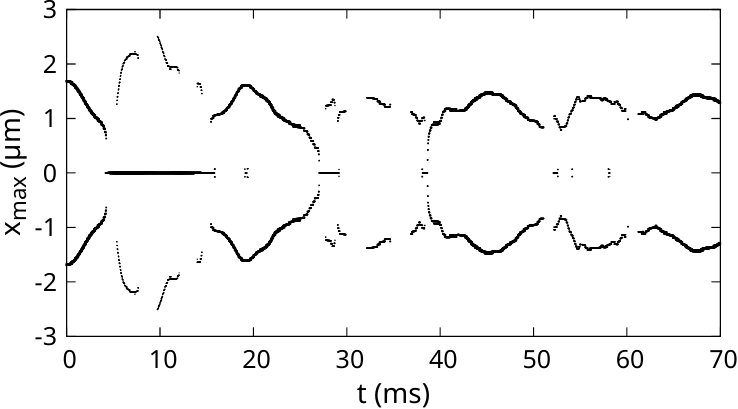}\put(-20,25){\color{black}(g)}

\caption{(a-f) Time evolution for a system with $2d=3\,\mu$m, $\varepsilon_{dd}=1.45$ and $N=10^4$. Panels show the cross-section of the density at $z=0$ at times after the removal of the inter-well barrier: 2.418 ms (a), 4.83 ms (b), 9.03 ms (c), 13.96 ms (d), 17.36 ms  (e) and 19.7 ms (f) in units of $1000/\mu$m$^3$. (g) Positions of the maxima of the density at the $y=0$ axis. The size of the dots is proportional to the density at the local maximum.} 
\label{1k451e4}
\end{figure}

Figure~\ref{1k451e4}(a)--(f) shows the dynamics for $N = 10^4$. In contrast to the $\varepsilon_{dd} = 1.5$ case, droplets merge and split during the evolution, and the number of density maxima is not conserved [Fig.~\ref{1k451e4}(g)]. The dynamics are irregular, with no clear periodicity.  

\begin{figure}[htbp]

%trim=left botm right top
\begin{tabular}{lll}
    \includegraphics[width=0.15\textwidth, trim={1.3cm 3cm 0 0}, clip]{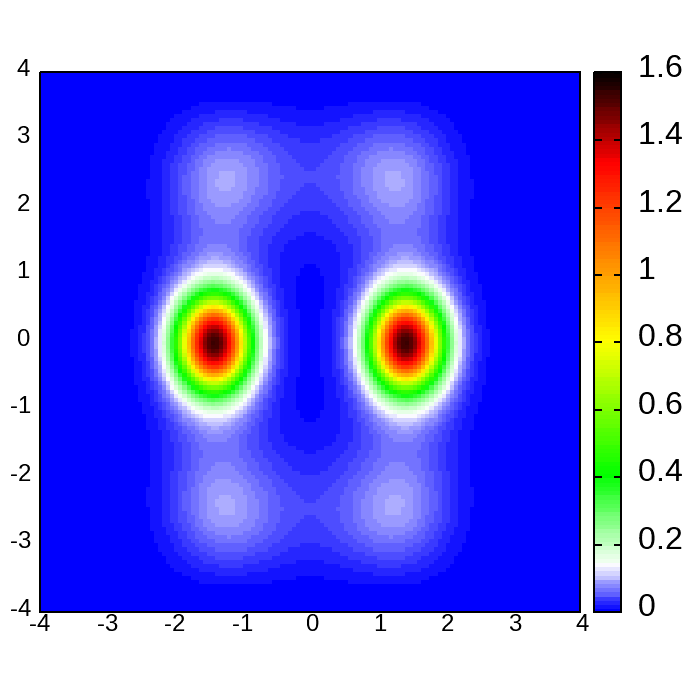} \put(-25,5){\color{yellow}(a)} &
  \includegraphics[width=0.15\textwidth, trim={1.3cm 3cm 0 0}, clip]{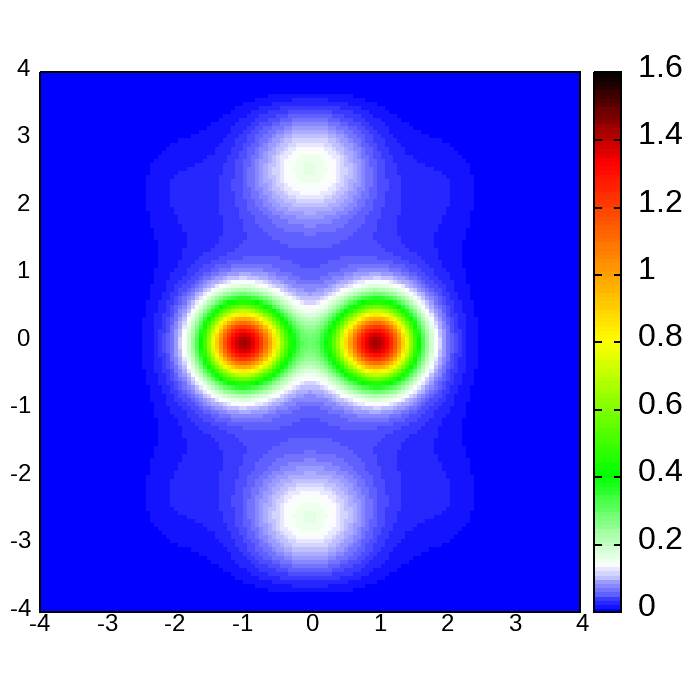} \put(-25,5){\color{yellow}(b)} &
  \includegraphics[width=0.15\textwidth, trim={1.3cm 3cm 0 0}, clip]{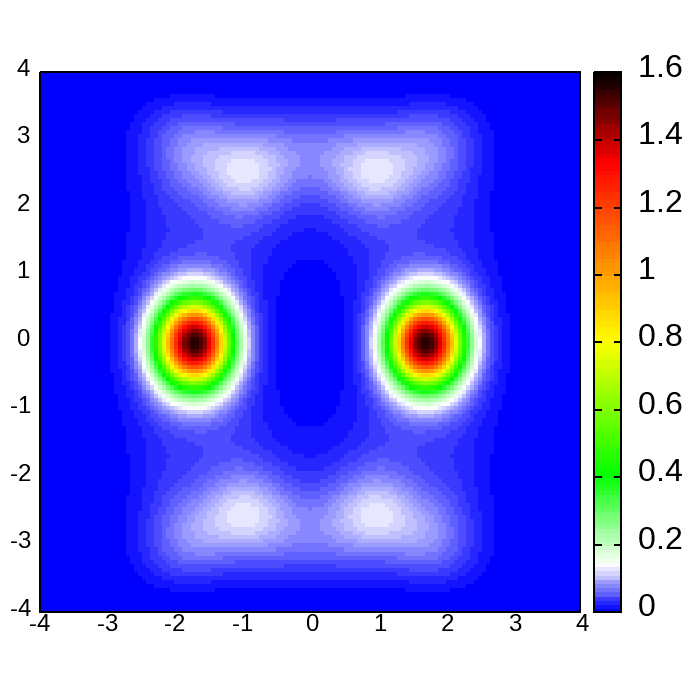} \put(-25,5){\color{yellow}(c)}\\
   \includegraphics[width=0.15\textwidth, trim={1.3cm 3cm 0 0}, clip]{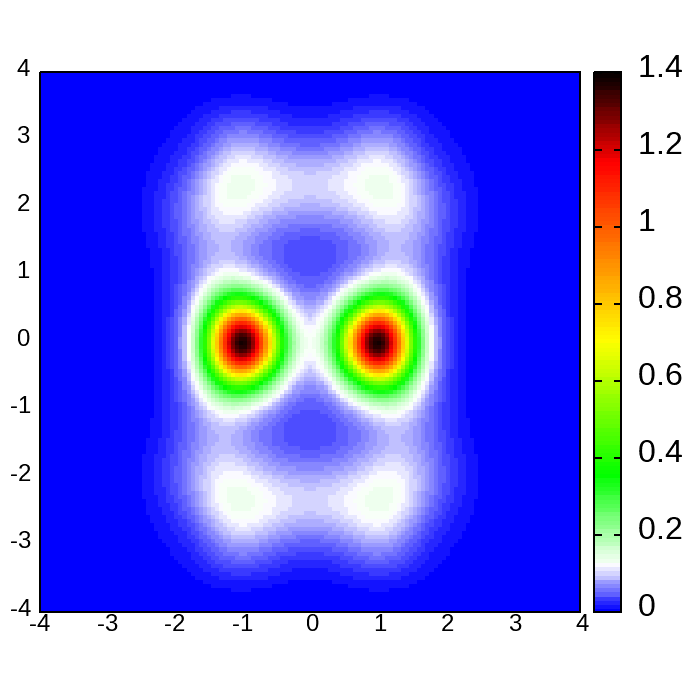}  \put(-25,5){\color{yellow}(d)}  &
    \includegraphics[width=0.15\textwidth, trim={1.3cm 3cm 0 0}, clip]{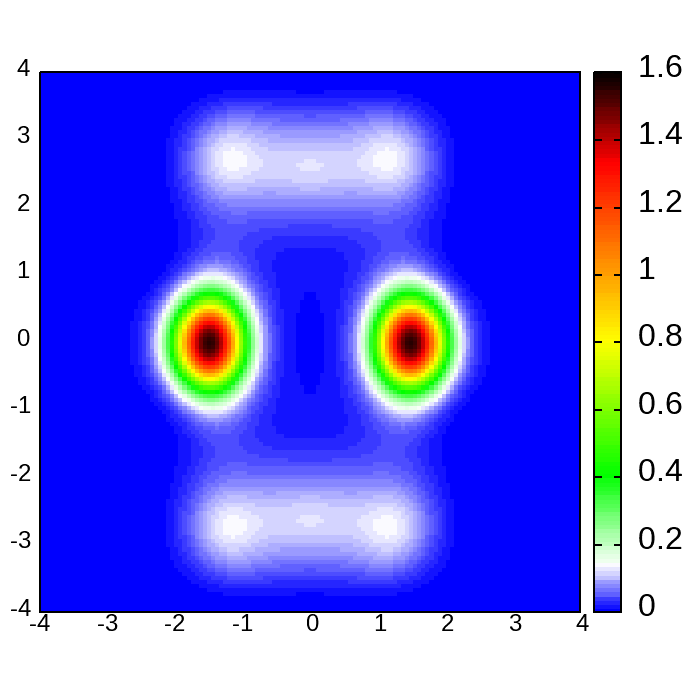}  \put(-25,5){\color{yellow}(e)} &
   \includegraphics[width=0.15\textwidth, trim={1.3cm 3cm 0 0}, clip]{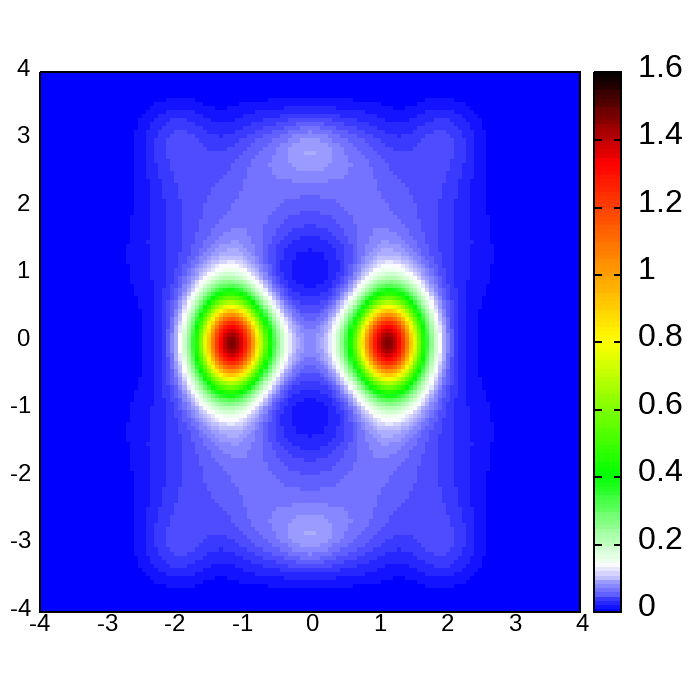} \put(-25,5){\color{yellow}(f)} \\
  \end{tabular}
  \begin{tabular}{ll}
 \includegraphics[width=0.22\textwidth]{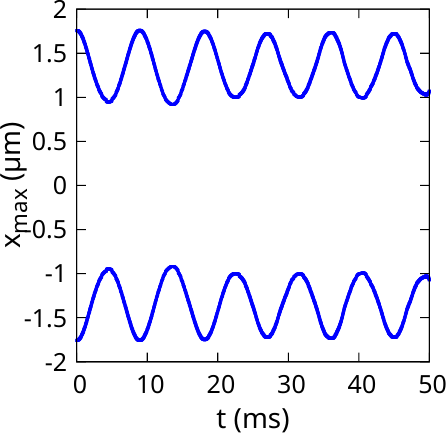}  \put(-25,25){(g)}&
  \includegraphics[width=0.22\textwidth]{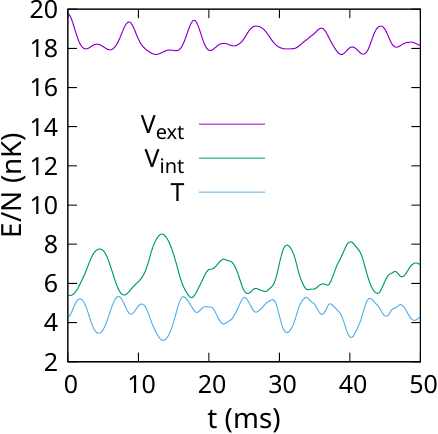} \put(-25,25){(h)}\\
  \end{tabular}
  \caption{
  (a-f) Time evolution for a system with $2d=3\mu$m, $\varepsilon_{dd}=1.45$ and $N=2\times 10^4$. Panels show the cross-section of the density at $z=0$ at times after the removal of the inter-well barrier 1.81 ms (a), 4.64 ms (b), 8.34 ms (c), 12.48 ms (d), 16.49 ms  (e) and 21.04 ms (f) in units of $1000/\mu$m$^3$. (g) Positions of the maxima of the density at the $y=0$ axis. (h) Contributions to the total energy. }
\label{1k452e4}
\end{figure}

For $N = 2 \times 10^4$, the system maintains two main droplets [Fig.~\ref{1k452e4}(a)--(f)], although complex dynamics develops in the surrounding halos.  
The droplet positions exhibit oscillations [Fig.~\ref{1k452e4}(g)] that can be fitted by Eq. (13) 
with $\tau = 215$ ms, $c = 0.432\,\mu\mathrm{m}$, $d = 1.335\,\mu\mathrm{m}$, and $T = 9$ ms.  
The energy contributions are less regular than the position oscillations [Fig.~\ref{1k452e4}(h)].  

\begin{figure}[htbp]
%trim=left botm right top
\begin{tabular}{lll}
\includegraphics[width=0.15\textwidth, trim={1.3cm 3cm 0 0}, clip]{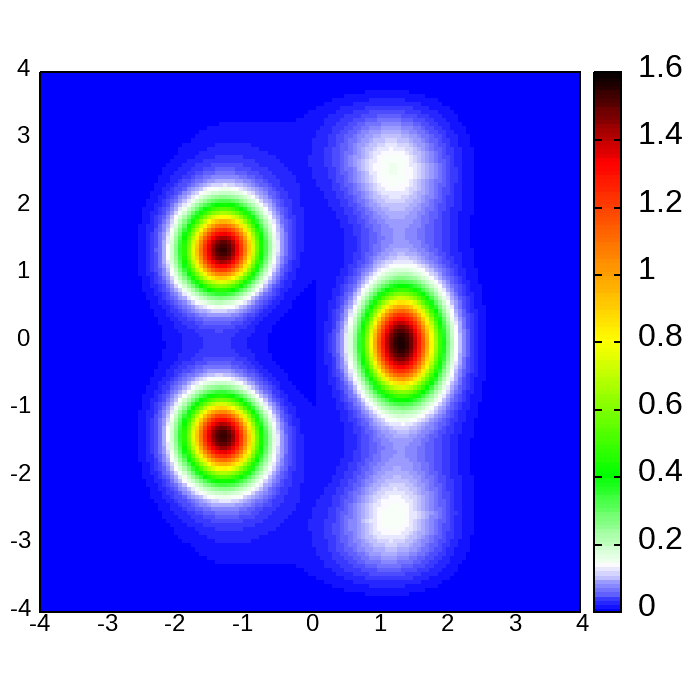}  \put(-25,5){\color{yellow}(a)}&
  \includegraphics[width=0.15\textwidth, trim={1.3cm 3cm 0 0}, clip]{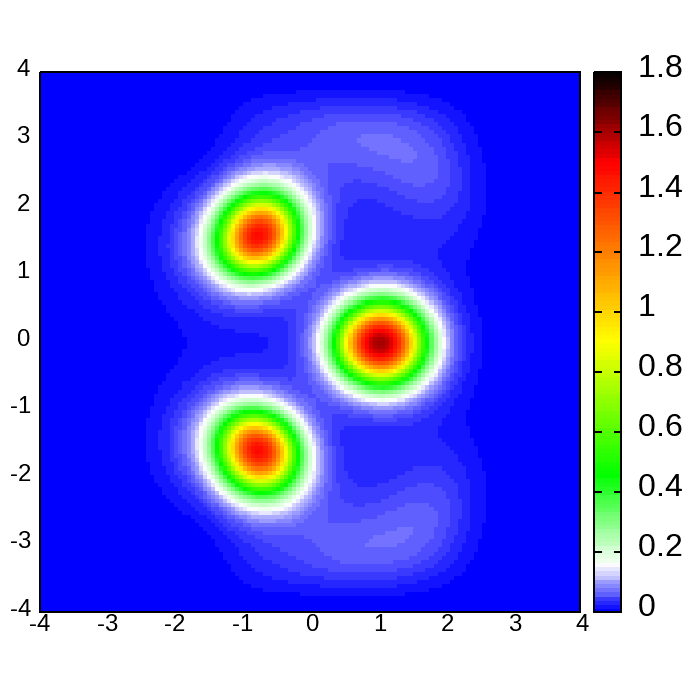} \put(-25,5){\color{yellow}(b)}&
  \includegraphics[width=0.15\textwidth, trim={1.3cm 3cm 0 0}, clip]{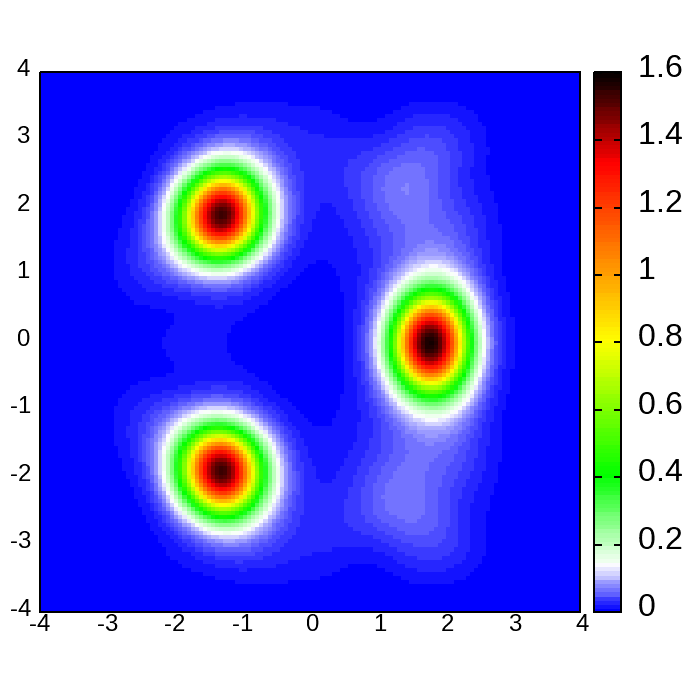}\put(-25,5){\color{yellow}(c)}\\
   \includegraphics[width=0.15\textwidth, trim={1.3cm 3cm 0 0}, clip]{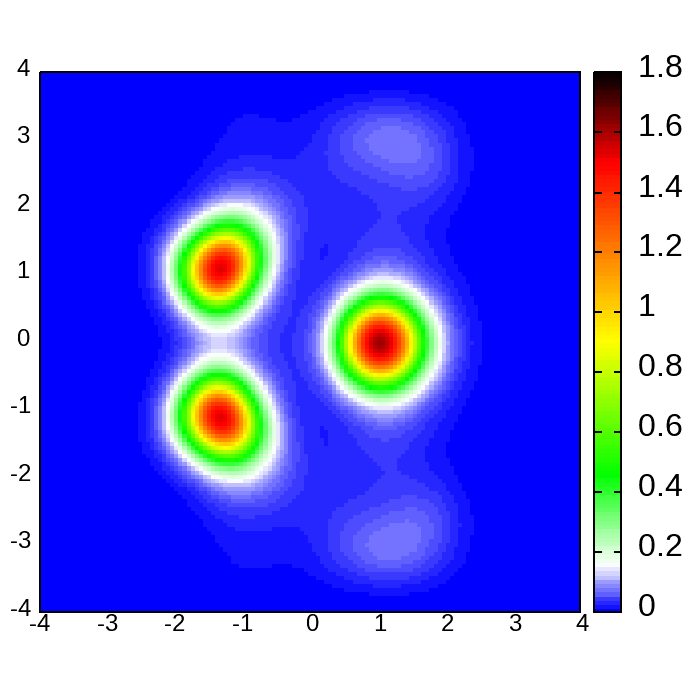} \put(-25,5){\color{yellow}(d)}  &
    \includegraphics[width=0.15\textwidth, trim={1.3cm 3cm 0 0}, clip]{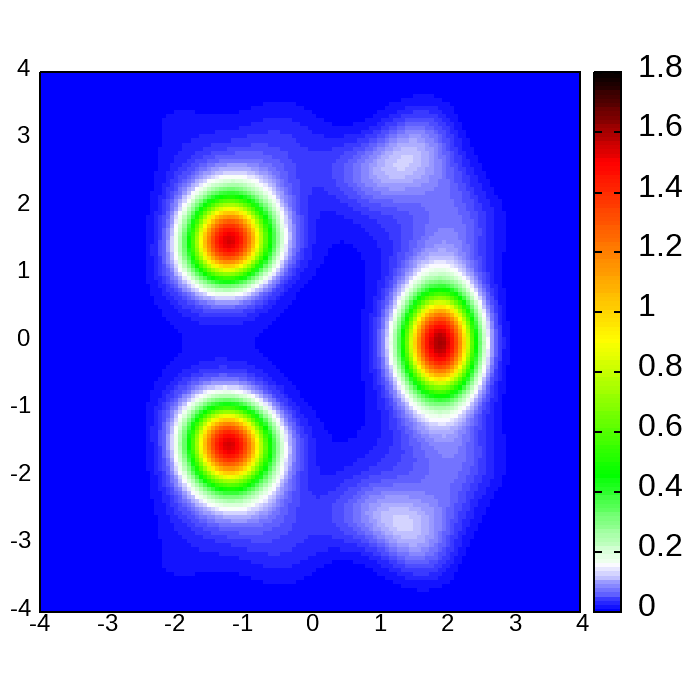} \put(-25,5){\color{yellow}(e)} &
   \includegraphics[width=0.15\textwidth, trim={1.3cm 3cm 0 0}, clip]{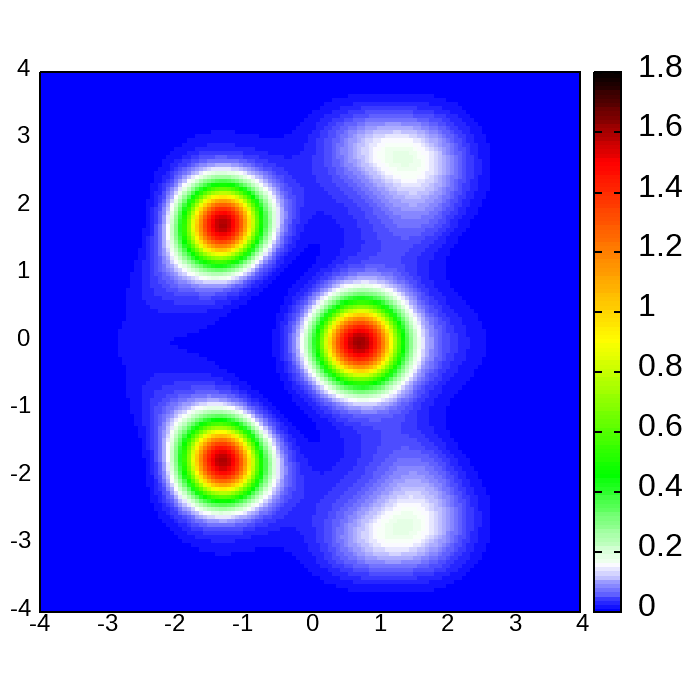} \put(-25,5){\color{yellow}(f)}\\
  \end{tabular}
   \begin{tabular}{ll}
 \includegraphics[width=0.22\textwidth]{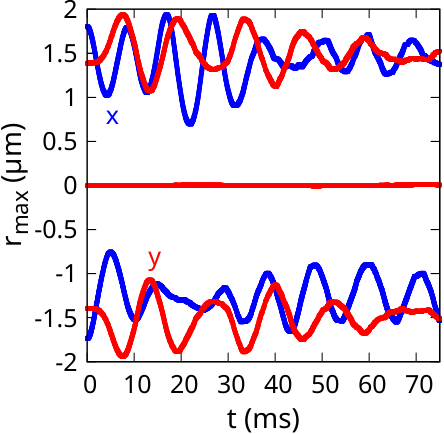} \put(-15,22){\color{black}(g)}&
\includegraphics[width=0.22\textwidth]{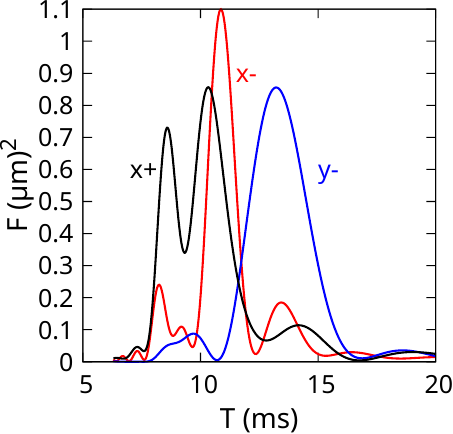} \put(-20,22){\color{black}(h)}
  \end{tabular}
\caption{
(a-f) Time evolution for a system with $2d=3\mu$m, $\varepsilon_{dd}=1.45$ and $N=2.75\times 10^4$. Panels show the cross-section of the density at $z=0$ at times after the removal of the inter-well barrier 2.18 ms (a), 4.64 ms (b), 8.34 ms (c), 12.48 ms (d), 16.49 ms  (e) and 21.04 ms (f) in units of $1000/\mu$m$^3$. (g) $x$ and $y$ coordinates of the three principal maxima of the condensate density of panels (a-f). (h) Fourier transform
of the positions of the maxima: $x+$  -- the $x$ position of the center of the droplet at the right-hand side of the origin, $x-$  and $y-$ -- the $x$ and $y$ positions of the 
center of one of the droplets at the left-hand side of the origin.}
\label{1k45275e4}
\end{figure}

For $N = 2.75 \times 10^4$, with an initial $2+1$ configuration [Fig. \ref{1k45275e4}] one finds   more complex dynamics with non-synchronized motion in the $x$ and $y$ directions.  
Fourier analysis reveals multiple characteristic periods: $\sim 13.2$ ms in $y-$ motion, $\sim 10.9$ ms in $x-$ motion (with additional components at $13.5$ ms and $8.25$ ms), and $\sim 10.4$ ms and $8.6$ ms for the solitary droplet.  

\begin{figure}[htbp]
%trim=left botm right top
\begin{tabular}{lll}
\includegraphics[width=0.15\textwidth, trim={1.3cm 3cm 0 0}, clip]{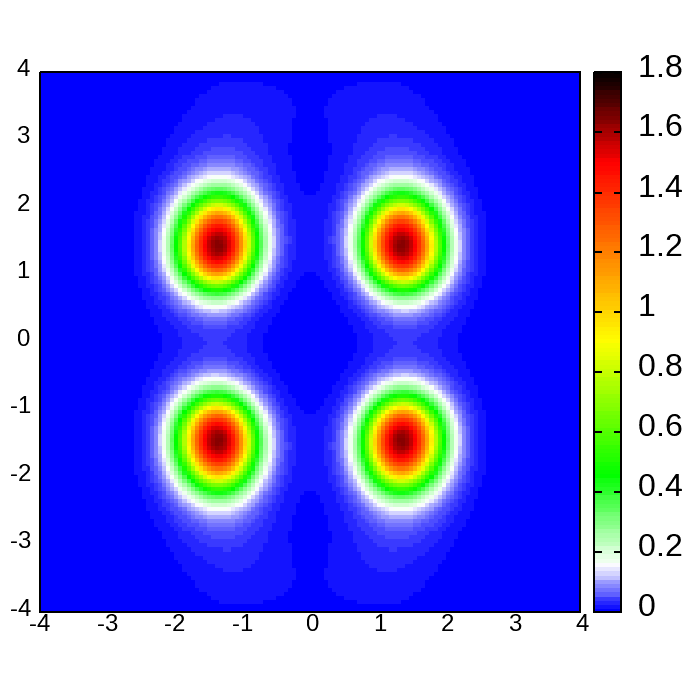}\put(-25,5){\color{yellow}(a)} &
  \includegraphics[width=0.15\textwidth, trim={1.3cm 3cm 0 0}, clip]{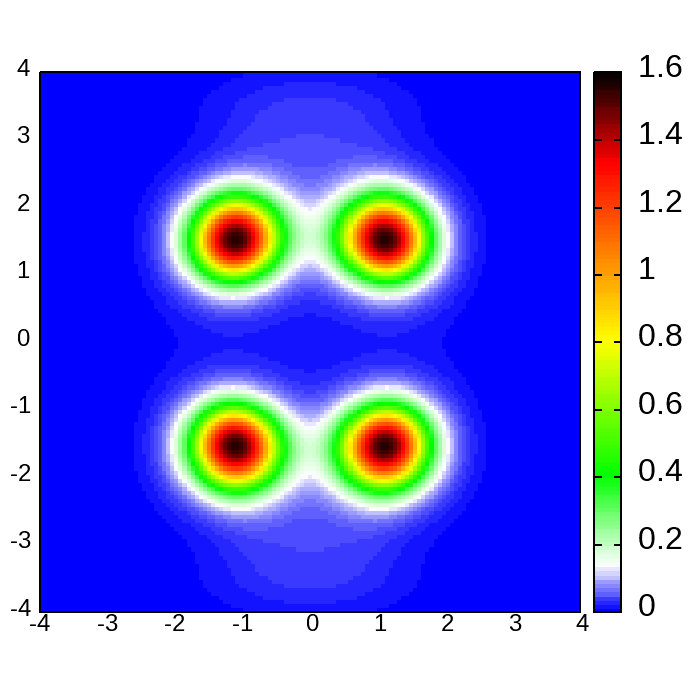}\put(-25,5){\color{yellow}(b)} &
  \includegraphics[width=0.15\textwidth, trim={1.3cm 3cm 0 0}, clip]{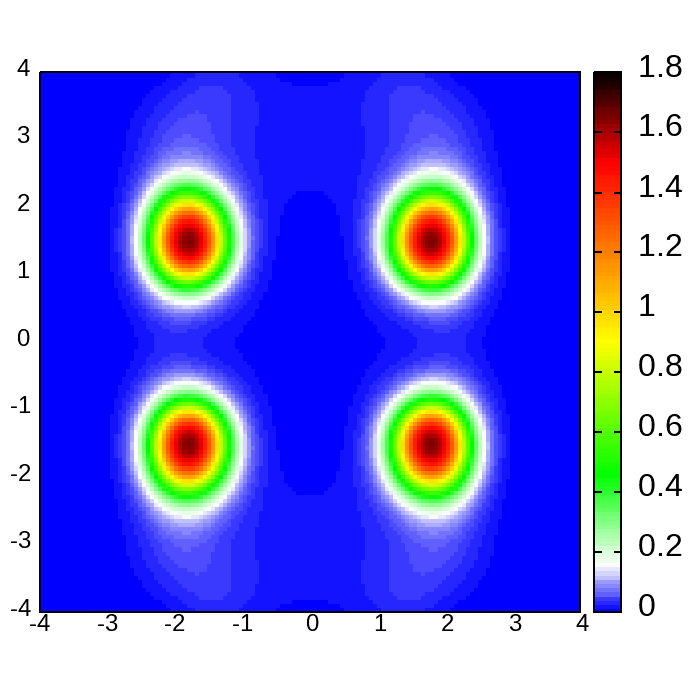} \put(-25,5){\color{yellow}(c)}\\
   \includegraphics[width=0.15\textwidth, trim={1.3cm 3cm 0 0}, clip]{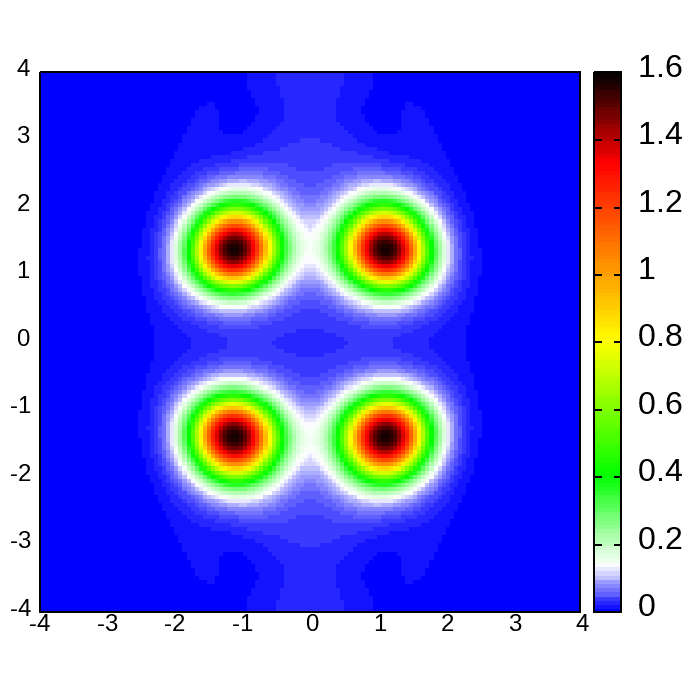} \put(-25,5){\color{yellow}(d)}  &
    \includegraphics[width=0.15\textwidth, trim={1.3cm 3cm 0 0}, clip]{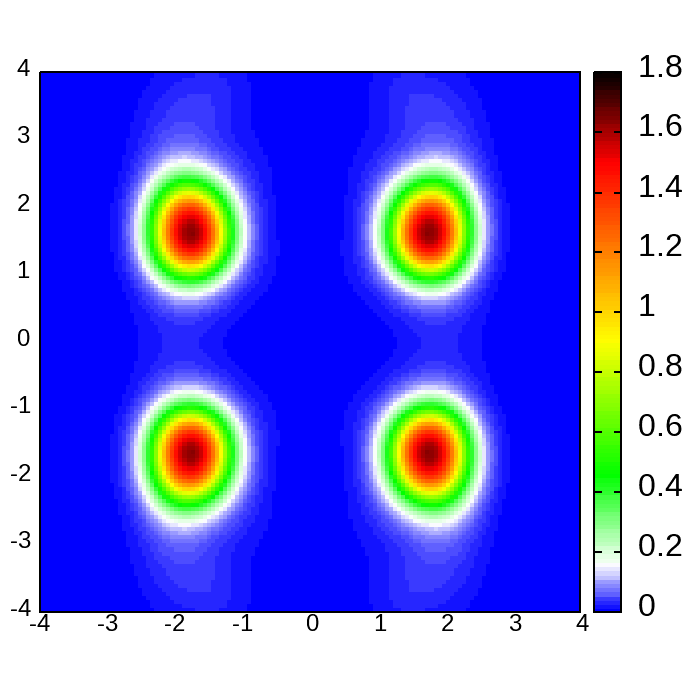}\put(-25,5){\color{yellow}(e)}  &
   \includegraphics[width=0.15\textwidth, trim={1.3cm 3cm 0 0}, clip]{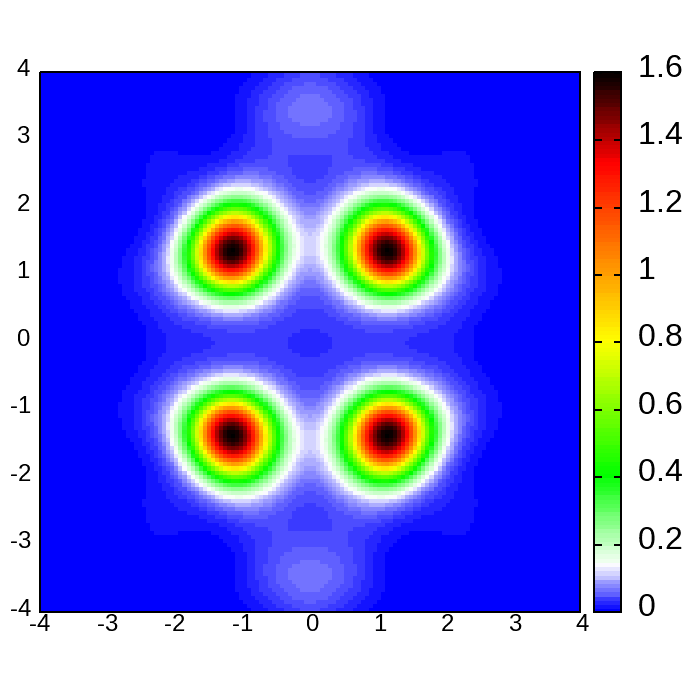}  \put(-25,5){\color{yellow}(f)}
  \end{tabular}
     \begin{tabular}{ll}
  \includegraphics[width=0.22\textwidth]{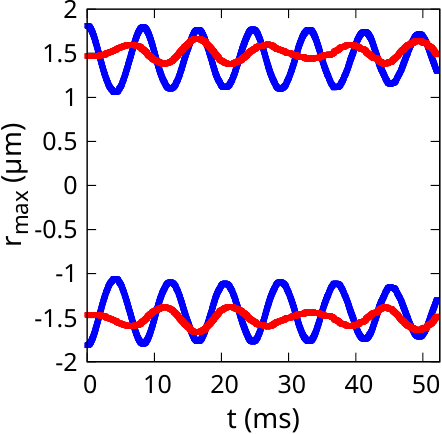}\put(-20,55){\color{black}(g)}&
\includegraphics[width=0.22\textwidth]{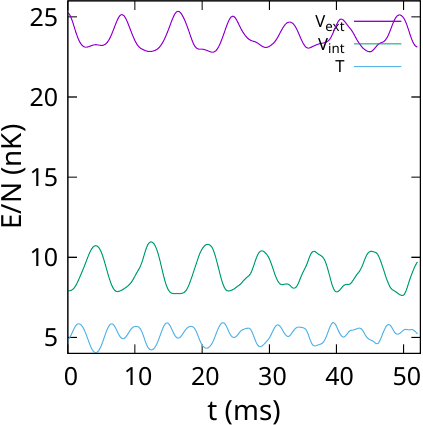} \put(-20,55){\color{black}(h)}
  \end{tabular}
\caption{
 (a-f) Time evolution for a system with $2d=3\mu$m, $\varepsilon_{dd}=1.45$ and $N=4\times 10^4$. Panels show the cross-section of the density at $z=0$ at times after the removal of the inter-well barrier 2.18 ms (a), 4.64 ms (b), 8.34 ms (c), 12.48 ms (d), 16.49 ms  (e) and 21.04 ms (f) in units of $1000/\mu$m$^3$. (g) Positions of the centers of the droplets in $x$ (blue) and $y$ (red) coordinates. (h) Contributions to the total energy.}
\label{1k454e4}
\end{figure}

For $N = 4 \times 10^4$, the four-droplet system exhibits more regular oscillations (Fig.~\ref{1k454e4}), with periods of $8.24$ ms in $x$ and $11$ ms in $y$.  
The oscillation amplitude is larger in the $x$ direction, with a fitted lifetime $\tau = 198$ ms, amplitude $0.374\,\mu\mathrm{m}$, and mean position $1.427\,\mu\mathrm{m}$.  
The energy contributions show nearly periodic behavior.  

\section{interwell distance $2d = 3\,\mu\mathrm{m}$ and $\varepsilon_{dd} = 1.4$}
\label{sec:edd140}

\subsection{Lowest-energy configurations}

\begin{figure}[htbp]
%trim=left botm right top
%\includegraphics[width=0.35\textwidth, trim={2.5cm 0.5cm 3.0cm 0.0cm}, clip]{1500/1k4faza/ef1k51k455e4.pdf} \put(-25,35){(a)} \\
%\includegraphics[width=0.35\textwidth, trim={2.5cm 0.5cm 3.0cm 0.0cm}, clip]{1500/1k4faza/rm1k451455e4.pdf} \put(-25,35){(b)} \\
\includegraphics[width=0.35\textwidth]{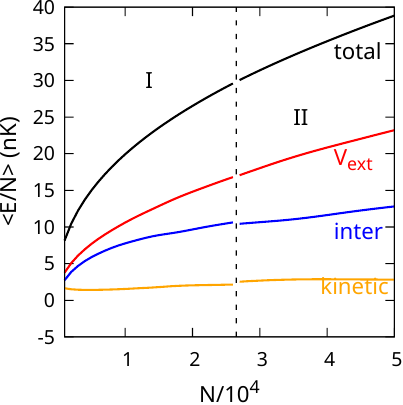} \put(-20,35){(a)}\\
\includegraphics[width=0.35\textwidth]{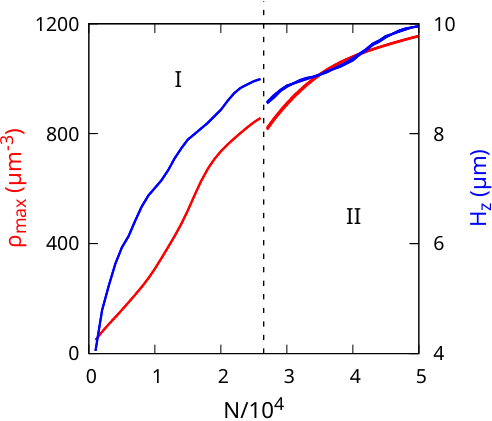} \put(-40,30){(b)}\\
\begin{tabular}{l l}
\includegraphics[width=0.22\textwidth, trim={3.2cm 0cm 0cm 0cm}, clip]{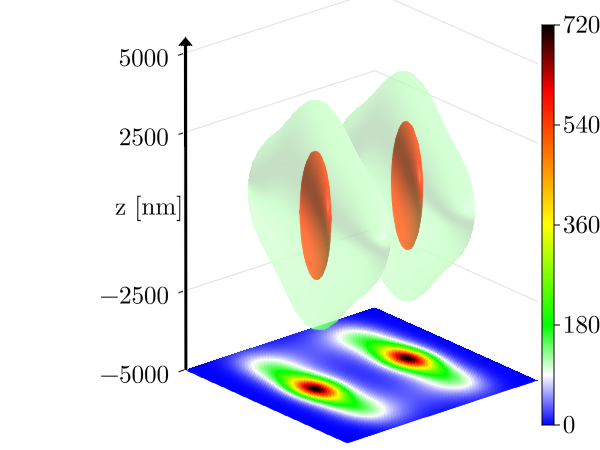} \put(-25,5.5){(c)}&
\includegraphics[width=0.22\textwidth, trim={3.5cm 0cm 0cm 0cm}, clip]{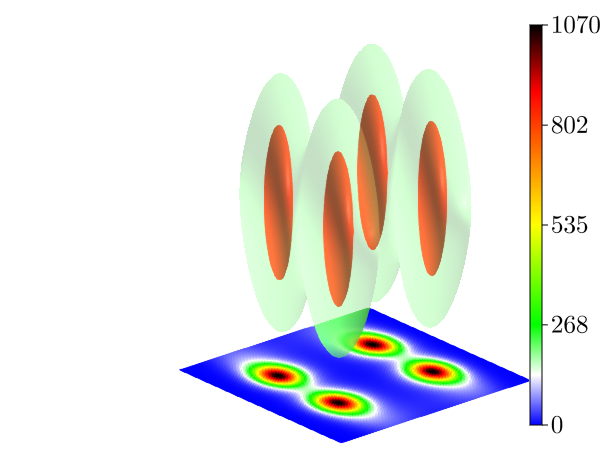} \put(-30,5.5){(d)}
\end{tabular}
\caption{ (a) Total energy and contributions to the energy for the ground-state 
at $\epsilon_{dd}=1.4$ as a function of the number of atoms. 
          (b) maximal atom density at $z=0$ plane.
          (c-d) Isosurface visualisations of the normalized BEC wavefunction for $N=20000$ (c), 40000 (d) for $\varepsilon=1.4$ and the lowest-energy configuration.
          The isosurfaces correspond to 20\% (transparent) and 80\% (opaque) of the maximum value.
          The plots span the area of $x$ (horizontal direction) and $y$ from -3.6 $\mu$m to 3.6 $\mu$m.
          At the base of the plots, below the  isosurfaces a cross section of the BEC density at $z=0$ is presented.
          The colorscale for the density is given in $\mu$m$^{-3}$ units.
          The plots (c-d) correspond to regions marked by I, II in (a) and (b).
          % WO: w obrazie (b) obliczenia wyszly mi trochę inne, można porównać, ale mniej więcej się zgadzają. Zmiana stanu w 26.5k atomów.
}
\label{sc4}
\end{figure}

Figure~\ref{sc4} shows results for $\varepsilon_{dd} = 1.4$. Only two types of ground-state configurations are found up to $N = 5 \times 10^4$: two-droplet (I) and four-droplet (II) states.  
No symmetry-broken configurations are observed, in contrast to the $\varepsilon_{dd} = 1.5$ and $1.45$ cases.  

\subsection{System dynamics}

\begin{figure}[htbp]
%trim=left botm right top
\begin{tabular}{lll}
\includegraphics[width=0.15\textwidth, trim={1.3cm 3cm 0 0}, clip]{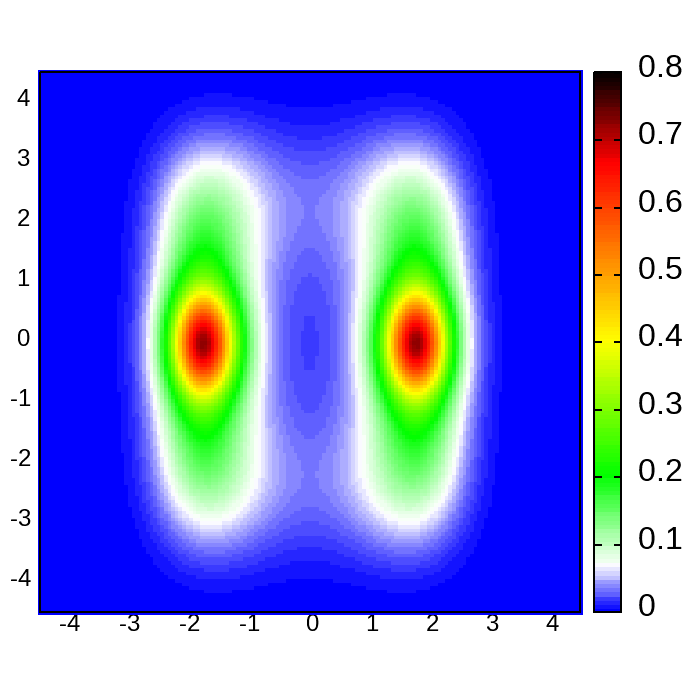}\put(-25,5){\color{yellow}(a)}&
\includegraphics[width=0.15\textwidth, trim={1.3cm 3cm 0 0}, clip]{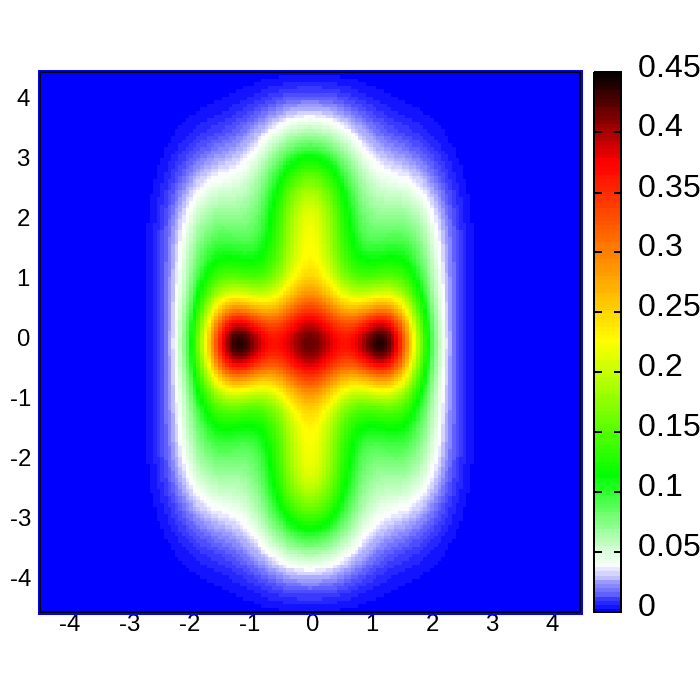}\put(-25,5){\color{yellow}(b)}&
\includegraphics[width=0.15\textwidth, trim={1.3cm 3cm 0 0}, clip]{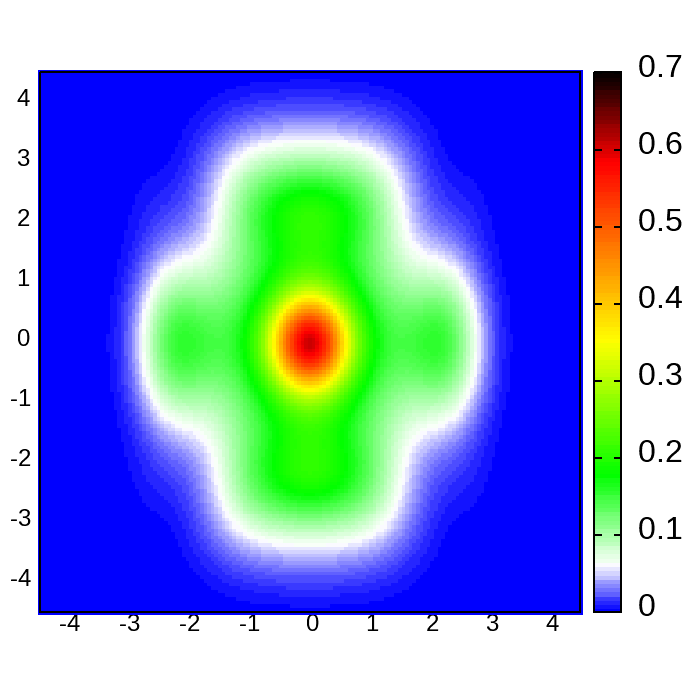}\put(-25,5){\color{yellow}(c)}\\
\includegraphics[width=0.15\textwidth, trim={1.3cm 3cm 0 0}, clip]{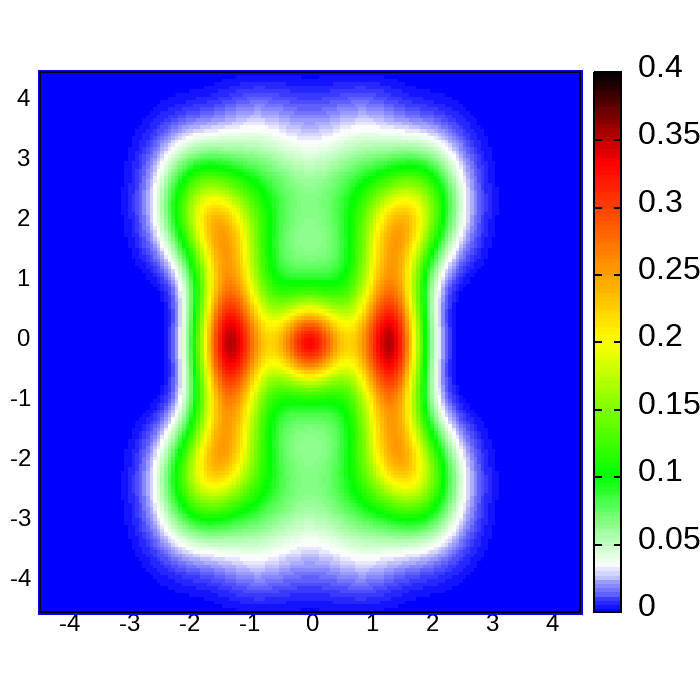}\put(-25,5){\color{yellow}(d)}&
\includegraphics[width=0.15\textwidth, trim={1.3cm 3cm 0 0}, clip]{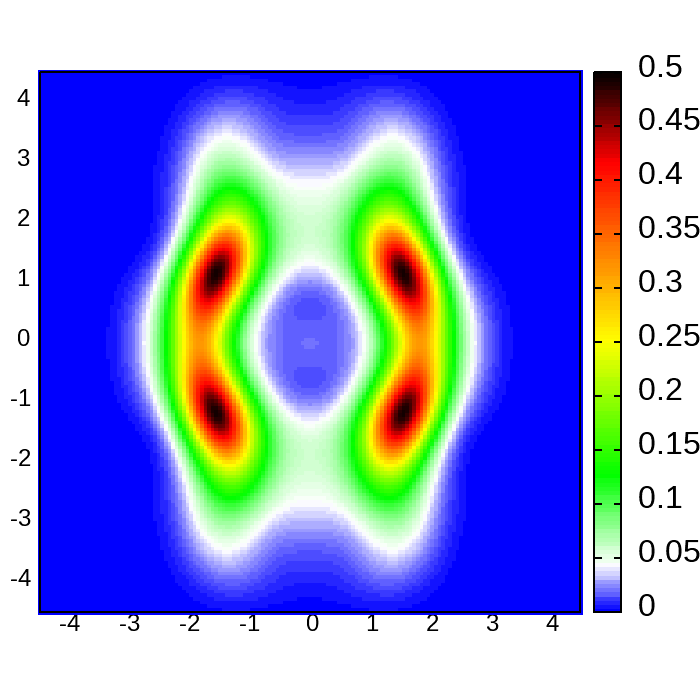}\put(-25,5){\color{yellow}(e)}&
\includegraphics[width=0.15\textwidth, trim={1.3cm 3cm 0 0}, clip]{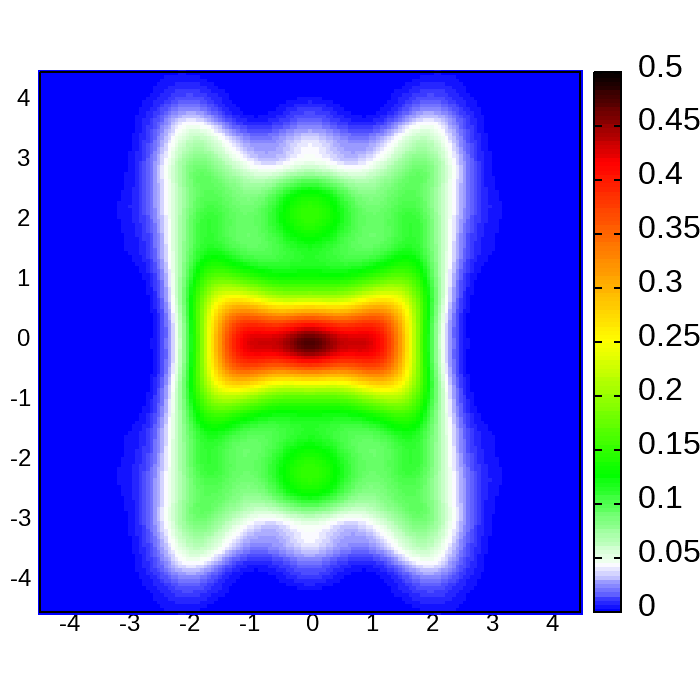}\put(-25,5){\color{yellow}(f)}\\
\hline
\includegraphics[width=0.15\textwidth, trim={1.3cm 3cm 0 0}, clip]{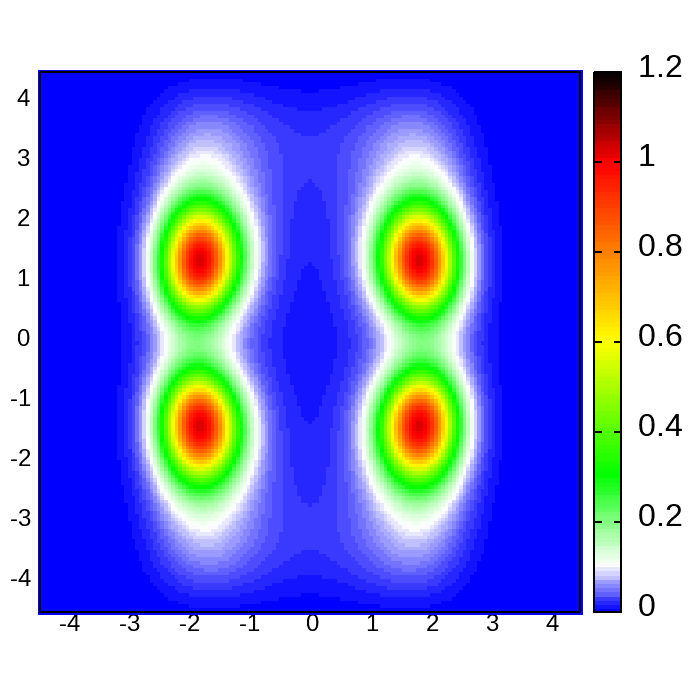}\put(-25,5){\color{yellow}(g)}&
\includegraphics[width=0.15\textwidth, trim={1.3cm 3cm 0 0}, clip]{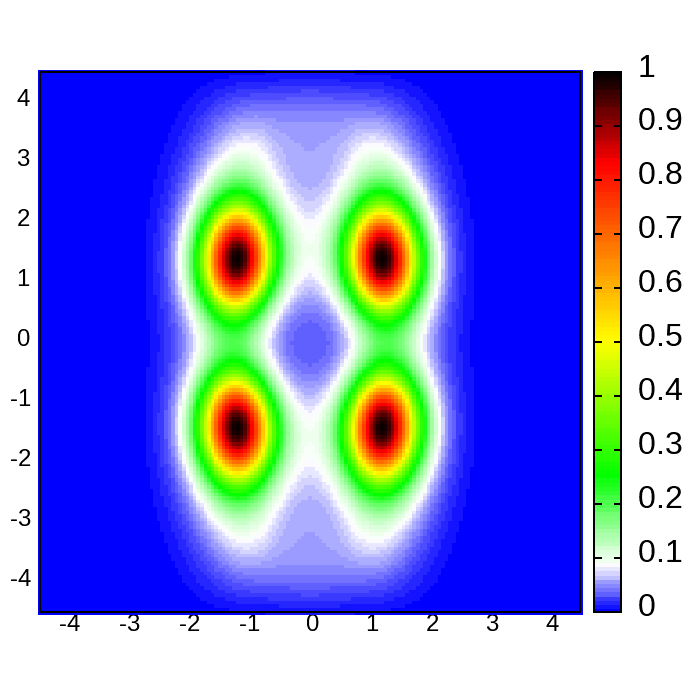}\put(-25,5){\color{yellow}(h)}&
\includegraphics[width=0.15\textwidth, trim={1.3cm 3cm 0 0}, clip]{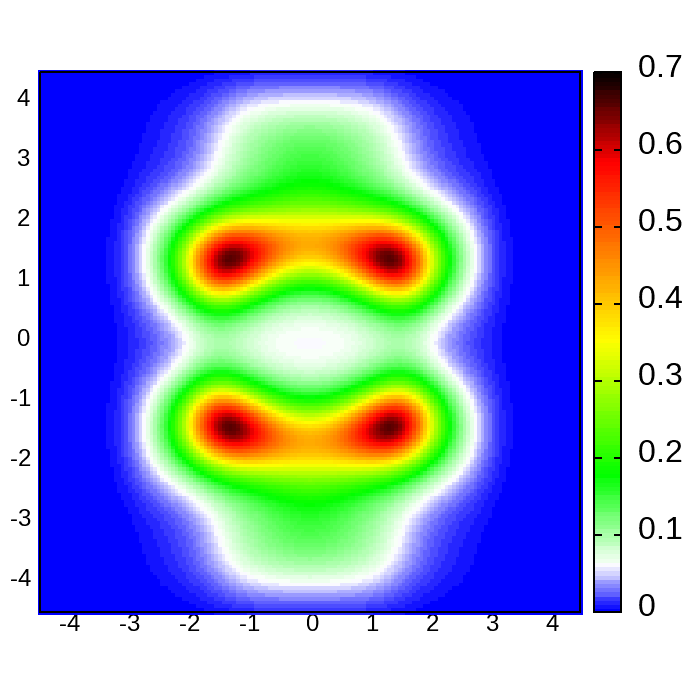}\put(-25,5){\color{yellow}(i)}\\
\includegraphics[width=0.15\textwidth, trim={1.3cm 3cm 0 0}, clip]{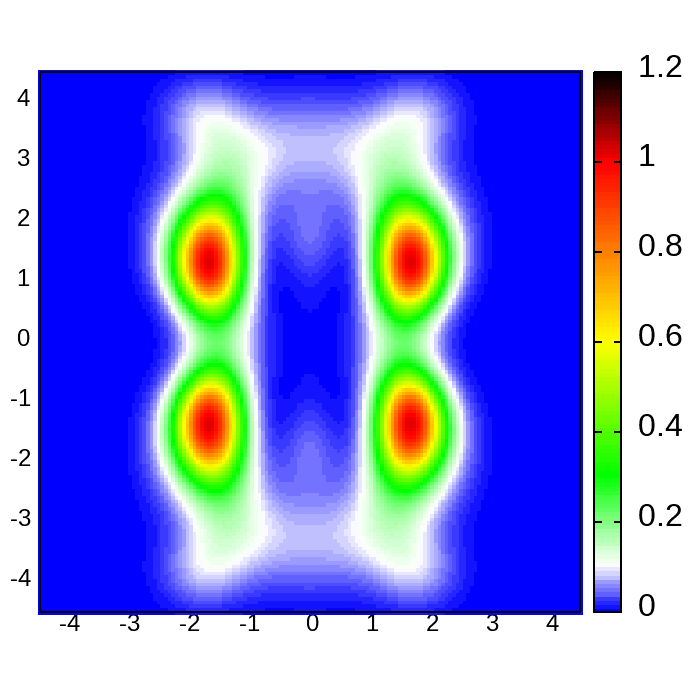}\put(-25,5){\color{yellow}(j)}&
\includegraphics[width=0.15\textwidth, trim={1.3cm 3cm 0 0}, clip]{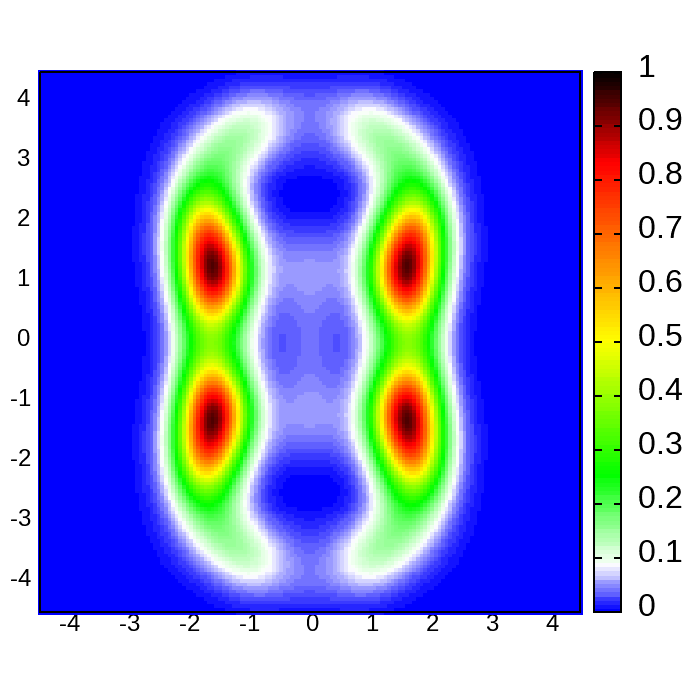}\put(-25,5){\color{yellow}(k)}&
\includegraphics[width=0.15\textwidth, trim={1.3cm 3cm 0 0}, clip]{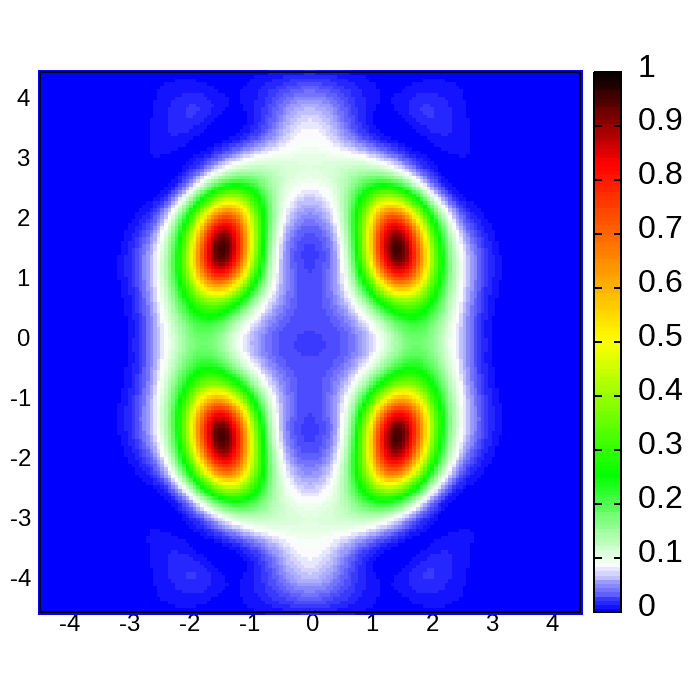}\put(-25,5){\color{yellow}(l)}\\
  \end{tabular}
  \begin{tabular}{l}
  \includegraphics[width=0.35\textwidth]{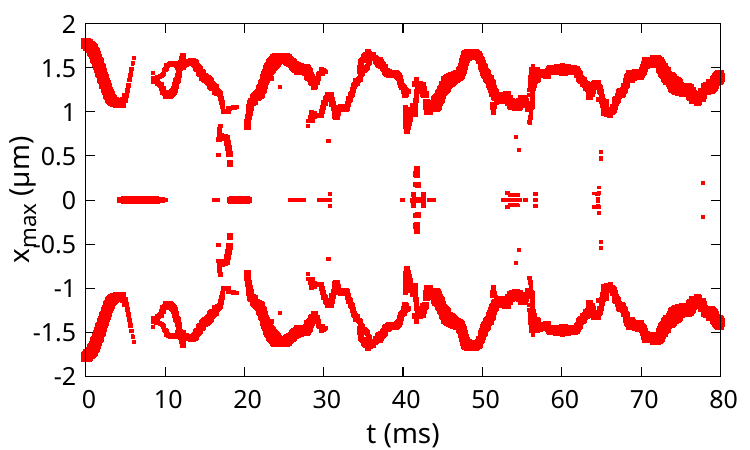}\put(-25,40){(m)} \\
  \includegraphics[width=0.35\textwidth]{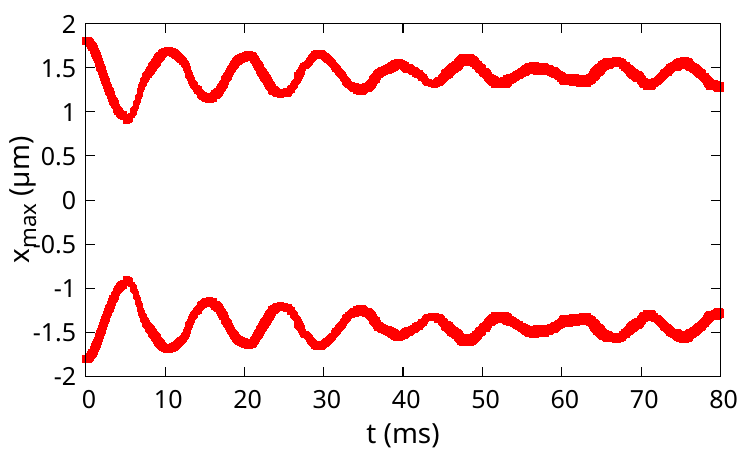}\put(-25,35){(n)}
  \end{tabular}
\caption{System dynamics for $2d=3\mu$m and $\varepsilon_{dd}=1.4$ for $N/10^4$ equal to 2 (a-f) and 4 (g-l). The cross sections of the condensate density given by the colorscale that is plotted in $1000/\mu$m$^3$ units at selected moments in time:  0 (a), 4.64 ms (b), 6.77 ms (c), 9.67 ms (d), 13.54 ms (e) and  20.31 ms (f) and 0 (g), 3.19 ms (h), 6.35 ms (i), 9.68 ms(j), 11.75 ms (k) and 18.19 ms(l).  The plots cover the interval of [$-4.5\mu$m,$4.5\mu$m] in both directions. (m) and (n) -- the $x$ positions of the maxima of the condensate density for $N=2\times 10^4$ and $N=4\times 10^4$, respectively. The size of the symbols is proportional to the density.}
\label{sc4d}
\end{figure}

Figure~\ref{sc4d} shows the dynamics for $N = 2 \times 10^4$ and $4 \times 10^4$.  
For $N = 2 \times 10^4$, the relatively weak dipolar interaction allows strong density redistribution, and the number of maxima is not conserved.  
For $N = 4 \times 10^4$, the number of droplets remains fixed, although their shape evolves significantly. The droplet positions exhibit approximately periodic oscillations with noticeable damping.  

\section{interwell distance $2d > 3\,\mu\mathrm{m}$ for $\varepsilon_{dd} = 1.5$}
\label{sec:largerd}

We now return to $\varepsilon_{dd} = 1.5$ and consider larger initial separations, corresponding to higher excess potential energy after barrier removal.  

For $2d = 5\,\mu\mathrm{m}$, the maximum atom number for which two droplets merge increases to $\sim 1.65 \times 10^4$, nearly twice the value for $2d = 3\,\mu\mathrm{m}$.  
For larger $N$, the droplets no longer merge but instead undergo oscillatory motion.  The oscillatory motion for $N=2\times 10^4$ is characterized by a lifetime of $\tau=125$ ms, the amplitude of $c=0.95\;\mu$m and a period of $T=10.5$ ms (see Eq.(13)). As compared to the case of   $2d = 3\,\mu\mathrm{m}$ for $N=2\times 10^4$ the lifetime of the oscillations is reduced by a factor of 6.2, which results
from a much stronger deformation of the droplets that move deeper inside the potential barrier using the excess energy. Consistent with the higher energy, the amplitude of the oscillations gets larger by a factor of 2.3 and in qualitative agreement with the above discussion, the period of the oscillations is increased, by a factor of 1.3.

\begin{figure}[htbp]
\begin{tabular}{lll}
\includegraphics[width=0.16\textwidth, trim={1.3cm 3cm 0 0}, clip]{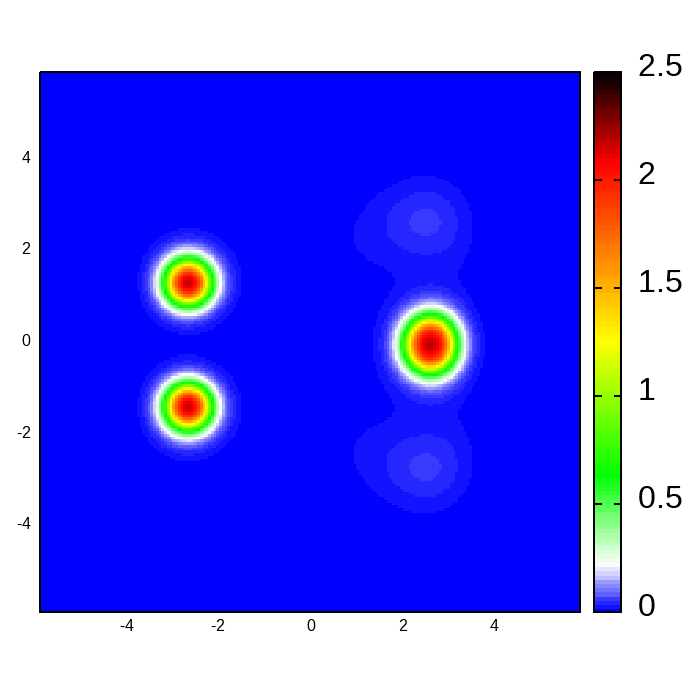}\put(-25,5){\color{yellow}(a)}&
\includegraphics[width=0.16\textwidth, trim={1.3cm 3cm 0 0}, clip]{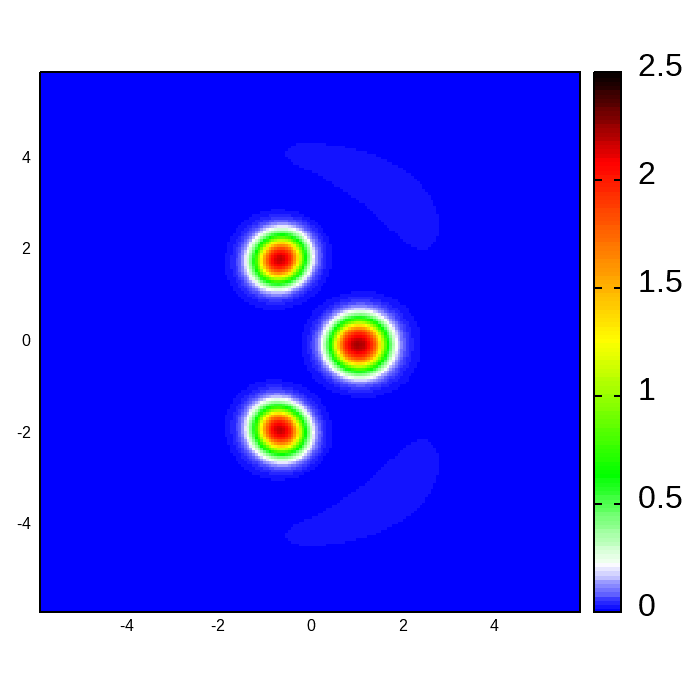}\put(-25,5){\color{yellow}(b)}&
\includegraphics[width=0.16\textwidth, trim={1.3cm 3cm 0 0}, clip]{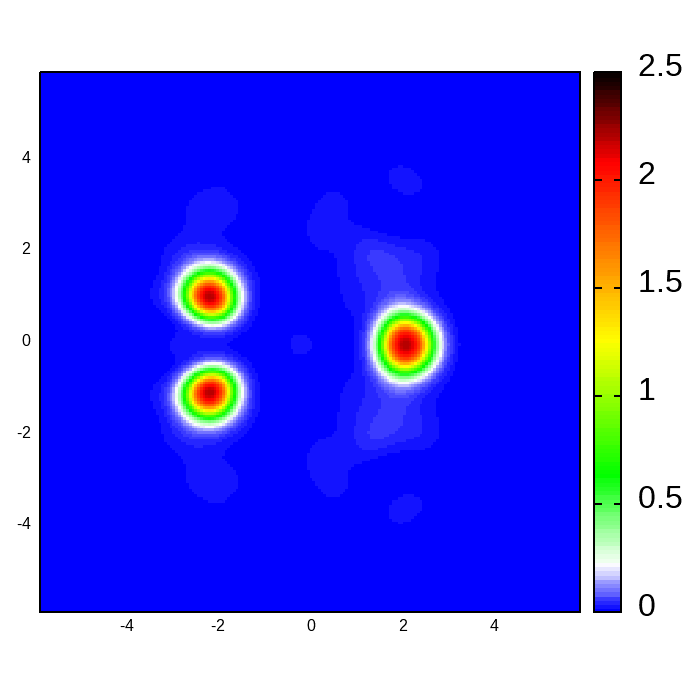}\put(-25,5){\color{yellow}(c)} \\
\includegraphics[width=0.16\textwidth, trim={1.3cm 3cm 0 0}, clip]{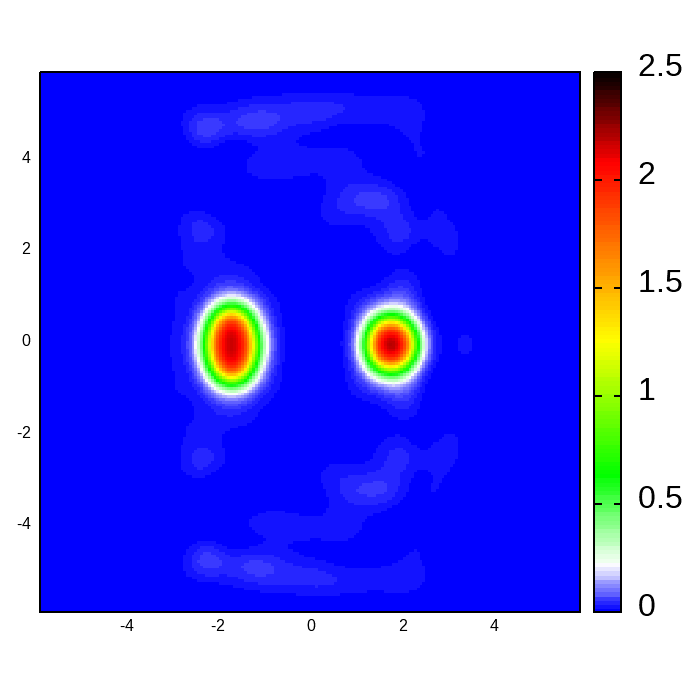}\put(-25,5){\color{yellow}(d)}&
\includegraphics[width=0.16\textwidth, trim={1.3cm 3cm 0 0}, clip]{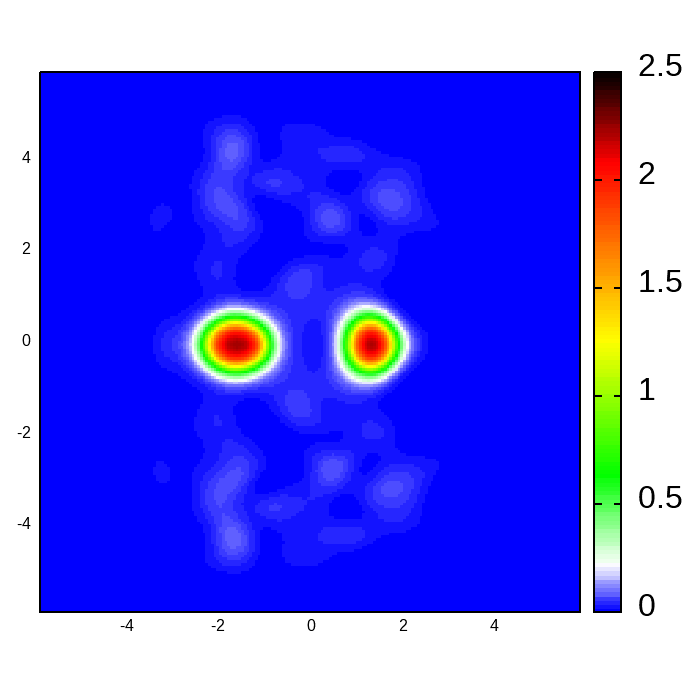}\put(-25,5){\color{yellow}(e)}&
\includegraphics[width=0.16\textwidth, trim={1.3cm 3cm 0 0}, clip]{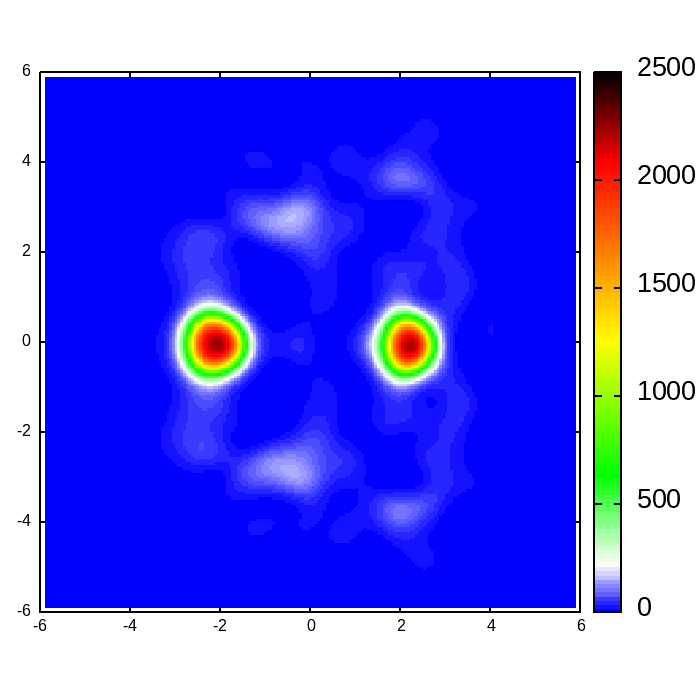}\put(-25,5){\color{yellow}(f)} \end{tabular}
\begin{tabular}{ll}
\includegraphics[width=0.24\textwidth]{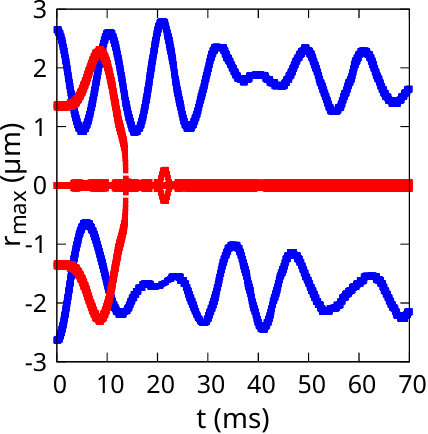}\put(-25,25){\color{black}(g)} &
\includegraphics[width=0.24\textwidth]{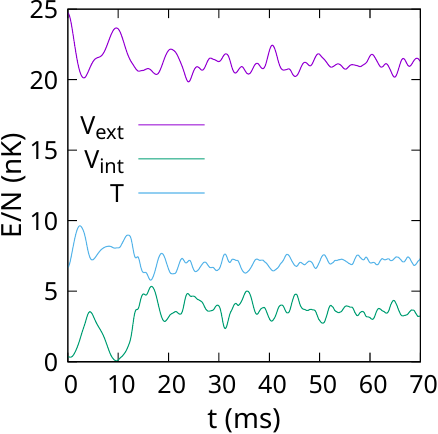} \put(-25,25){\color{black}(h)}\\
\end{tabular}
\caption{(a-f)
Snapshots of the condensate density at $z=0$ for $2d=5\mu$m, $\varepsilon_{dd}=1.5$ and $N=3\times 10^4$. The color scale  is plotted in $1000/\mu$m$^3$ units at selected moments in time 0 (a), 6.14 ms (b), 12.29 ms (c), 18.38 ms (d), 24.57 ms (e) and  30.72 ms (f). The plots cover the interval of [$-5.9\mu$m,$5.9\mu$m] in both directions.  (g) Positions of the maxima of the condensate density in $x$ (blue) and $y$ (red). (h) Contributions to the energy. }\label{2k5_3e4}
\end{figure}

For asymmetric $1+2$ configurations (Fig.~\ref{2k5_3e4}), the two droplets on one side merge after the first collision with the solitary droplet, at $t \sim 14$ ms.  
In this case, the interaction energy exceeds its initial value after merging, in contrast to the lower-energy regime.  

For symmetric $2 \times 2$ configurations, the dynamics differ from the $2d = 3\,\mu\mathrm{m}$ case.  
At $2d = 5\,\mu\mathrm{m}$, oscillations along $y$ become more pronounced, and the oscillation periods in $x$ and $y$ become more similar, indicating stronger coupling between directions.  

For $2d = 7\,\mu\mathrm{m}$, similar behavior is observed.  
However, for $1+1$ configurations with $N = 2 \times 10^4$ (Fig.~\ref{3k5_2e4}), a transient giant droplet forms at $t \sim 5$ ms and subsequently decays into two droplets aligned along the $y$ axis.  
This reorientation reflects internal dynamics of the excited droplet and contrasts with the decay along the $x$ axis observed for smaller separations.

\begin{figure}[htbp]
\begin{tabular}{ll}
\includegraphics[width=0.25\textwidth]{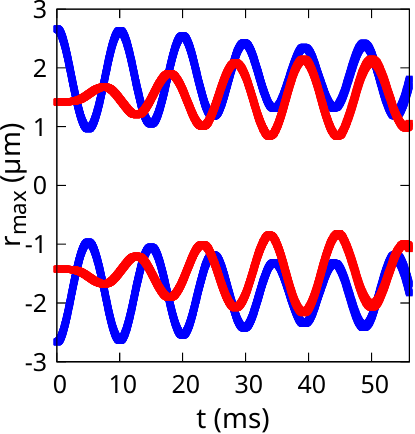} \put(-25,70){(a)}&
\includegraphics[width=0.25\textwidth]{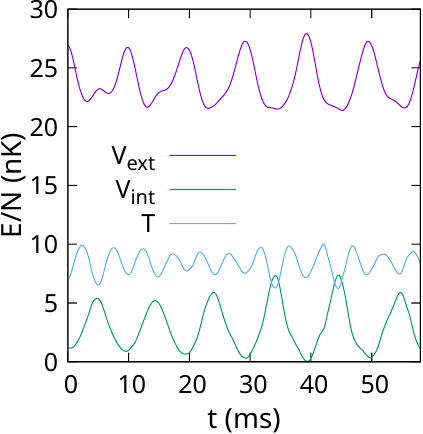} \put(-25,70){(b)}\end{tabular} 
\includegraphics[width=0.4\textwidth]{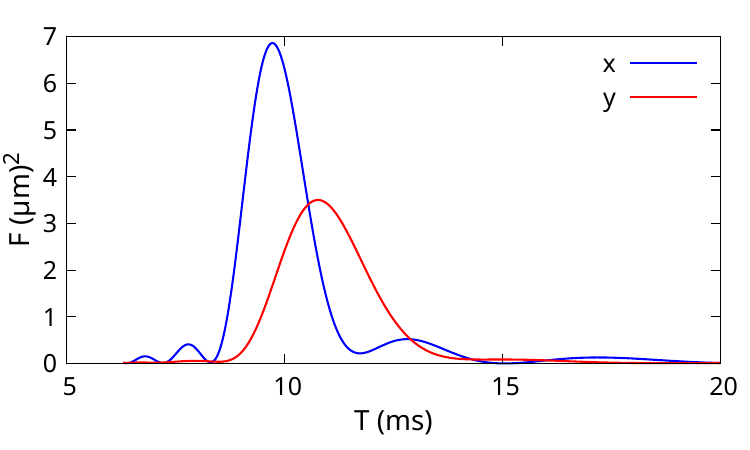} \put(-25,50){(c)}
\caption{Evolution of the system of 2x2 droplets for $\varepsilon_{dd}=1.5$ and $2d=5\mu$m and $N=4\times 10^4$. (a) The positions
of the centers of the droplets in $x$ (blue) and $y$ (red) directions.
(b) Contributions to the total energy. (c) Fourier transform of $x$ and $y$ positions of the centres of one of the droplets. }\label{2k5_4e4}
\end{figure}

\begin{figure}[htbp]
\begin{tabular}{lll}
\includegraphics[width=0.16\textwidth, trim={1.3cm 3cm 0 0}, clip]{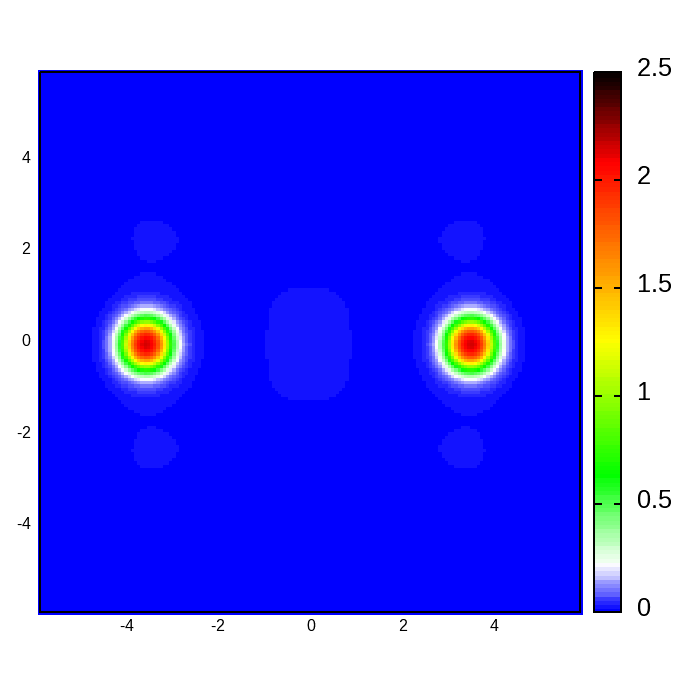}\put(-25,5){\color{yellow}(a)}&
\includegraphics[width=0.16\textwidth, trim={1.3cm 3cm 0 0}, clip]{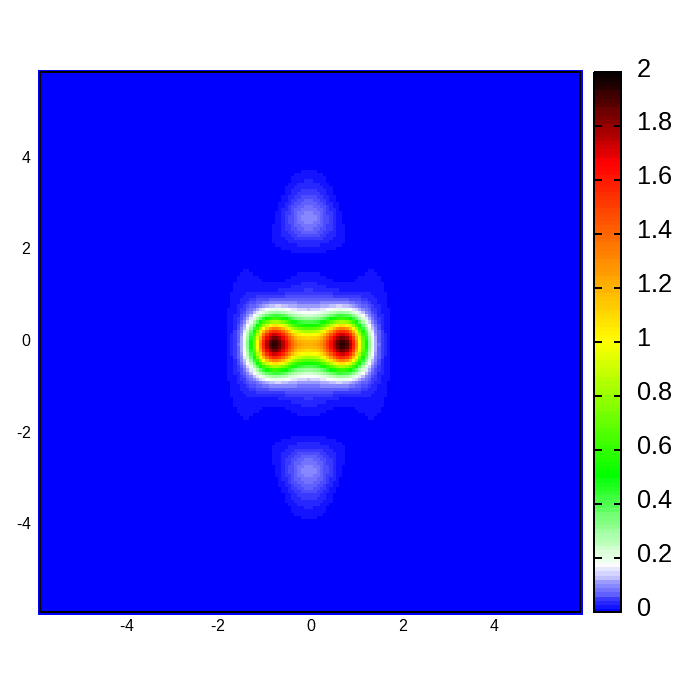}\put(-25,5){\color{yellow}(b)}&
\includegraphics[width=0.16\textwidth, trim={1.3cm 3cm 0 0}, clip]{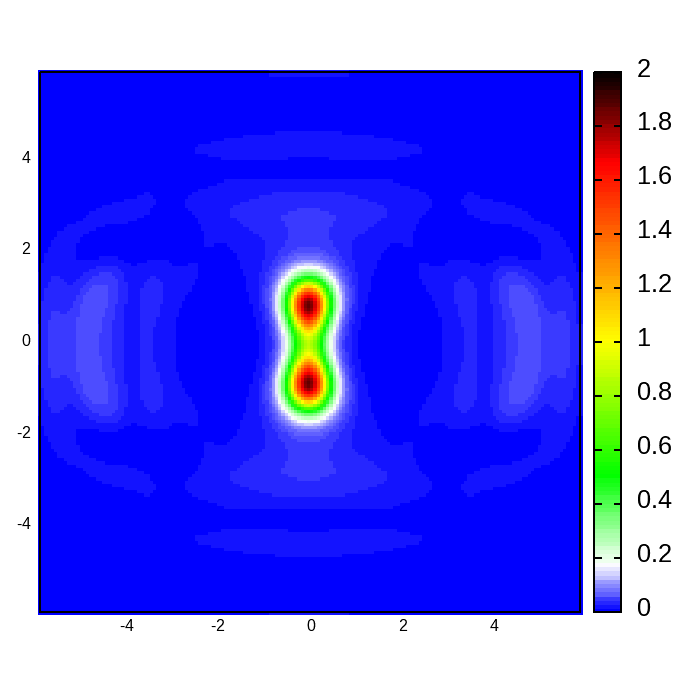}\put(-25,5){\color{yellow}(c)} \\
\includegraphics[width=0.16\textwidth, trim={1.3cm 3cm 0 0}, clip]{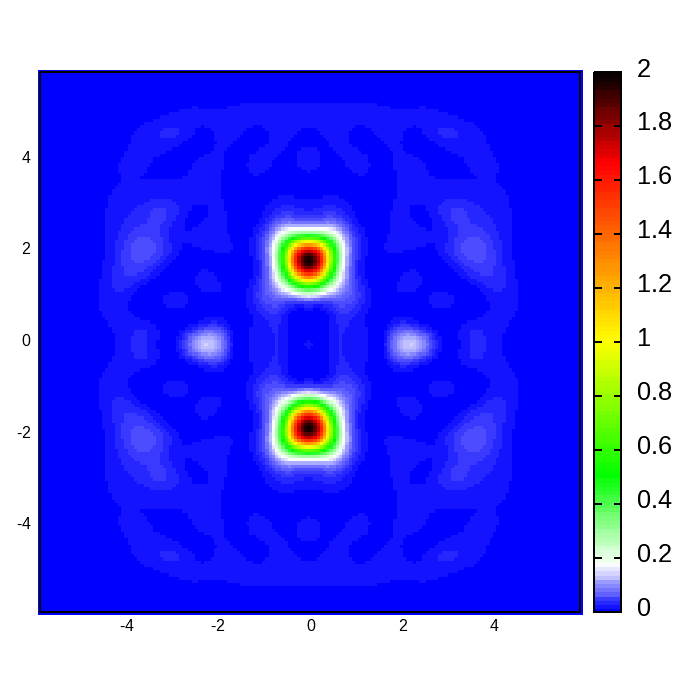}\put(-25,5){\color{yellow}(d)}&
\includegraphics[width=0.16\textwidth, trim={1.3cm 3cm 0 0}, clip]{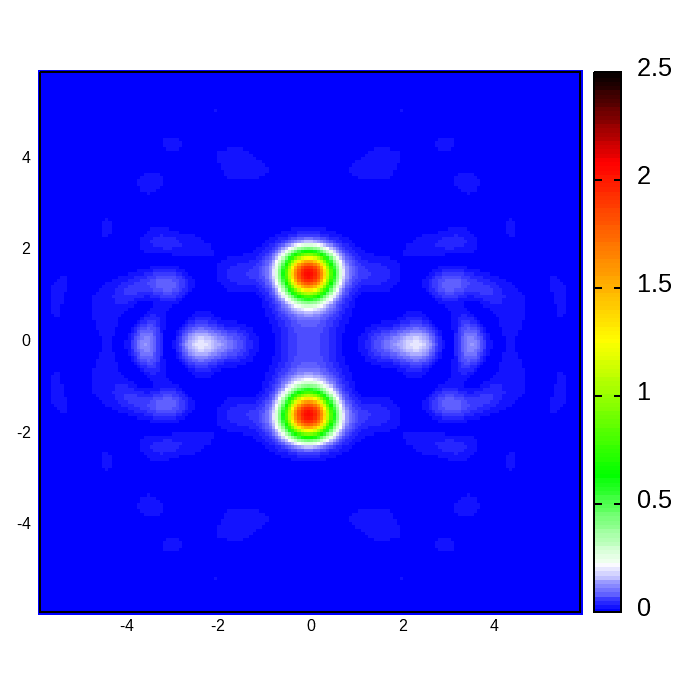}\put(-25,5){\color{yellow}(e)}&
\includegraphics[width=0.16\textwidth, trim={1.3cm 3cm 0 0}, clip]{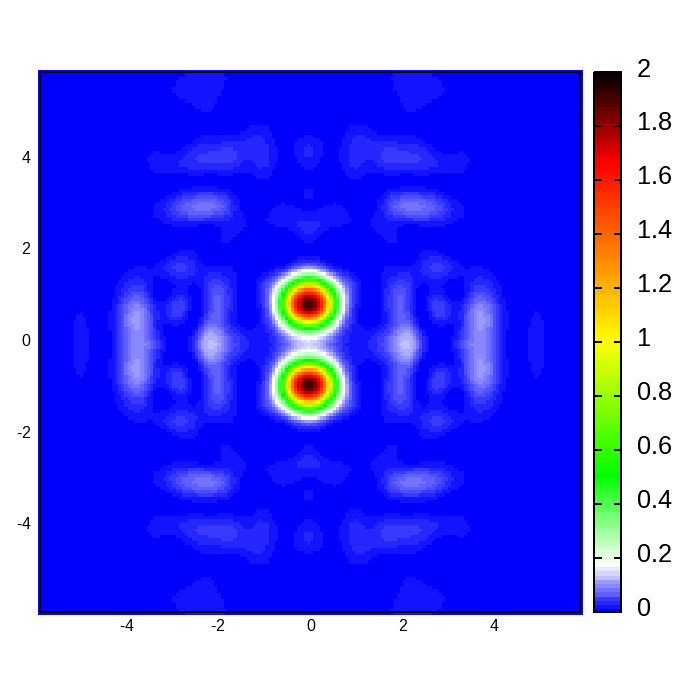}\put(-25,5){\color{yellow}(f)} \end{tabular}
\begin{tabular}{ll}
\includegraphics[width=0.25\textwidth]{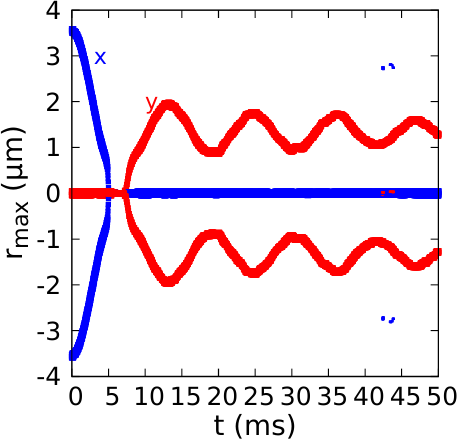} \put(-25,30){(g)}&
\includegraphics[width=0.25\textwidth]{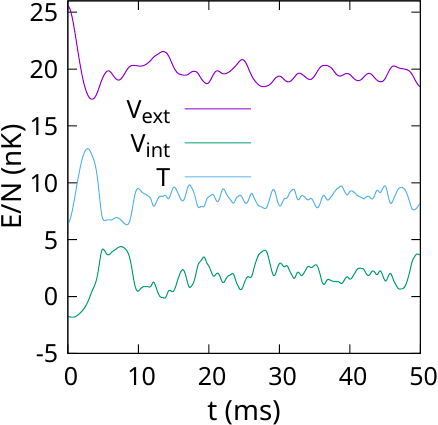} \put(-25,30){(h)}\end{tabular} 

\caption{Evolution of the system of two droplets for $\varepsilon_{dd}=1.5$ and $2d=7\;\mu$m and $N=2\times 10^4$. (a-f) Condensate density at $z=0$ given by the colorscale in $1000\;\mu\,$m$^3$ units for $t=0$ (a), 4.84 ms (b), 8.47 ms (c),  12.09 ms (d), 15.82 ms (e), 19.35 ms (f). (g) The positions
of the centers of the droplets in $x$ (blue) and $y$ (red) directions.
(h) Contributions to the total energy. }\label{3k5_2e4}
\end{figure}

\section{small asymmetry}
\label{sec:tilt}

We introduce a slight asymmetry into the system by implementing a linear potential in the $x$ direction,
\[
V_{x,\mathrm{lin}} = C \cdot x,
\]
with $C = \frac{0.25\,\mathrm{nK}}{d}$ where $2d$ represents the distance between the potential well minima.
The resulting small potential shift of 0.5 nK between the wells is likely to be present in the experiments.
The linear potential is kept after removal of the interwell barrier for evaluation of the time-dependent dynamics.

\subsection{Phase diagram}

\begin{figure*}[tbp]
%trim=left botm right top
\centering
\begin{tabular}{ll}
\includegraphics[height=0.35\textwidth]{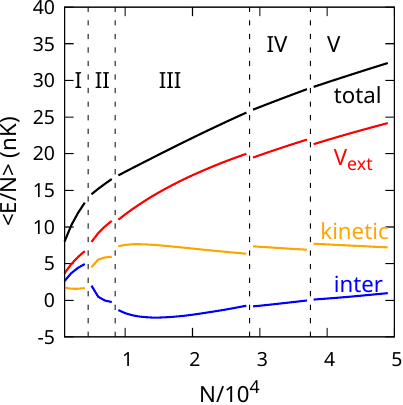} \put(-25,35){(a)}
\includegraphics[height=0.35\textwidth]{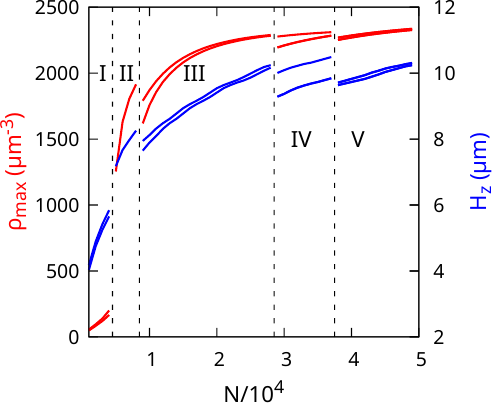} \put(-50,35){(b)} 
\end{tabular}
\begin{tabular}{lllll}
        \includegraphics[width = 0.15\linewidth, trim={2.5cm 0cm 0cm 0cm}, clip]{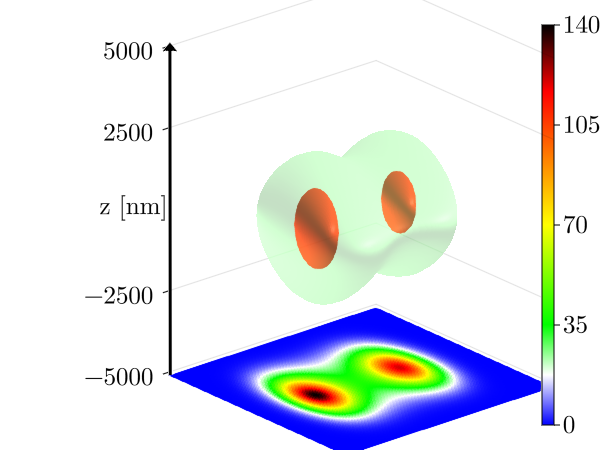} \put(-20.5, 1){(c)}&
        \includegraphics[width = 0.15\linewidth, trim={3.2cm 0cm 0cm 0cm}, clip]{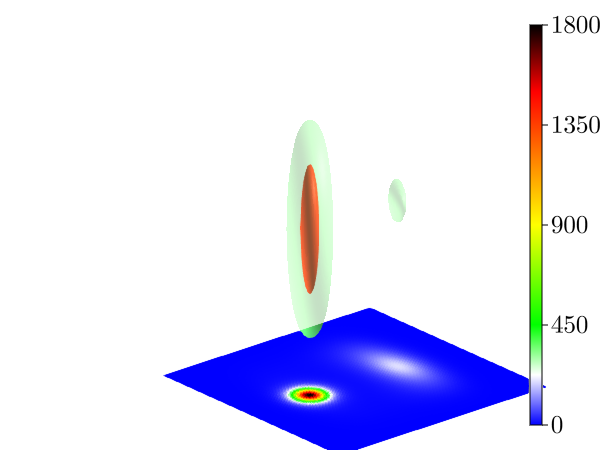} \put(-20.5, 1){(d)}&
        \includegraphics[width = 0.15\linewidth, trim={3.2cm 0cm 0cm 0cm}, clip]{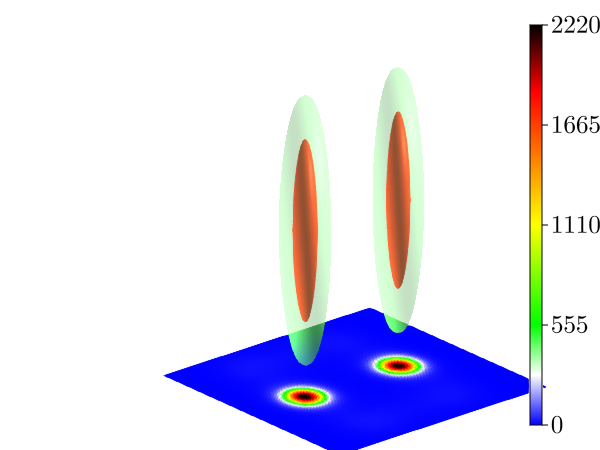} \put(-20.5, 1){(e)}&
        \includegraphics[width = 0.15\linewidth, trim={3.2cm 0cm 0cm 0cm}, clip]{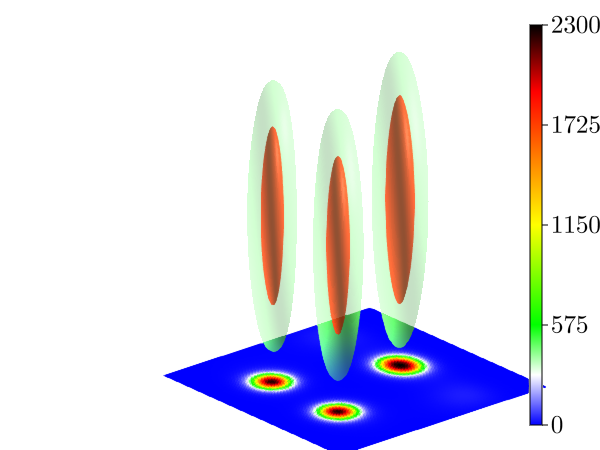} \put(-20.5, 1){(f)}&
        \includegraphics[width = 0.15\linewidth, trim={3.2cm 0cm 0cm 0cm}, clip]{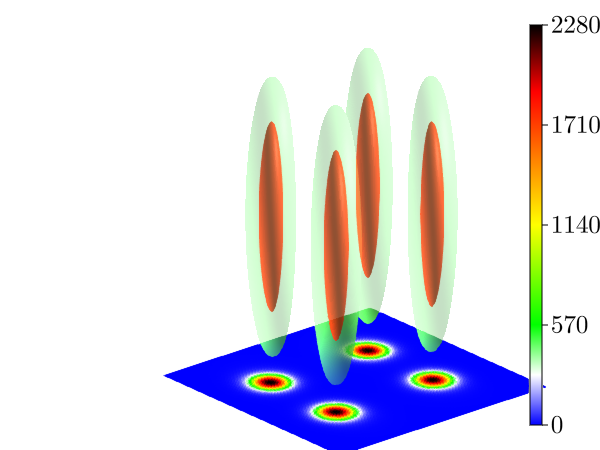} \put(-20.5, 1){(g)}
\end{tabular}
\caption{
        (a) The  contributions to the total energy for the ground-state at $\varepsilon_{dd}=1.5$ as a function of $N$ for the double-well  potential with minima spaced by $2d=3 \,\mu$m and additionally applied linear potential;
        (b) maximal atom density at $z=0$ plane (red lines) and droplet height $H_z$ in $z$ direction (blue lines).
        The vertical lines mark the ranges of a varied localization type:
        I -- a delocalized system (no droplet formation), II -- a system with a fully developed droplet in one of the wells, III -- a system with droplets in both wells, IV --  two droplets in one well and one in the other, V -- a system with 4 droplets. 
        (c-h) Isosurfaces  of the BEC density for $N=3\times10^3$ (c), $7\times 10^3$ (d), $2\times 10^4$ (e), $3.3\times 10^4$ (f) and $4\times 10^4$ (g) in the lowest-energy configuration.
        The isosurfaces correspond to 20\% (transparent) and 80\% (opaque) of the maximum density value.
        The plots span the area of $x$ (horizontal direction) and $y$ from -3.6 $\mu$m to 3.6 $\mu$m.
        At the base of the plots, below the isosurfaces, a cross-section of the BEC density at $z=0$ is presented.
        The colorscale for the density is given in $\mu$m$^{-3}$ units.
        The plots correspond to regions marked by I (c), II (d), III (e),  IV (f) and V (g).
}
\label{dfaz_asymmetric_edd15}
\end{figure*}

Figure~\ref{dfaz_asymmetric_edd15} shows the phase diagram of a system with a slight asymmetry.
It does not differ significantly from the phase diagram of a system with a fully symmetric potential [Fig. \ref{difaz1k5_1500}].
The asymmetric states, which also appear in the symmetric potential, are more frequently energetically favorable.
Moreover, a smaller number of atoms suffices for the system to condense into a single droplet.

The droplets in configurations that are symmetric in the symmetric (labeled as III and V in Fig. \ref{dfaz_asymmetric_edd15}) system are now slightly asymmetric.
The masses of droplets in the symmetric configurations differ by as little as tenths of a percent (e.g., for $N = 2.8\times10^4$) and by as much as about 10\% (the case $N = 9\times10^3$).
The heights of the droplets, on the other hand, differ by at most about 4\% (the case $N = 9\times10^3$).

\subsection{Dynamical evolution}

\begin{figure*}[htp]
    \centering
\begin{tabular}{l  l}
    \includegraphics[width=0.35\linewidth]{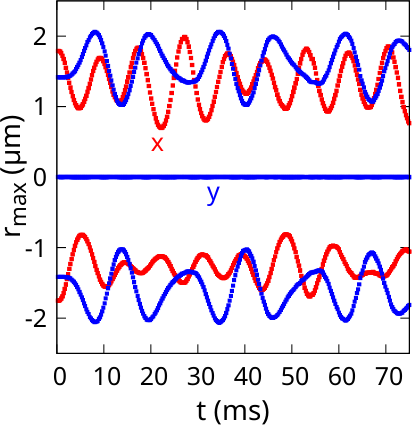}\put(-20,35){\color{black}(a)}
    \includegraphics[width=0.35\linewidth]{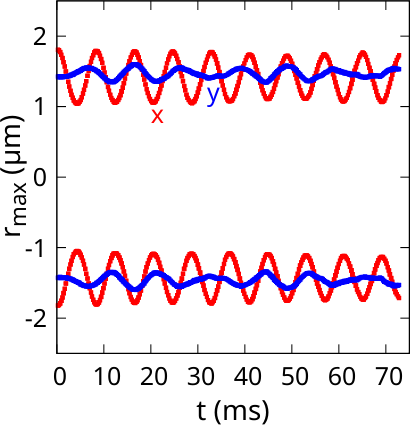}\put(-20,35){\color{black}(b)}&
\end{tabular}
    \caption{Trajectory of droplets within an asymmetrical potential after removal of interwell barrier for (a) $N = 3\times10^4$, (b) $N = 4\times10^4$.}
    \label{asy_time}
\end{figure*}

Figure~\ref{asy_time} shows the time evolution of the droplet density after the central barrier is suddenly removed.
Compared with the static differences discussed in the previous section, the dynamical response of the two wells is remarkably similar [see Fig. \ref{3e41k51k5} and \ref{4e41k51k5}] as the initial separation of the droplets along the $x$ axis is essentially unchanged.
In the near-symmetric configuration (two droplets per well, but with slightly different densities and heights), the left-hand droplets settle at $x \approx -1821\,\mathrm{nm}$ and the right-hand ones at $x \approx 1805\,\mathrm{nm}$, corresponding to an asymmetry of only about $1\%$ in their distance from the center of the trap.


\clearpage
\begin{thebibliography}{00}

\bibitem{naturedroplet} M. Schmitt, M. Wenzel, F. B\"ottcher, I. Ferrier-Barbut, and T. Pfau, 
"Self-bound droplets of a dilute magnetic quantum liquid", 
Nature {\bf 539}, 259 (2016) .

\bibitem{prxdroplet} L. Chomaz, S. Baier, D. Petter, M. J. Mark, F. W\"achtler, L. Santos, and F. Ferlaino,
"Quantum-Fluctuation-Driven Crossover from a Dilute Bose-Einstein Condensate to a Macrodroplet in a Dipolar Quantum Fluid",
Phys. Rev. X {\bf 6}  041039 (2016).

\bibitem{silamu}
Fabian Bottcher, Jan-Niklas Schmidt, Matthias Wenzel, Jens Hertkorn, Mingyang Guo, Tim Langen, and Tilman Pfau,
"Transient Supersolid Properties in an Array of Dipolar Quantum Droplets",
Phys. Rev. X {\bf 9}, 011051 (2019).%– Published 22 March, 2019

\bibitem{yahoo} Y. Cau, M. Rosenkranz, Z. Lei and W. Bao,
"Mean-field regime of trapped dipolar Bose-Einstein condensates in one and two dimensions"
Phys. Rev. A {\bf 82}, 043623 (2010).

\bibitem{santos} 
F. W\"achtler and L. Santos,
"Ground-state properties and elementary excitations of quantum droplets in dipolar Bose-Einstein condensates",
Phys. Rev. A {\bf 94}, 043618 (2016).

\bibitem{prldroplet} I. Ferrier-Barbut, H. Kadau, M. Schmitt, M. Wenzel, and Tilman Pfau,
"Observation of Quantum Droplets in a Strongly Dipolar Bose Gas",
Phys. Rev. Lett. {\bf 116}, 215301 (2016).
\bibitem{rosendroplet} H. Kadau, M. Schmitt, N. Wenzel, C. Wink, T. Maier, I. Ferrier-Barbut, and T. Pfau, 
"Observing the Rosensweig instability of a quantum ferrofluid",
Nature {\bf 530}, 194 (2016).

\bibitem{santos2} 
F. W\"achtler and L. Santos,
"Quantum filaments in dipolar Bose-Einstein condensates",
 Phys. Rev. A {\bf 93}, 061603(R) (2016).
 
% w tej drugiej pracy są filary chyba jako pierwsze pokazane
\bibitem{multidroplets} 
M.  Wenzel, F. B\"ottcher, T. Langen, I. Ferrier-Barbut, and T. Pfau, ''Striped states in a many-body system of tilted dipoles''
Phys. Rev. A {\bf 97}, 05360 (2017).

\bibitem{striped}
L. Tanzi, E. Lucioni, F. Fama, J. Catani, A. Fioretti, C. Gabbanini, R.N. Bisset, L. Santos, and G. Modugno,
"Observation of a Dipolar Quantum Gas with Metastable Supersolid Properties",
Phys. Rev. Lett. {\bf 122}, 130405 (2019).

\bibitem{striped2}
M. Wenzel, F. Boettcher, Tim Langen, Igor Ferrier-Barbut, and Tilman Pfau, 
"Striped states in a many-body system of tilted dipoles",
Phys. Rev. A {\bf 96}, 053630 (2017).

\bibitem{dropletcrystal}
D. Baillie and P. B. Blakie,
"Droplet Crystal Ground States of a Dipolar Bose Gas",
Phys. Rev. Lett. {\bf 121}, 195301 (2018).

\bibitem{dropletcr2}
J. Sanchez-Baena, R. Bombin, and J. Boronat,
"Ring solids and supersolids in spherical shell-shaped dipolar Bose-Einstein condensates",
Phys. Rev. Res. {\bf 6}, 033116 (2024).

\bibitem{dropletcr3}
Hari Sadhan Ghosh, Soumyadeep Halder, Subrata Das, and Sonjoy Majumder, 
"Induced supersolidity and hypersonic flow of a dipolar Bose-Einstein condensate in a rotating bubble trap",
Phys. Rev. A {\bf 110}, 033322 (2024).

\bibitem{dr0}
E. Poli, T. Bland, C. Politi, L. Klaus, M. A. Norcia, F. Ferlaino, R. N. Bisset, and L. Santos,
"Maintaining supersolidity in one and two dimensions",
Phys. Rev. A {\bf 104} 063307 (2021).

\bibitem{drm1}
M.A. Norcia, C. Politi,  L. Klaus, et al,
"Two-dimensional supersolidity in a dipolar quantum gas",
Nature {\bf 596}, 357–361 (2021). 
\bibitem{dropletcr4}
Luis E. Young-S. and S.K. Adhikari,
"Mini droplet, mega droplet and stripe formation in a dipolar condensate",
Physica D {\bf 455}, 133910 (2023).
 
\bibitem{drc1}
D. Baillie, R. M. Wilson,  R. N. Bisset,  and P. B. Blakie,
"Self-bound dipolar droplet: A localized matter wave in free space",
Phys. Rev. A {\bf 94}, 021603(R) (2016).

\bibitem{drcr4} 
Matthias Schmidt, Lucas Lassablière, Goulven Quéméner, and Tim Langen
"Self-bound dipolar droplets and supersolids in molecular Bose-Einstein condensates",
Phys. Rev. Res. {\bf 4}, 013235 (2022).
%PHYSICAL REVIEW RESEARCH 4, 013235 (2022)

 \bibitem{general}
 R. N. Bisset, R. M. Wilson, D. Baillie, and P. B. Blakie,
% PHYSICAL REVIEW A 94, 033619 (2016)
"Ground-state phase diagram of a dipolar condensate with quantum fluctuations",
Phys. Rev. A {\bf 94},  033619 (2016).

\bibitem{supersolid2}
T. Bland, E. Poli, C. Politi, L. Klaus, M.A. Norcia, F. Ferlaino, L. Santos, and R.N. Bisset, 
"Two-Dimensional Supersolid Formation in Dipolar Condensates",
Phys. Rev. Lett. {\bf 128}, 195302 (2022).

\bibitem{supersolidlate} A. Alana, I.L. Eguisquiza, and M. Modugno,
"Supersolid formation in a dipolar condensate by roton instability",
Phys. Rev. A {\bf 108}, 033316 (2023).
%PHYSICAL REVIEW A 108, 033316 (2023).
%https://doi.org/10.1038/s41586-021-03725-7

\bibitem{supersolid}
L. Chomaz et al.,
"Long-Lived and Transient Supersolid Behaviors in Dipolar Quantum Gases"
Phys. Rev. X {\bf 9}, 021012 (2019).

\bibitem{review}
L. Chomaz, I. Ferrier-Barbut, F. Ferlaino, B.  Laburthe-Tolra, B. L. Lev and T. Pfau,
"Dipolar physics: a review of experiments with magnetic quantum gases",
Rep. Prog. Phys. {\bf 86}, 026401 (2023).

\bibitem{review2}
K. Mukherjee, T. Arnone Cardinale, L. Chergui, P. St\"urmer and S. M. Reimann
"Droplets and supersolids in ultra-cold atomic quantum gases",
Eur. Phys. J. Spec. Top. {\bf 232}, 3417 (2023).

\bibitem{review3}
Fabian Böttcher, Jan-Niklas Schmidt, Jens Hertkorn, Kevin S H Ng, Sean D Graham, Mingyang Guo, Tim Langen, and Tilman Pfau,
"New states of matter with fine-tuned interactions: quantum droplets and dipolar supersolids",
Rep. Prog. Phys. {\bf 84}, 012403 (2021).

\bibitem{merging_1d}
{
G. A. Bougas, T. Bland, H. R. Sadeghpour, S. I. Mistakidis,
"Signatures of rigidity and second sound in dipolar supersolids",
Phys. Rev. A {\bf 113}, L041305 (2026).
}

\bibitem{gauss_sweep}
{
E. L. Brakensiek, G. A. Bougas, S. I. Mistakidis
"Generation of dipolar supersolids through a barrier sweep in droplet lattices",
arXiv:2603.29203 (2026).
}
\bibitem{164Dy}
M. Lu,  Nathaniel Q. Burdick, Seo Ho Youn,  and Benjamin L. Lev,
"Strongly Dipolar Bose-Einstein Condensate of Dysprosium",
Phys. Rev. Lett. {\bf 107}, 190401 (2011).
\bibitem{death}
Maximilian Sohmen, Claudia Politi, Lauritz Klaus, Lauriane Chomaz, Manfred J. Mark, Matthew A. Norcia, and Francesca Ferlaino,
"Birth, Life, and Death of a Dipolar Supersolid",
Phys. Rev. Lett. {\bf 126} 233401 (2021).

\bibitem{merging}
H. Li, E. Halperin, S. Ronen, and J.L. Bohn,
"Merging dipolar supersolids in a double-well potential",
Phys. Rev. A 109, 013307 (2024).
%https://journals.aps.org/pra/pdf/10.1103/PhysRevA.109.013307

\bibitem{multicond} 
Pranay Nayak, Ratheejit Ghosh, and Rejish Nath
"Density engineering via intercondensate dipole-dipole interactions",
Phys. Rev. A {\bf 110}, 053319 (2024).
%https://journals.aps.org/pra/pdf/10.1103/PhysRevA.110.053319

\bibitem{oscila2} K. Mukherjee and S. M. Reimann,
"Classical-linear-chain behavior from dipolar droplets to supersolids",
Phys. Rev. A {\bf 107}, 043319 (2023).

\bibitem{forced} 
L. Tanzi, S. M. Roccuzzo, E. Lucioni, F. Famà, A. Fioretti, C. Gabbanini, G. Modugno, A. Recati, and  S. Stringari, 
"Supersolid symmetry breaking from compressional oscillations in a dipolar quantum gas",
Nature {\bf 574}, 382 (2019).

\bibitem{oscila}
Luis E. Young-S. and S. K. Adhikari,
"Dipole-mode and scissors-mode oscillations of a dipolar supersolid",
Phys. Rev.  A {\bf 107}, 053318 (2023).



\bibitem{dropletsall}
S. I. Mistakidis, K. Mukherjee, S. M. Reimann, and H. R. Sadeghpour,
"Tunneling dynamics of Dy164 supersolids and droplets", 
Phys. Rev. A {\bf 110}, 013323 (2024).

\bibitem{chiny} Q. Zhu and C. Kong,
"Fundamental and dipole-mode quantum droplets in one-dimensional
dipolar Bose-Einstein condensates",
Phys. Lett. A {\bf 536}, 130291 (2025).

\bibitem{spania}
G. E. Astrakharchik  and B. A. Malomed,
"Dynamics of one-dimensional quantum droplets",
Phys. Rev. A {\bf 98}, 013631 (2018).

\bibitem{dropmix1}
C. R. Cabrera, L. Tanzi, J. Sanz, B. Naylor, P. Thomas, P. Cheiney, and L. Tarruell,
"Quantum liquid droplets in a mixture of Bose-Einstein condensates",
Science {\bf 359}, 301 (2018).

\bibitem{dropmix2} 
P. Cheiney, C. R. Cabrera, J. Sanz, B. Naylor, L. Tanzi, and L. Tarruell, 
"Bright Soliton to Quantum Droplet Transition in a Mixture of Bose-Einstein Condensates"
Phys. Rev. Lett. {\bf 120}, 135301 (2018).

\bibitem{dropmix3} 
G. Semeghini, G. Ferioli, L. Masi, C. Mazzinghi, L. Wolswijk, F. Minardi, M. Modugno, G. Modugno, M. Inguscio, and M. Fattori,
"Self-Bound Quantum Droplets of Atomic Mixtures in Free Space",
Phys. Rev. Lett. {\bf 120}, 235301 (2018).

\bibitem{dropzd}
G. Ferioli, G. Semeghini, L. Masi, G. Giusti,  G. Modugno, M. Inguscio,  A. Gallemí,  A. Recati, and M. Fattori, 
"Collisions of Self-Bound Quantum Droplets",
Phys. Rev. Lett. {\bf 122}, 090401 (2019).

\bibitem{dropzd2}
M. Pylak, F. Gampel, M. Płodzien, and M. Gajda,
"Manifestation of relative phase in dynamics of two interacting Bose-Bose droplets",
Phys. Rev. Res. {\bf 4}, 013168 (2022).

\bibitem{lhy1}
T. D. Lee, Kerson Huang and C. N. Yang,
"Eigenvalues and Eigenfunctions of a Bose System of Hard Spheres and Its Low-Temperature Properties",
Phys. Rev. {\bf 106}, 1135  (1957).

\bibitem{lhy0}
R. Schuetzhold, M. Uhlmann, Y. Xu, and U.R. Fischer,
"Mean Field expansion in Bose-Einstein condensates with finite-range interactions",
Int. J. Modern. Phys. B {\bf 20}, 3555 (2006).

\bibitem{lhy2}
Aristeu R. P. Lima and Axel Pelster,
"Quantum fluctuations in dipolar Bose gases"
Phys. Rev. A {\bf 84}, 041604(R) (2011).% – Published 14 October, 2011

\bibitem{prl25}
S. Da and V.W. Scarola,
"Unveiling Supersolid Order via Vortex Trajectory Correlations",
Phys. Rev. Lett. {\bf 134}, 163401 (2025).

\bibitem{fes1}
C. Chin, R. Grimm, P. Julienne, and E. Tiesinga,
"Feshbach resonances in ultracold
gases",
Rev. Mod. Phys. {\bf 82}, 1225 (2010).

\bibitem{fes2}
Y. Tang, A. Sykes, N.Q. Burdick, J.L. Bohn, and B.L. Lev,
"s-wave scattering lengths of the strongly dipolar bosons 162 Dy and 164 Dy",
Phys. Rev. A {\bf 92}, 022703 (2015).

\bibitem{fes3}
T Maier et al.,
"Emergence of chaotic scattering in ultracold Er and Dy",
Phys. Rev. X {\bf 5}, 041029 (2015).

\bibitem{github}
{ The numerical codes developed for the purpose of the present study are available on GitHub \href{https://github.com/OrlowskiWojtek/GPESolver}{https://github.com/OrlowskiWojtek/GPESolver}}.
 % ================ poniżej są nieużyte w tekście ================

\bibitem{datas}
The data that support the findings of this article are openly available: https://zenodo.org/records/19396221 


\end{thebibliography}
\end{document}